\documentclass[twocolumn,showpacs,pre,eqsecnum,citeautoscript,amsmath,amssymb,floatfix,superscriptaddress]{revtex4-1}
\usepackage{bm,xcolor,amsmath,amssymb,mathrsfs,latexsym,graphicx,psfrag,float}
\usepackage{dcolumn}
\usepackage{hyperref}

\usepackage{leftidx}

\usepackage{tabularx}

\usepackage[english]{babel}
\usepackage[utf8]{inputenc}
\usepackage{bbold}

\usepackage{tikz}
\usetikzlibrary{arrows,decorations.pathmorphing,backgrounds,positioning,fit,matrix}

\usepackage[normal]{subfigure}		

\newcommand{\lt}{\left(}
\newcommand{\rt}{\right)}
\newcommand{\lqq}{\left[}
\newcommand{\rqq}{\right]}
\newcommand{\lan}{\left\langle}
\newcommand{\ran}{\right\rangle}
\newcommand{\abs}[1]{\left| #1 \right|}
\newcommand{\eval}[1]{\left.\right|_{ #1 }}
\newcommand{\av}[1]{\lan #1 \ran}
\newcommand{\avb}[1]{\lan #1 \ran_\xi}

\newcommand{\trace}[1]{\text{tr}\left\{ #1 \right\}}
\newcommand{\matb}{\left(\begin{array}}
\newcommand{\mate}{\end{array}\right)}
\newcommand{\sysb}{\left\{\begin{array}}
\newcommand{\syse}{\end{array}\right.}

\newcommand{\wh}{\widehat}
\newcommand{\ha}{\frac{1}{2}}
\newcommand{\mal}{\mathcal}
\newcommand{\rmd}{{\rm{d}}}

\newcommand{\rme}[1]{{\rm{e}}^{#1}}

\newcommand{\uar}{\uparrow}
\newcommand{\dar}{\downarrow}

\newcommand{\ket}[1]{\left| #1 \ran}
\newcommand{\bra}[1]{\lan #1 \right|}

\newcommand{\proj}[1]{\ket{#1} \bra{#1}}
\newcommand{\projl}[2]{\ket{#1}_{#2} \bra{#1}_{#2}}
\newcommand{\comm}[2]{\left[ #1, #2 \right]}
\newcommand{\acomm}[2]{\left\{ #1, #2 \right\}}

\newcolumntype{L}[1]{>{\raggedright\arraybackslash}p{#1}} 
\newcolumntype{C}[1]{>{\centering\arraybackslash}p{#1}} 
\newcolumntype{R}[1]{>{\raggedleft\arraybackslash}p{#1}} 

\newcommand{\be}{\begin{equation}}
\newcommand{\ee}{\end{equation}}

\newcommand{\prodl}[2]{\prod\limits_{#1}^{#2}}
\newcommand{\suml}[2]{\sum\limits_{#1}^{#2}}

\newcommand{\changer}[1]{\textcolor{black}{#1}}

\newcommand{\comma}{\quad , \quad}
\newcommand{\mand}{\quad \text{and} \quad}

\newcommand{\nol}{\nonumber \\}

\newcommand{\eq}[2]{\begin{align}\label{#1} #2 \end{align}}

\newcommand{\auxn}{\text{nn}}

\newcommand{\si}{\wh{\sigma}}
\newcommand{\n}{\wh{n}}
\newcommand{\p}{\wh{p}}
\newcommand{\x}{\wh{\xi}}
\newcommand{\pp}{\Pi}
\newcommand{\PP}{\wh{\pp}}
\newcommand{\dd}{\wh{d}}
\newcommand{\bb}{\wh{b}}
\newcommand{\cc}{\wh{c}}

\newcommand{\R}{\mathbb{R}}

\newcommand{\freq}{\varpi}
\newcommand{\freqd}{\nu}
\newcommand{\wa}{\wh{a}}

\begin{document}

\title{Non-equilibrium effective field theory for absorbing state phase transitions in driven open quantum spin systems}

\author{Michael Buchhold}
\affiliation{Institut f\"ur Theoretische Physik, Universit\"at zu K\"oln, D-50937 Cologne, Germany}
\author{Benjamin Everest}
\affiliation{School of Physics and Astronomy, University of Nottingham, Nottingham, NG7 2RD, United Kingdom}
\affiliation{Centre for the Mathematics and Theoretical Physics of Quantum Non-Equilibrium Systems, University of Nottingham, Nottingham, NG7 2RD, UK}
\author{Matteo Marcuzzi}
\affiliation{School of Physics and Astronomy, University of Nottingham, Nottingham, NG7 2RD, United Kingdom}
\affiliation{Centre for the Mathematics and Theoretical Physics of Quantum Non-Equilibrium Systems, University of Nottingham, Nottingham, NG7 2RD, UK}
\author{Igor Lesanovsky}
\affiliation{School of Physics and Astronomy, University of Nottingham, Nottingham, NG7 2RD, United Kingdom}
\affiliation{Centre for the Mathematics and Theoretical Physics of Quantum Non-Equilibrium Systems, University of Nottingham, Nottingham, NG7 2RD, UK}
\author{Sebastian Diehl}
\affiliation{Institut f\"ur Theoretische Physik, Universit\"at zu K\"oln, D-50937 Cologne, Germany}

\date{\today}

\begin{abstract}

Phase transitions to absorbing states are among the simplest examples of critical phenomena out of equilibrium. The characteristic feature of these models is the presence of a fluctuationless configuration which the dynamics cannot leave, which has proved a rather stringent requirement in experiments. Recently, a proposal to seek such transitions in highly tuneable systems of cold atomic gases offers to probe this physics and, at the same time, to investigate the robustness of these transitions to quantum coherent effects. Here we specifically focus on the interplay between classical and quantum fluctuations in a simple driven open quantum model which, in the classical limit, reproduces a contact process, which is known to undergo a continuous transition in the ``directed percolation'' universality class. We derive an effective long-wavelength field theory for the present class of open spin systems and show that, due to quantum fluctuations, the nature of the transition changes from second to first order, passing through a bicritical point which appears to belong instead to the ``tricritical directed percolation'' class.

\end{abstract}





\maketitle
\section{Introduction}
The dynamics of many-body systems is typically too complex to admit a complete description. It is well-known, however, that for systems at thermal equilibrium time-averaged, macroscopic quantities (i.e., quantities which do not react to fluctuations on microscopic time- and length-scales) can be equivalently extracted from appropriate statistical ensembles \cite{Huang_book, Bowley_book}. Statistical mechanics provides a very powerful simplification which recasts all the relevant physics in terms of a few thermodynamic parameters and potentials independently of the initial state of the system, although one could envision cases in which some initial state information is kept due to an extensive amount of symmetries, and the ensembles would have to be generalized accordingly \cite{Jaynes1957, Jaynes1957_2, RigolGGE}. 

Equilibrium systems, however, are but a portion of what nature has in store. Despite significant efforts, a thorough, systematic understanding of non-equilibrium phenomena has yet to be developed. As in equilibrium, though, there are cases in which collective behaviors supersede the minute details of the microscopic dynamics, allowing their description in terms of few coarse-grained variables and rules. One example is given by cooperative relaxation at the onset of glassiness \cite{Ritort03, Keys2011} in which, e.g., it is not possible to change the local configuration of particles without an extensively growing number of rearrangements in the neighborhood taking place. Another relevant instance relies on the presence of continuous phase transitions \cite{Ma_book, QPT, Tauber-book, NEQ_PT1}. These are associated to a diverging length in the correlations of fluctuations \cite{Ma_book, Goldenfeld_book, Mussardo}. Hence, fluctuations encompass larger and larger portions of the system as the critical point is approached, so that they end up being governed only by general features which do not depend on the scale, such as dimensionality and symmetries. This idea lies at the basis of the concept of \emph{universality}; simply put, all systems sharing these scale-insensitive features will display quantitatively identical behavior at asymptotic distances, and studying one instance will provide information on all of them. It is therefore a relevant task to identify and study phase transitions as they provide a natural classification scheme.

In dynamical systems, a crucial distinction must be made depending on whether detailed balance conditions -- or the associated symmetry, \emph{microreversibility} \cite{Microrev1, Microrev2, Microrev3, Sieberer15} -- hold or not. In the former case, the system will evolve towards a stationary equilibrium state. Examples of this kind are systems subject to an external thermal bath, which have been extensively investigated and classified \cite{HH}. It is worth remarking that this symmetry might be absent from the microscopic description, but be effectively recovered under coarse-graining at long times and long wavelengths \cite{Sieberer2013, Eq_vs_NonEq, SiebererRev}. If this is not the case, the system will instead remain out of equilibrium even in the long-time limit, being typically described by \emph{flux equilibrium} states \cite{Racz2002}. Phase transitions in this regime will have no equilibrium counterpart but are genuinely non-equilibrium in nature \cite{NEQ_lattice, NEQ_PT1}. 

In classical physics, a paradigmatic class of systems displaying the latter kind of transitions is given by models with absorbing states \cite{DP_Hinrichsen}. These are stochastic processes whose dynamical rules are built in such a way that there is a configuration (or set thereof) which, once reached, cannot be left under the evolution (hence the term ``absorbing'', to be contrasted with the remainder of the phase space, dubbed ``transient''). The absorbing property of this subspace survives coarse-graining and thus prevents detailed balance from being recovered at any scale. The most characteristic universality class in this set of systems is probably \emph{directed percolation} (originally studied in \cite{DP_original}, see \cite{DP_Hinrichsen} for a review) which, for our present purposes, is more easily introduced via the so-called ``contact process'' (CP) \cite{CP_original, DP_Hinrichsen, Janssen2005}. The CP is defined on a lattice of classical Ising variables (either $\uar$ or $\dar$) with rules which mimic an epidemic spreading: an $\uar$ site is \emph{active} (sick) and can either decay to $\dar$ with a certain rate $\gamma$ or produce another active site in its neighborhood with another rate $\kappa$. Sites in the $\dar$ state are \emph{inactive} (healthy) and can only be activated (infected) via the aforementioned mechanism, which we will refer to in the following as \emph{branching}. This can be summarized as
\be
	\uar \stackrel{\gamma}{\longrightarrow}   \dar \quad \quad \uar \dar \stackrel{\kappa}{\longrightarrow} \uar \uar.
	\label{eq:updown}
\ee
As their name suggests, inactive sites do not produce any dynamics. The configuration where all sites are inactive thus cannot be left and constitutes the unique absorbing state of the model. Note that the two processes in \eqref{eq:updown} are competing: decay tends to deplete the system of $\uar$s, whereas branching tries to fill it up. In the thermodynamic limit, depending on the ratio $\kappa / \gamma$ between the rates the dynamics starting from an active configuration can end up in two distinct phases: for $\kappa \ll \gamma$ decay dominates and the system at long times invariably falls into the absorbing state. For $\kappa \gg \gamma$, instead, a finite density of active sites persists for arbitrarily long times and the dynamics survives in the transient portion of the phase space. Note that this is only strictly true in the thermodynamic limit: for any finite size, there is always a finite probability of a (rare) fluctuation trapping the system into its absorbing state. In the active phase, however, the time required for such a fluctuation to take place increases with the system size \cite{Lubeck2003}. The active and absorbing phases are separated by a critical point $\kappa_c / \gamma_c$, marking the directed percolation (DP) transition \cite{DP_Hinrichsen}, with the stationary density $n$ of active sites acting as an order parameter (i.e., $n = 0$ in the absorbing phase versus $n > 0$ in the active one).

The directed percolation class is conjectured \cite{DP_Janssen, Grassberger1982} to encompass all systems featuring a one-component order parameter, short-range interactions, no additional symmetries, and a unique, fluctuationless absorbing state. This last condition is crucial; the difficulty in having a perfectly fluctuation-free state in real systems has made it a challenge to identify experimental setups undergoing a phase transition in this class \cite{Hinrichsen_exp}. The first clear examples have only recently been highlighted in two-dimensional nematic liquid crystals \cite{DP_exp, DP_explong} and one- \cite{DP_exp_1D} and two-dimensional \cite{Sano2016} turbulent flows. \changer{In addition, a recent numerical study links DP to the onset of turbulence in quantum fluids (such as superfluids) \cite{Takahashi2017}.}
Upon relaxing the other assumptions, different transitions, alongside their universality classes, have been identified and investigated: for instance, the introduction of quenched spatial randomness \cite{DP_quench1, DP_quench2, DP_quench3} makes the DP critical point unstable (it constitutes a ``relevant'' perturbation in the renormalization group sense) and generates non-universal power laws; the presence of multiple absorbing states often leads to the appearance of discontinuous transitions \cite{Domany_Kinzel, Bassler1996}; other symmetries, such as preservation of the parity of active sites \cite{Jensen1993, Cardy_Tauber1996}, also change the critical properties, as does introducing long-range processes (L{\'e}vy flights) \cite{Janssen1999}. As in equilibrium systems, multicritical behavior can emerge when higher-order processes take over the simple ones in Eq.~\eqref{eq:updown} \cite{TriDP_FT1, TriDP_FT2}. 
A simple example studied in the literature is the so-called tricritical directed percolation \cite{Grassberger1982, Lubeck2006, Grassberger2006}, obtained e.g. by adding processes involving pairs such as $\uar \uar \dar \to \uar \uar \uar$ or $\uar \uar \to \dar \dar$. Depending on the relative rates of these processes compared to the ordinary DP ones, the transition may become first-order by crossing a bicritical point \footnote{The term ''tricritical directed percolation'' has been established in the literature. The critical point, however, separates two distinct stable phases of matter and is thus according to the statistical mechanics definition a bicritical point.}. 

Recently, a proposal \cite{Marcuzzi2015} has been made to realize DP with cold atomic gases excited to high-lying electronic orbitals (so-called \emph{Rydberg states} \cite{Gallagher84, Rydberg2}). This yields greatly enhanced dipolar or van-der-Waals mutual interactions \cite{Ryd-QI, Beguin2013}, which can easily produce strong correlations and in fact induce several examples of collective behaviors \cite{Carr2013, Urvoy2015, Rydberg1, AF-num1, Sibalic2016}. In particular, they make it possible to engineer a facilitation mechanism \cite{Ates07, Amthor2010, PRL-KinC}, where atoms lying at a certain distance from already-excited ones have a much higher probability of getting excited, thereby reproducing a branching process (rate $\kappa$ above) \cite{Lesanovsky14, Valado2015}. Spontaneous radiative decay provides the competing process (rate $\gamma$ above). A strong dephasing noise projects the dynamics onto an effective classical master equation \cite{Degenfeld14, Marcuzzi14}, although the microscopic dynamics is properly described by a quantum master equation. \emph{Quantum driven-dissipative systems} such as this one currently attract great theoretical interest \cite{dallatorre10, Dallatorre2012, Dalla13, DPT4, Sieberer15, SiebererRev, Overbeck16} and have been investigated in a broad spectrum of experimental set-ups, including, e.g., light-driven semiconductor heterostructures \cite{Carusotto13}, arrays of driven microcavities \cite{Hartmann2008, Tomadin2010} and cold atoms in optical lattices \cite{Schaub2015} and cavities \cite{Ritsch2013, Brennecke2012}.
These systems share in common that the microscopic processes governing the driving and dissipation explicitly break detailed balance, pushing these systems out of equilibrium \cite{Sieberer15,SiebererRev}. However, as mentioned above, equilibrium conditions can be recovered on mesoscopic time- and length-scales upon coarse-graining and this turns out to be indeed the case in several instances \cite{mitra06,dallatorre10,Dalla13,Sieberer2013}; from a physical perspective, this is due to the ``fast'' degrees of freedom acting as an effective thermal bath for the ``slow'' ones \cite{Th-Mitra1, Th-Mitra2}. However, examples have been identified in which not only the non-equilibrium nature \cite{Altman15, Prosen2010}, but also the quantum coherent aspects \cite{Lee13,Marino2015} of the dynamics persist under rescaling to arbitrarily long wavelengths. The proposal outlined above then opens up a new path to explore -- i.e., to check the robustness of DP under the influence of \emph{quantum} fluctuations in regimes which are not dominated by the dephasing noise. This question was addressed in Ref.~\cite{Marcuzzi2016} via an effective action approach; it was found that, while the nature of the transition from the absorbing phase to the active one does not change when the quantum terms are small compared to the classical ones, it instead switches to discontinuous (first-order) in the opposite regime.
\subsection{Overview and key results}

\emph{From the microscopic action to an effective field theory} -- In order to perform the transition from the microscopic physics described in terms of a quantum master equation for the underlying spin model, we have devised a procedure which incorporates the qualitatively crucial short distance physics in terms of suitable mean field theory, and allows us to systematically construct the long wavelength excitation dynamics on top of it. \changer{An indispensable part is played in the latter by the finiteness of the local spin Hilbert space, or equivalently by the fact that the magnetization is bounded. This constraint is accounted for in the present approach based on noisy Heisenberg-Langevin equations. It is a necessary requirement for the implementation of the microscopic dynamical rules of the contact process: when lifted, it produces strikingly different behaviors (see, e.g., \cite{Bosonic_CP}). Other widely-employed approaches, such as bosonization via a Holstein-Primakoff transformation (see e.g.~Ref.~\cite{maghrebi2015}), do not preserve this constraint and thus do not constitute a viable option in our case.} Technically, we first recast the quantum master equation into noisy Heisenberg equations for the three onsite spin operators $\sigma_i^{x,y,z}$. The resulting equations of motion are decoupled at the mean field level, which accounts for the short distance physics of the problem. 
These equations feature two gapped and one potentially gapless variable, the latter being associated with density fluctuations. We then map the problem into a Martin-Siggia-Rose-Janssen-De Dominicis (MSRJD) functional integral. 

After elimination of the gapped fluctuations, we end up with a description in terms of a dynamical action for the density variable alone. In the limit where the coherent microscopic processes vanish, we reproduce the action governing the DP universality class, so that our procedure represents one of the rare instances where the DP action is derived from a concrete microscopic model. 
Incorporating the ``quantum scale'' associated to the coherent branching process introduces a new relevant parameter in the problem on the microscopic level, and leads to important structural modifications of the DP action -- among them, an interaction parameter may cross zero and change sign, signaling a critical endpoint of a second order phase transition, which afterwards turns into a first order one. Moreover, the defining rapidity inversion symmetry of DP is broken in our model. On the other hand, the second key structural property of DP -- the existence of an inactive state, signaled by a noise level that scales to zero with the density -- is still present in our long wavelength theory. 

\emph{Structure and key properties of the phase diagram} -- The phase diagram of our model is depicted in Fig. 1. The additional quantum scale, describing coherent branching, adds a relevant parameter to the problem and is thus expected to give rise to an additional phase transition in the problem. Indeed, a new first order phase transition is found in the absence of incoherent branching. Increasing the incoherent branching gives rise to a critical endpoint, manifestly characterized by different symmetries than DP, and therefore giving rise to a distinct universality class. The corresponding long-wavelength action in the vicinity of the bicritical point resembles the effective action for the so-called ``tricritical directed percolation'' class.
 In order to assess the physics of the new (bi-)critical point and the first order transition, we elaborate as follows: \\
(i) \emph{Nature of the bicritical point phase transition} -- We compute a set of critical exponents, determining the universality class, of the bicritical point. To this end, we develop a background-field functional RG method approaching the phase transition from the active side, based on previous work benchmarked for the DP universality class \cite{Buchhold2016}. This approach is capable of effectively incorporating higher loop effects, which turns out to be crucial at the bicritical point. We furthermore deliver an exact RG argument for the protection from the generation of an \emph{additive} Markovian noise level term.  Remarkably, this substitutes the usual symmetry based argument due to the presence of rapidity inversion symmetry, which we cannot rely on in the presence of coherent branching. Finally, we estimate the Ginzburg scale for the extent of the critical domain near the phase transition. As expected, its range increases substantially when lowering the dimension.  \\
(ii) \emph{First order transition} -- We investigate the properties of the first order non-equilibrium phase transition in a homogeneous optimal path approximation. A remarkable trait of the analysis is that, despite the problem is manifestly out of equilibrium due to the special nature of the noise, we are still able to construct a stationary, non-Gibbsian probability distribution within our approximations. The role of the density-dependent noise is to stabilize the inactive phase with respect to what an analogous, but field-independent noise would do. In addition, we estimate finite size effects, and find that systems of around 5000 lattice sites should suffice to see a clear discontinuity, evidencing the first order nature of the phase transition. We note that this approach gives a rough idea on the physics of the first order transition only.  It discards explicitly instanton-like, spatially inhomogeneous field configurations, that should play a role at least close to the transition. Surprisingly little is known on non-equilibrium first order transitions, and we reserve an in-depth study of this problem for future research.

\emph{Physical implementation} -- 
We furthermore discuss an idea for implementing the considered physics with the help of atomic lattice systems in which interacting Rydberg states are excited both coherently and incoherently. This could provide a guide for current experiments to address the competition between classical and coherent processes in non-equilibrium phase transitions.

This paper is structured as follows: in Sec.~\ref{sec:micro} we introduce the microscopic model and derive the effective  functional integral description for its dynamical properties; in Sec.~\ref{sec:results} we analyze the phase diagram and highlight the nature of the phase transitions encountered. The more detailed discussion of the properties of the bicritical point can be subsequently found in Sec.~\ref{BiCri}, while Sec.~\ref{SecFO} is devoted to the features of the first-order line. The connection with current experiments with Rydberg atoms is finally established in Sec.~\ref{ExpSec}, where we also report a numerical study carried over with quantum-jump Monte Carlo techniques before providing our concluding remarks (Sec.~\ref{sec:concl}).

\section{Microscopic model and density action functional}
\label{sec:micro}
We consider here the \emph{quantum contact process} originally introduced in Ref.~\cite{Marcuzzi2016}, which is defined on a $d$-dimensional square lattice with spacing $a$. For simplicity, we label the sites with a single index $l = 1 \ldots N$; each individual site is a two-level quantum system which can be either active ($\ket{\uar}_l$) and contribute to the dynamics or inactive ($\ket{\dar}_l$) and remain inert until activated. In the Rydberg-atom language of Ref.~\cite{Marcuzzi2015} these would correspond to an excited atom and a ground state one, respectively. Note that, in contrast to the classical contact process, here we admit generic coherent superpositions $\alpha_l|\uar \rangle_l+\beta_l| \dar \rangle_l$ with $|\alpha_l|^2+|\beta_l|^2=1$. The dynamics is defined in terms of the following processes:
\begin{itemize}
\item[(i)] \textbf{decay}: active sites are spontaneously inactivated at a rate $\gamma$ ($\ket{\uar} \stackrel{\gamma}{\rightarrow} \ket{\dar}$).
\item[(ii)] \textbf{classical branching/coagulation}: to mimic the facilitated dynamics introduced above, we consider incoherent activation at rate $\kappa$ of sites neighboring an excitation ($\ket{\uar \dar} \stackrel{\kappa}{\rightarrow} \ket{\uar \uar}$), but we also account for the time-reversed process, i.e., facilitated inactivation or \emph{coagulation} occurring at the same rate ($\ket{\uar \uar} \stackrel{\kappa}{\rightarrow} \ket{\uar \dar}$). In our current conventions, the actual rates are proportional to the number $N_A$ of active neighbors, e.g., $\ket{\uar \dar \uar} \stackrel{2\kappa}{\longleftrightarrow} \ket{\uar \uar \uar}$.
\item[(iii)] \textbf{quantum branching/coagulation}: we introduce a Hamiltonian $H$ (see further below) which connects precisely the same states connected by classical branching and coagulation, i.e., such that $\bra{a} H \ket{b} = N_A \Omega   $ if $\ket{a} \stackrel{N_A \kappa}{\longleftrightarrow} \ket{b}$ (and, in particular, $\bra{a} H \ket{b} = 0$ if $N_A =0$), $\Omega$ being an overall coefficient fixing its amplitude. The example above translates here to $\bra{\uar \dar \uar} H \ket{\uar \uar \uar} = 2 \Omega$.
\end{itemize}
The third process is the minimal quantum equivalent of the second one and provides the quantum competition to the purely classical process. It is also important to remark that (i)-(iii) preserve the fundamental property of DP, i.e., the presence of a unique absorbing state corresponding to the fully-inactive one $\ket{abs} = \otimes_l \ket{\dar}_l$. In order to describe the dynamics of this quantum contact process, we will discuss the corresponding microscopic Heisenberg-Langevin equations and derive an effective long-wavelength non-equilibrium path integral description. The latter is particular well suited to describe the dynamics close to the active-to-inactive -- i.e. the absorbing-state -- phase transition.

\subsection{Microscopic model}
The ideal model presented above is a driven open quantum lattice of spin-$\tfrac{1}{2}$ variables. In order to define it formally, it is convenient to introduce here a complete set of spin operators acting on site $l$,
\be
\begin{split}
	\si_l^+   = &\ket{\uar}_l \bra{\dar}_l \comma  \si_l^- = \ket{\dar}_l \bra{\uar}_l \mand \\ 
	&\si_l^z = \projl{\uar}{l} - \projl{\dar}{l} 
\end{split}
\ee
or, equivalently,
\be 
\begin{split}
	\si_l^x = & \si_l^+ + \si_l^- \comma \si_l^y = -i\si_l^+ + i\si_l^-  \mand  \\ 
	&\n_l = \si_l^+ \si_l^- = \projl{\uar}{l} . 
	\label{eq:set2}
\end{split}
\ee
In particular, $\n_l$ is the local projector onto an active site, i.e., $\n_l \ket{\uar}_l = \ket{\uar}_l$ and $\n_l \ket{\dar}_l = 0$. Its global expectation value $n = (1/N) \sum_l \av{\n_l}$ will constitute our order parameter. 
As we are considering only Markovian processes as appropriate for these systems \cite{SiebererRev}, the time evolution of the system's density matrix $\rho$ is given by a quantum master equation \cite{Lindblad76, Breuer_P}
\eq{Eq1}{
\partial_t \rho = \mal{S} \rho=-i\left[H,\rho\right] + \sum_{l}\mathcal{L}^{(d)}_{l}\rho+\sum_{l}\mathcal{L}^{(b)}_{l}\rho+\sum_{l}\mathcal{L}^{(c)}_{l}\rho.
}
We have introduced the shorthand $\mathcal S$ for the superoperator acting on the density matrix $\rho$ for future reference. The coherent part (iii) is encoded in the Hamiltonian
\eq{Eq2}{
H=\Omega\sum_l\PP_l \,\si^x_l \,\text{ with }\, \PP_l = \sum_{m \text{ nn } l} \n_m,
}
where ``nn'' denotes a summation restricted to nearest neighbors only. The operator $\PP_l$ ``counts'' the number of active nearest neighbors of $l$ and enforces the constraint of at least one excitation being present for being able to flip site $l$. Processes (i) and (ii) are instead accounted for via the Liouvillians $\mathcal{L}^{(i)}, i = d,b,c$, with the apices distinguishing between those contributing to decay (d), classical branching (b), and coagulation (c). The Liouvillians are each generated by a set of Lindblad or quantum jump operators $L^{(i)}_m$, and  take the standard Lindblad form \cite{Lindblad76, Breuer_P},  
\be
	\mathcal{L}^{(i)} \rho = \sum_m  \lqq  L^{(i)}_m \rho L^{(i)\dag}_m - \ha \acomm{L^{(i)\dag}_m L^{(i)}_m}{\rho}
	\rqq,
	\label{eq:struct}
\ee
which ensures preservation of probability and positivity. 
For the dissipative processes considered, the jump operators read
\begin{subequations}
\begin{align}
	L^{(d)}_{l,m} &\equiv L^{(d)}_{l} = \sqrt{\gamma} \, \si_l^- , \label{eq:Ld} \\
	L^{(b)}_{l,m} &= \sqrt{\kappa} \, \n_m \si_l^+ , \label{eq:Lb} \\
	L^{(c)}_{l,m} &= \sqrt{\kappa} \, \n_m \si_l^-,  \label{eq:Lc}
\end{align}
\end{subequations}
so that we find
\eq{Eq3}{
\mathcal{L}^{(d)}_{l}\rho= \gamma \left(\si^-_l\rho\si^+_l - \ha\left\{\n_l,\rho\right\}\right)
}
for spontaneous decay,
\eq{Eq4}{
\mathcal{L}^{(b)}_{l}\rho = \kappa \sum_{m \text{ nn } l} \left( \n_m \si^+_l \rho \n_m \si^-_l  - \ha  \left\{ \n_m (1-\n_l) ,\rho\right\}\right)
}
for classical branching and
\eq{Eq5}{
\mathcal{L}^{(c)}_{l}\rho=\kappa  \sum_{m \text{ nn } l} \left( \n_m\si^-_l\rho \n_m\si^+_l -  \ha \left\{ \n_m \n_l,\rho\right\}\right).
}
for classical coagulation. We remark here that, according to the formalism outlined in Refs.~\cite{Degenfeld14, Marcuzzi14}, in the presence of strong decoherence noise (acting with a rate $\gamma_{deph} \gg \Omega$), the evolution under the Hamiltonian \eqref{Eq2} effectively reduces, up to leading order in $\Omega / \gamma_{deph}$, to a classical master equation which can be described via a set of jump operators
\be
	L^{(H)}_{l} = \sqrt{\frac{4\Omega^2}{\gamma_{deph}}} \, \PP_l \lt \si_l^- + \si_l^+ \rt,
\ee
which only differ from the ones in Eqs.~\eqref{eq:Lb} and \eqref{eq:Lc} by the fact that, in the presence of $N_A$ active neighbors, the rate of ``facilitated flipping'' is enhanced quadratically ($4 N_A^2 \Omega^2 / \gamma_{deph}$), instead of linearly ($N_A \kappa$). However, at the critical point the density of active sites $n$ vanishes; therefore, the critical properties are dominated by configurations in which $N_A$ remains low on average. In particular, if $\av{N_A} \lesssim 1$, then $N_A^2 \approx N_A$, since typically it mostly takes the discrete values $0$ and $1$. Therefore, this difference can at most shift the critical point and change the profile of $n$ in the active phase, but cannot modify the universal properties.

\subsection{Heisenberg-Langevin equations}
\changer{In order to derive a path integral description for the current model, we will determine the Heisenberg-Langevin equations for the spin operators in this section. These equations represent the equations of motion for the spins in the presence of Hamiltonian and dissipative dynamics and by construction preserve the local spin algebra (see App.~\ref{AppLast} some general aspects of Heisenberg-Langevin equations). The latter is crucial for the correct implementation of the contact process dynamics in terms of local quantum operators. Afterwards, we will perform a mean-field decoupling, which approximates the spin operators by local, stochastically fluctuating fields, obeying Langevin equations of motion. The Langevin dynamics will then be recast in terms of a non-equilibrium path integral, which is discussed below. One should note that a Holstein-Primakoff approximation of the master equation and a subsequent mapping of the master equation to a Keldysh path integral, as e.g. performed in Ref.~\cite{maghrebi2015}, typically replaces the strong constraint on the spin Hilbert space via a soft constraint, which implements the spin algebra not exactly but only on average. This is not sufficient in order to derive a field theory for the contact process and thus the present approach via the Heisenberg-Langevin equations is required instead.}

In order to keep our order parameter explicit, we write the Heisenberg-Langevin equations in terms of the set \eqref{eq:set2} of one-spin observables (alongside the identity, they span the entire local Hilbert space of operators). For convenience, we introduce the shorthand $\wh{s}_l=\sum_{m \text{ nn }l}\si^x_m$ and the coordination number $z = 2d$ of the lattice, i.e. the number of nearest neighbors per site, where we recall that $d$ is the number of spatial dimensions. The equations of motion (EOM) are derived by applying \eqref{eq:full_O}, which leads to
\begin{eqnarray}
\partial_t \n_l&=&-\gamma \n_l+\left[\Omega\si^y_l-\kappa(2\n_l-1)\right] \PP_l + \x_l^n,\label{Eq8}\\
\partial_t\si^x_l&=&   - \Omega\si^y_l\wh{s}_l-\frac{\kappa z+\gamma}{2}\si^x_l-\kappa\si^x_l\PP_l+\x_l^x,\label{Eq9}\\
\partial_t\si^y_l&=&\Omega\si^x_l\wh{s}_l-\frac{\kappa z+\gamma}{2}\si^y_l+\x^y_l  \nonumber\\
&& -\left[2\Omega(2\n_l-1)+\kappa\si^y_l\right] \PP_l.\label{Eq10}
\end{eqnarray}
Note that Eq.~\eqref{Eq9} differs from Eq.~(3) of Ref.~\cite{Marcuzzi2016} by the sign of the first addend (which reads $+\Omega \si^y_l\wh{s}_l$ there, once translated in our present notation). This constitutes a typo which we correct here; the discussion of the phase diagram and critical properties, however, remains completely unaffected, as we shall show in the following.
As anticipated above, in order to fix the noise operators we consider a system-bath coupling which, once the bath variables are integrated out in a Born-Markov approximation, yields the same deterministic part of the equations \eqref{Eq8}-\eqref{Eq10}. 
We assume that the spatial correlations of the bath are much shorter than the lattice constant $a$, such that noise correlations between different lattice sites are absent and every lattice site is effectively coupled to its own independent (but identical) environment. This allows us to focus our subsequent analysis on a single site; for simplicity, in the derivation of the noise we will be dropping the position index. The discussion for the general case can be straightforwardly recovered by adding a subscript $l$ to all system and bath operators. We need three terms to separately account for decay, branching and coagulation, which will generate contributions $\x_d$, $\x_b$ and $\x_c$ to the noise, respectively. We thus introduce the three local Hamiltonians $H_d$, $H_b$ and $H_{c}$. The former reads
\eq{Eq12}{
H_d=\sum_q\lambda_q \left(\si^+\dd_q+\dd^{\dagger}_q\si^-\right)+\sum_q\omega_q \dd^{\dagger}_q \dd_q,
}
where the $\dd_q$ operators represent bosonic modes ($\comm{\dd_q}{\dd_k^\dag} = \delta_{qk}$), $\omega_q$ their dispersion relation and $\lambda_q$ their respective coupling with the spin. Since decay corresponds to photon emission into the vacuum, we assume these modes to be in a state $\rho_d^0$ at zero temperature and sufficiently numerous so that the action of the system on them can be considered negligible (i.e., they can be approximated as a continuum of modes).
In order to reproduce the branching and coagulation dynamics above, we actually have to impose a further constraint, i.e., that there are two independent baths of harmonic oscillators for every pair of neighboring spins. 
The system-bath Hamiltonians will read, for a generic (neighboring) pair, 
\be
\begin{split}
\label{Eq13}
H_{b}=\sum_k \alpha_k\ \n_{\auxn} \, \left(\si^- \bb_k  + \bb^{\dagger}_k \si^+ \right)+\sum_k\nu_k \bb^{\dagger}_k \bb_k, \\
H_{c}=\sum_k \alpha_k\ \n_{\auxn} \, \left(\si^+ \cc_k + \cc^{\dagger}_k  \si^- \right)+\sum_k\nu_k \cc^{\dagger}_k \cc_k,
\end{split}
\ee
where the $\text{nn}$ denotes a given neighbor of the site considered, and correspondingly
\be
\begin{split}
\label{Eq13-2}
H_{b,\auxn} = \sum_k \alpha_k\ \n \, \left(\si^-_\auxn \bb_{k,\auxn}  + \bb^{\dagger}_{k,\auxn} \si^+_\auxn \right) + \sum_k\nu_k \bb^{\dagger}_{k,\auxn} \bb_{k,\auxn}, \\
H_{c,\auxn} = \sum_k \alpha_k\ \n \, \left(\si^+_\auxn \cc_{k,\auxn} + \cc^{\dagger}_{k,\auxn}  \si^-_\auxn \right)+\sum_k\nu_k \cc^{\dagger}_{k,\auxn} \cc_{k,\auxn},
\end{split}
\ee
and the $\bb_k$s and $\cc_k$s are bosonic modes with equal dispersions $\nu_k$ and coupling $\alpha_k$ to the spin. These baths are initialized in equal states $\rho_{b/c}^0$, to allow excitation and de-excitation of the spin at the same rate.

We start by considering spontaneous decay. The (ordinary) Heisenberg equations under the action of $H_d$ read
\begin{eqnarray}
\partial_t\si^-&=&i[H_d,\si^-]=i\sum_q\lambda_q \dd_q\si^z,\label{Eq14}\\
\partial_t \n &=& i[H_d,\n]=i\sum_q\lambda_q \left(\dd^{\dagger}_q \si^- - \si^+ \dd_q\right),\label{Eq15}\\
\partial_t \dd_q & = & i[H_d, \dd_q] = -i \lambda_q \si^- - i\omega_q \dd_q. \label{Eq16}
\end{eqnarray}
Equation \eqref{Eq16} can be formally integrated yielding
\eq{Eq17}{
\dd_q(t)=\dd_q(0)e^{-i\omega_qt}-i\lambda_q\int_0^t \rmd t' \, \si^-(t')e^{-i\omega_q(t-t')}.
}
Inserting \eqref{Eq17} into \eqref{Eq15} gives
\begin{eqnarray}
\partial_t \n(t) & = & -\sum_q  \lambda^2_q \int_0^t \rmd t' (\si^+(t')\si^-(t)e^{i\omega_q(t-t')} + \text{h.c.}) + 
\nonumber\\
&&i\sum_q\lambda_q \left(\dd_q^{\dagger}(0) \si^-(t)e^{i\omega_qt}  -  \si^+(t) \dd_q(0) e^{-i\omega_qt}\right)\nonumber\\
&\approx &-\gamma \n(t) + \underbrace{i\sum_q\lambda_q\left(\dd_q^{\dagger}(0)\si^-e^{i\omega_qt}-\text{h.c.}\right)}_{\x^n_d}.\label{Eq18}
\end{eqnarray}
The first term of Eq.~\eqref{Eq18} is obtained by applying the Born-Markov approximation for a bath, which is fluctuating rapidly on typical system time scales \cite{Scully}. The effective coupling strength $\gamma=2\pi\lambda^2(0)D(0)$ is proportional to the bath density of states $D(\omega)=\sum_q\delta(\omega-\omega_q)$ and the coupling constants $\lambda(\omega)=\sum_q\lambda_q\delta(\omega-\omega_q)$ both evaluated at zero frequency. The second term contains information on the initial state of the bath and is nothing but the desired noise operator. This clarifies the meaning of the noise average $\avb{\cdot}$, which is nothing else than the trace over the bath degrees of freedom
\be
	\avb{\cdot} = \trace{(\cdot) \, \rho_d^0}.
\ee
Since $\rho_d^0$ is a definite-particle-number state, the noise has zero mean $\avb{\x^n_d} = 0$. The variance, however, does not vanish and reads
\begin{eqnarray}
\avb{ \x^n_d(t)\x^n_d(t') } & \stackrel{\text{B-M}}{=} & \sum_q \lambda_q^2 \left[ n_q e^{i\omega_q(t-t')}\si^-(t)\si^+(t')\right. \nonumber\\
&&\left.+(1 + n_q) e^{-i\omega_q(t-t')}\si^{+}(t)\si^-(t')\right] \ \ \ \ \nonumber\\
&\overset{\text{B-M}}{=} & \gamma (N_d + \n(t))\delta(t-t')  \\
& \overset{T=0}{=} &  \gamma \n(t)\delta(t-t').\ \ \label{Eq19}
\end{eqnarray}
Here, we have applied the Born-Markov approximation to commute system and bath variables at different times and employed the shorthand $n_q = \avb{ \dd^{\dagger}_q \dd_q}$ and $N_d = \sum_q  n_q$. The second (approximate) equality comes, as mentioned above, from assuming that the bath fluctuates much faster than the typical system timescales, implying that both spectral densities $D(\omega)$ and $\lambda(\omega)$ are slowly-varying functions of their arguments, such that the summation effectively yields a time-local result. The final equality is exact and comes from the fact that the bath is at zero temperature, hence $\avb{\dd^{\dagger}_q \dd_q} = 0 \,\, \forall \, q$ and $N_d = 0$.
Equation \eqref{Eq19} highlights the multiplicative nature of the density noise ($\xi^n_d\sim\sqrt{n}$). This property leads to a noiseless density channel for the empty state $n=0$, and ensures the absence of density fluctuations in the absorbing state. A small but non-zero temperature of the bath $T\neq0$ will instead lead to a non-vanishing bath photon number $N_d \sim T^d$ (valid for relativistic bosonic particles in $d$ spatial dimensions) and modify the absorbing-state nature of the transition on timescales $\tau > (\gamma N_d)^{-1} \sim T^{-d}$ and distances $x > (\gamma N_d)^{-1/2}\sim T^{-d/2}$. For sufficiently low temperatures, as achieved by current cold atom experiments, these scales are much larger than the system's and these effects can thus be ignored.

The remaining equation of motion for $\si^-$ can be solved in the same spirit of Eqs.~\eqref{Eq18}, \eqref{Eq19}, which yields
\eq{Eq20}{
\partial_t\si^- (t) = -\frac{\gamma}{2}\si^- (t) + \underbrace{i\sum_q \lambda_q \, \dd_q(0) \, \si^z (t) \, e^{i\omega_q t}}_{\x^-_d(t)}.
}
This defines the noise operator $\xi^-_d(t)$ and, via conjugation, $\xi^{+}_d(t)=\left(\xi^{-}_d(t)\right)^{\dagger}$, as well as $\xi^x_d = \xi^+_d  +  \xi^-_d$ and $\xi^y_d = i(\xi^-_d  -  \xi^+_d )$. The complete noise correlations can be determined from Eqs.~\eqref{Eq18}, \eqref{Eq20}, which are straightforwardly extended to the entire lattice by re-instating the position indices. In the $(x,y,n)$ basis (i.e., set \eqref{eq:set2}) the noise correlations can be expressed as
\eq{Eq21}{
\avb{ \bm{\x}_{d,l}(t)\bm{\x}^{\dagger}_{d,l'}(t') } = \gamma\delta(t-t')\delta_{l,l'}\left(\begin{array}{ccc} 1 & -i & \si^-_l\\ i & 1 & i\si^-_l  \\ \si^+_l  &  -i \si^+_l & \n_l
\end{array}\right),
}
where $\bm{\x}^{\dagger}_{d,l}(t)=( \x^x_{d,l}(t), \x^y_{d,l}(t) , \x^n_{d,l}(t))$. As pointed out above, the noise average $\avb{\cdot}$ represents a quantum mechanical average over the bath degrees of freedom, such that the entries in \eqref{Eq21} remain operator valued. We remark again that the noise is only additive in the $\sigma^{x,y}$ channels, while it remains multiplicative in the density channel. In the limit $\Omega\ll\kappa$ the coupling of the density field to the $\si^y$ matrix can be eliminated and leads to a modification of the branching rate $\kappa$, as mentioned above. In this limit, the Heisenberg-Langevin equation for the density \eqref{Eq8} has an absorbing configuration for $\{n_l\}=0$. We will show in the following, that the latter feature persists for all values of $\Omega$ and that the $\{n_l\}=0$ configuration remains an absorbing state for the density channel.
Note that, due to our choice of the bath state $\rho_d^0$, the noise is Gaussian and therefore entirely defined in terms of its mean expectation value and two-point correlations. Due to the Markov approximation, the noise is white (time-local) as well.

So far, we have not considered the noise contribution from the branching and coagulation dynamics stemming from the Hamiltonian \eqref{Eq13}. Interestingly, there is no need to: as long as we are only interested in the critical properties, we can safely neglect higher orders in $n$, as they will just provide subleading corrections. Due to the factors $\n$ and $\n_\auxn$ in the Hamiltonians $H_{b/c}$ and $H_{b/c,\auxn}$ we are guaranteed that the noise terms $\x_b$ and $\x_c$ will never dominate, at low densities, over the decay noise $\x_d$. Therefore, for simplicity, we can safely neglect their presence and set $\xi \equiv \xi_d + \xi_b + \xi_c \to \xi_d$. For completeness, we provide a discussion on the discarded terms in Appendix \ref{app:B}

Together with the noise kernel \eqref{Eq21}, the Heisenberg Langevin equations \eqref{Eq8}-\eqref{Eq10} represent the starting point for our analysis of the absorbing state phase transition in terms of a non-equilibrium path integral framework. While the deterministic part of the Heisenberg Langevin equations is exact, we have approximated the noise kernel up to leading order in the density according to the previous discussion and kept only the decay contribution which still generates all relevant terms.

\subsection{Non-equilibrium path integral description}
In order to investigate the dynamics close to the absorbing state, we derive a non-equilibrium path integral description for the density variable $n$. Our method is based on the Martin-Siggia-Rose-Janssen-de Dominicis (MSRJD) approach \cite{MSR, MSR+J, MSR+D, Tauber-book}, a well-established mapping of Langevin equations into effective field theory actions. In order to do this, we therefore first need to reduce our Heisenberg-Langevin equations \eqref{Eq8}-\eqref{Eq10} to semiclassical ones. This is achieved via a mean-field, site-decoupling approximation (i.e., averages involving operators acting on different sites are factorized, $\av{O_l O_m} \to \av{O_l} \av{O_m}$ for $l \neq m$), which yields a set of (deterministic) equations
\be 
\begin{split}
	\partial_t n_l & = -\gamma n_l + \lqq  \Omega \sigma_l^y - \kappa (2n_l - 1)  \rqq \pp_l, \\
	\partial_t \sigma_l^x & = -\Omega \sigma_l^y s_l - \frac{\kappa z + \gamma}{2} \sigma_l^x - \kappa \sigma_l^x  \pp_l, \\
	\partial_t \sigma_l^y & = \Omega \sigma_l^x s_l - \frac{\kappa z + \gamma}{2} \sigma_l^y - \lqq 2\Omega (2n_l - 1) + \kappa \sigma_l^y  \rqq \pp_l,
	\label{eq:MFeqs}
\end{split}
\ee
as we recall that both $\wh{s}_l$ and $\PP_l$ act non-trivially only on the nearest-neighbors of $l$, and not on site $l$ itself. Relying on translational invariance to make an uniform assumption ($n_l = n_m \equiv n$ and $\sigma_l^{x/y} = \sigma_m^{x/y} \equiv \sigma^{x/y}$ $\forall \,\, l,m$) one can further reduce them to
\be 
\begin{split}
	\partial_t n & = \lqq -\gamma +z \lqq \Omega \sigma^y - \kappa (2n-1)   \rqq \rqq n  , \\
	\partial_t \sigma^x  & = - \lqq  \Omega z \sigma^y   + \frac{\kappa z + \gamma}{2}  + \kappa z  n \rqq \sigma^x   , \\
	\partial_t \sigma^y &  = \Omega z (\sigma^x)^2 - \frac{\kappa z + \gamma}{2} \sigma^y - zn \lqq   2\Omega (2n-1) + \kappa \sigma^y   \rqq  .
\end{split}
\ee 
The stationary solutions are found by setting the time derivatives to $0$. Introducing the dimensionless constants $\chi = z \kappa / \gamma$ and $\omega = z \Omega / \gamma$, we find that the stationary density of active sites obeys
\be
	n \lqq   \lt 4 \omega^2  + 2 \chi^2   \rt n^2 - 2 \lt   \omega^2 - \chi  \rt  n +  \ha \lt 1 - \chi^2 \rt \rqq   = 0.
\ee
The solution $n = 0$ corresponds to the absorbing state, which is always present, but is only dynamically stable for $\chi < 1$ (see Appendix \ref{app:MF}). For $\chi > 1$, instead, any perturbation away from it will grow to reach one of the other solutions, marking the active phase. For $\chi < 1$, the non-absorbing solutions still exist as long as the discriminant is positive ($\omega^4 + \chi^4 + 2\omega^2 (\chi^2 - \chi-1) \geq 0$). If additionally $\omega^2 > \chi$, a saddle-node bifurcation takes place, corresponding to a first-order phase transition to a coexisting, bistable regime. 

The mean-field equations \eqref{eq:MFeqs} constitute our starting point. 
We \changer{stress} here again that the present approach respects the constrained nature of the local spin Hilbert space, which is crucial for the correct description of the contact process. This is advantageous over a bosonic Holstein-Primakoff \cite{Holstein1940} approximation, which introduces a much larger bosonic Hilbert space and does not preserve the local spin constraint, i.e. it allows for an arbitrary number of bosonic excitations being present on each site.
In the classical case, it turns out that the latter produces a completely different behavior \cite{Bosonic_CP}, e.g., if the branching and decay rates are equal, the average density of excitations remains constant. \changer{As a consequence, the representation of the spins in terms of Holstein-Primakoff bosons excludes any absorbing dark state, unless the local Hilbert space has a strict upper bound on the number of bosons per lattice site.} It is therefore important to keep the ``hard-core'' (or ``exclusion'') as a fundamental property of the dynamics. \changer{This hard-core constraint promotes any bosonic field theory to a formidable problem to solve and is conveniently avoided by the present approach.}
Furthermore, the fluctuations induced by the environment must be taken into account in a way that is consistent with the discussion above. In particular, it is crucial to maintain the multiplicative nature of the noise on $n$. The variables $n_l$, $\sigma^{x/y}_l$ are now real-valued and the noise must be as well. We thus introduce a Gaussian, white noise $\bm{\xi}_l^\intercal = (\xi_l^x, \xi_l^y, \xi_l^n)$ with vanishing mean and a covariance matrix extracted from the Hermitian part of the operatorial one in Eq.~\eqref{Eq21} (see Eq.~\eqref{Eq25} below). 

Since continuous phase transitions involve collective modes, we can adopt at this point a mesoscopic description for our system, i.e., we take the continuum limit. This corresponds to sending the lattice spacing $a \to 0$ while appropriately rescaling the coupling constants. We thus replace our quantities by the corresponding local densities
\be
	n_l(t) \to n_X \comma \sigma^{x/y} (t )  \to   \sigma^{x/y}_X,
\ee
where we denote $X \equiv (\vec{x}, t)$, $\vec{x}^\intercal = (x_1 , x_2 , \ldots , x_d)$ being the continuous spatial coordinate. 
In the spirit of a low frequency effective field theory, the corrections introduced by fluctuations over the site-decoupling approximation are taken into account by terms which contain higher powers of the variables or derivatives. Close to a (second order) phase transition, this procedure becomes particularly efficient, as each of the coupling constants can be classified according to canonical power counting, and both higher-order derivatives and densities lower the degree of relevance of the couplings.
With this in mind, we discard higher order spatial derivatives and set $\pp_l (t) \rightarrow (a^2\nabla^2+z)n_X$, $s_l (t) \rightarrow (a^2\nabla^2+z)\sigma^x_X$. 
For simplicity, we also rescale time according to $t\rightarrow\gamma t$ and define the dimensionless couplings $\chi=z\kappa/\gamma$, $\omega=z\Omega/\gamma$ and the diffusion constant $D=\chi a^2/z$. The continuum Langevin equations now read
\begin{widetext}
\begin{eqnarray}
\partial_tn_X&=&(\chi-1+D\nabla^2) n_X+(\frac{\omega}{\chi}\sigma^y_X-2n_X)(D\nabla^2+\chi)n_X+\xi^n_X,\label{Eq22}\\
\partial_t\sigma^x_X&=&-\frac{\chi+1}{2}\sigma^x_X  - \frac{\omega}{\chi}\sigma^y_X(D\nabla^2+\chi)\sigma^x_X    -   \sigma^x_X(D\nabla^2+\chi)n_X+\xi^x_X,\label{Eq23}\\
\partial_t\sigma^y_X&=&-\frac{\chi+1}{2}\sigma^y_X    -  \sigma^y_X(D\nabla^2+\chi)n_X+\frac{\omega}{\chi}\sigma^x_X(D\nabla^2+\chi)\sigma^x_X  -  \lqq   \frac{2\omega}{\chi} (2n_X-1) \rqq (D\nabla^2+\chi)n_X+\xi^y_X,\label{Eq24}
\end{eqnarray}
\end{widetext}
with Markovian noise kernel 
\eq{Eq25}{
\langle \bm{\xi}_X\bm{\xi}^{\dagger}_Y\rangle=\frac{\delta(X-Y)}{2}\left(\begin{array}{ccc}2 & 0& \sigma^x_X\\ 0& 2& \sigma^y_X\\  \sigma^x_X  & \sigma^y_X & 2n_X
\end{array}\right) \equiv \bm{M}_{XY}.
}
We note that, in Eq.~\eqref{Eq22}, the linear term in $n_X$ changes sign at $\chi = 1$. This indicates a closing gap and corresponds, at the mean-field level, to a continuous phase transition taking place at this point. Conversely, at $\chi = 1$ the equations for $\sigma^{x/y}$ remain gapped, and these variables play the role of spectator modes at the transition, which will allow us to integrate them out in the MSRJD path-integral framework.

Due to its Gaussian nature, the properties of the noise are entirely determined by the matrix \eqref{Eq25}; the full distribution can be expressed as
\be
	p [\bm{\xi}] =  \mal{N}  \rme{- \ha \bm{\xi}^\dag \ast \bm{M}^{-1} \!\ast \bm{\xi}}
	\label{eq:Gauss}
\ee
with $\mal{N}$ a suitable normalization ensuring 
\be 
	\int \mal{D} [\bm{\xi}] \, p[\bm{\xi}] = 1,
\ee
$\mal{D} [\bm{\xi}] = \mal{D} [\xi^x, \xi^y, \xi^n]$ a suitable functional measure, and ``$\ast$'' denoting convolution over the spatial and temporal coordinates, i.e., 
\be
	A \ast B = \int \rmd X \, A_X B_X \equiv \int \rmd^d x \rmd t \, A (\vec{x}, t) B (\vec{x}, t) .
\ee
In Eq.~\eqref{eq:Gauss}, $\bm{M}$ depends on $\sigma^{x/y}$ and $n$. These are to be interpreted here as the solutions $\sigma^{x/y}_\xi$ and $n_\xi$ of the Langevin equations \eqref{Eq22}-\eqref{Eq24} at fixed realization $\bm{\xi}$ of the noise. By definition, we have $\avb{\cdot} = \int \mal{D}\bm{\xi} \, (\cdot) p [\bm{\xi}]$.

We proceed now with the standard MSRJD construction \cite{Tauber-book}: we shall introduce here the vectorial shorthand $\bm{\sigma}^\intercal = (\sigma^x, \sigma^y, n)$ for the variables and $\bm{\sigma}^\intercal_\xi = (\sigma^x_\xi, \sigma^y_\xi, n_\xi)$ for the solutions at fixed $\bm{\xi}$. In principle, all correlation and response properties of the system are encoded in the system's generating functional
\be 
	Z [\tilde{h}^n, \tilde{h}^x, \tilde{h}^y  ]  \equiv \avb{\rme{\bm{\tilde{h}} \ast \bm{\sigma}_\xi}} =  \avb{\rme{\tilde{h}^n \ast n_{\xi} + \tilde{h}^x \ast \sigma^x_{ \xi} + \tilde{h}^y \ast \sigma^y_{\xi} } },
	\label{eq:gen_func}
\ee
where $\bm{\tilde{h}}_X^\intercal = (\tilde{h}^{x}_X, \tilde{h}^{y}_X, \tilde{h}^{n}_X)$ are the conjugated fields (sources) to the variables. Generic correlations can then be found via functional differentiation:
\be
\begin{split}
	\avb{ n_{X_1} \ldots n_{X_k} \sigma^x_{X_{k+1}} \ldots \sigma^x_{X_{m}} \sigma^y_{X_{m+1}} \ldots \sigma^x_{X_{q}}    }  =  \prodl{i_n = 1}{k}  \frac{\delta}{\delta \tilde{h}^n_{X_{i_n}}} \\  \prodl{i_x = k+1}{m}   \frac{\delta}{\delta \tilde{h}^x_{X_{i_x}}}    \prodl{i_y = m+1}{q}     \frac{\delta}{\delta \tilde{h}^y_{X_{i_y}}}   Z [\tilde{h}^n, \tilde{h}^x, \tilde{h}^y  ] \eval{\bm{\tilde{h}} = 0}.
\end{split}
\ee
We recall that the average in the definition of the generating functional \eqref{eq:gen_func} can be expressed as
\be
	Z [\bm{\tilde{h}}  ] = \int \mal{D} [\bm{\xi}] \, \rme{\bm{\tilde{h}} \ast \bm{\sigma}_\xi}  \,  p[\bm{\xi}].
\ee
We multiply the integrand by
\be
\begin{split}
	1 &= \int \mal{D} [\bm{\sigma}] \, \delta (\bm{\sigma} - \bm{\sigma}_\xi) =  \\ 
	&= \int \mal{D} [\sigma^x, \sigma^y, n] \, \delta (\sigma^x - \sigma^x_\xi)  \delta (\sigma^y - \sigma^y_\xi)  \delta (n - n_\xi),
\end{split}
\ee
where $\delta$ denotes here a functional Dirac delta function such that, e.g.,
\be
	\int \mal{D} [n] \,  F(n) \delta (n - n_\xi) = F(n_\xi)   
\ee
for every test functional $F$. Assuming that the integrations over $\bm{\sigma}$ and $\bm{\xi}$ can be exchanged, this yields
\be
	Z [\bm{\tilde{h}}]  =   \int \mal{D}[\bm{\sigma}] \, \rme{\bm{\tilde{h}} \ast \bm{\sigma}}    \int \mal{D}[\bm{\xi}] \, \delta (\bm{\sigma} - \bm{\sigma}_\xi) p[\bm{\xi}]
\ee
Note that the generating exponential factor does not depend now on the noise $\bm{\xi}$; correspondingly, the integral over $\bm{\sigma}$ is performed over \emph{all} possible trajectories for the variables, while it is the $\delta$ function which ensures that only those which represent valid solutions of the Langevin equations actually contribute to its result.
Denoting now for brevity the r.h.s. of Eqs.~\eqref{Eq22}-\eqref{Eq24} with $\bm{\mal{R}}_X^\intercal = (\mal{R}_X^x , \mal{R}_X^y, \mal{R}_X^n )$, such that
\begin{eqnarray}
\partial_tn_X=\mathcal{R}^n_X, \ \partial_t\sigma^x_X=\mathcal{R}^x_X, \ \partial_t\sigma^y_X=\mathcal{R}_X^y,\label{Eq26}
\end{eqnarray}
we can rewrite the $\delta$ functions as
\be
	\delta (\bm{\sigma} - \bm{\sigma}_\xi)  =  \mal{J} \delta (\partial_t \bm{\sigma} - \bm{\mal{R}}),
\ee
where $\mal{J}$ is the Jacobian accounting for the corresponding change of variables. This Jacobian is a functional of the integration variables $\bm{\sigma}^\intercal = (\sigma^x, \sigma^y, n)$ and in principle it could not be neglected. However, it can be shown \cite{Tauber-book} that it produces a term $\propto \theta(0)$, where
\be 
	\theta(t) = \sysb{lc} 1 & (t>0) \\ 0 & (t <0)  \syse
\ee 
is the Heaviside step function, and its role is exactly to remove the ambiguity in the definition of $\theta(0)$ in expectation values. Setting $\theta(0) = 0$, we can thereby proceed as if $\mal{J} = 1$.
The integration over $\bm{\sigma}$ now takes the form
\begin{widetext}
\be
\begin{split}
&\int\mathcal{D}[n_X,\sigma^x_X,\sigma^y_X] \, \rme{\bm{\tilde{h}} \ast \bm{\sigma}} \, \delta(\partial_tn_X-\mathcal{R}^n_X)\delta(\partial_t\sigma^x_X-\mathcal{R}^x_X)\delta(\partial_t\sigma^y_X-\mathcal{R}^y_X) =
 \nonumber\\
\label{Eq27}&\int\mathcal{D}[n_X,\sigma^x_X,\sigma^y_X,\tilde{n}_X,\tilde{\sigma}^x_X,\tilde{\sigma}^y_X] \, \rme{\bm{\tilde{h}} \ast \bm{\sigma}}  \,  \rme{-\tilde{n} \ast (\partial_t n-\mathcal{R}^n)} \rme{-\tilde{\sigma}^y \ast (\partial_t\sigma^y - \mathcal{R}^y)} \rme{-\tilde{\sigma}^x \ast (\partial_t\sigma^x-\mathcal{R}^x)},
\end{split}
\ee
\end{widetext}
where we have introduced the imaginary \emph{response fields} $\bm{\tilde{\sigma}}_X^\intercal = (\tilde{\sigma}^x_X, \tilde{\sigma}^y_X, \tilde{n}_X)$ and applied the integral representation of the $\delta$-function, $\delta(x)=\int_y \rmd y \, e^{-iyx} / 2\pi$ where the ``response variable'' $\tilde{x}$ would correspond in this case to $iy$. The denomination ``response fields'' comes from the fact that, if one introduces source terms in the Langevin equations $\bm{\mal{R}}_X \to \bm{\mal{R}}_X + \bm{h}_X$, not to be confused with the effective sources $\bm{\tilde{h}}$ which appear in the definition of $Z$, one sees that the linear response of any observable $O$ to one of these fields is (with the slight abuse of notation $\sigma^n \equiv n$)
\be
	\frac{\delta \avb{O}}{\delta h^i_X}\eval{\bm{h} = 0} = \avb{O \tilde{\sigma}^i_X}. 
	\label{eq:response}
\ee

Since $(\mal{R}^x, \mal{R}^y, \mal{R}^n)$ are linear in $(\xi^x, \xi^y, \xi^n)$, respectively, the integration over the noise can be now computed according to the standard Gaussian identity
\be
	\mal{N} \int \mal{D} [\bm{\xi}] \rme{-\ha \bm{\xi}^\dag \ast \bm{M}^{-1}\! \ast \bm{\xi} \, + \, \bm{\tilde{\sigma}}^\dag \ast \bm{\xi}}  = \rme{\ha \bm{\tilde{\sigma}}^\dag \ast \bm{M} \ast \bm{\tilde{\sigma}} } .
\ee
The exponent above is straightforwardly re-expressed in terms of the variable and response fields via the definition \eqref{Eq25} of the covariance matrix. Since at this point it is the only part which comes from the noise, we shall refer to minus it as the ``fluctuating part'' of the action $S_{\text{fluc}}$. It reads
\begin{widetext}
\be
\begin{split}
	S_{\text{fluc}} =  -\ha \bm{\tilde{\sigma}}^\dag \ast \bm{M} \ast \bm{\tilde{\sigma}} = - \ha \int \rmd X \, \lqq \lt \tilde{\sigma}^x_X \rt^2 + \lt \tilde{\sigma}^y_X \rt^2 + n_X \tilde{n}_X^2 + (\sigma^x_X \tilde{\sigma}^x_X + \sigma^y_X \tilde{\sigma}^y_X) \tilde{n}_X \rqq
\end{split}
\ee
\end{widetext}
The remainder comes instead from the conservative portion of the Langevin equations and constitutes (minus) the ``deterministic part'' of the action $S_{\text{det}}$. The generating functional now has the form
\eq{Eq30}{
Z[\bm{\tilde{h}}] = \int\mathcal{D}[\bm{\sigma} , \bm{\tilde{\sigma}}] \rme{\bm{\tilde{h}} \ast \bm{\sigma} - S_{\text{det}}-S_{\text{fluc}}}.}
The total action of the system is defined as the sum $S=S_{\text{det}}+S_{\text{fluc}}$. Its full expression is reported below, where we employ the additional abbreviation $P_X=(D\nabla^2+\chi)$ to make it more compact:
\begin{widetext}
\begin{eqnarray}
S&=&\int_X\tilde{n}_X\left[\left(\partial_t-D\nabla^2+1-\chi\right)n_X+\left(2n_X-\frac{\omega}{\chi}\sigma^y_X\right)P_X n_X-\frac{1}{2}\left(\tilde{n}_X n_X-\tilde{\sigma}^x_X\sigma^x_X-\tilde{\sigma}^y_X\sigma^y_X\right)\right]\nonumber\\
&&+\int_X\tilde{\sigma}^x_X\left[\left(\partial_t+\frac{\chi+1}{2} +\left(\frac{\omega}{\chi}\sigma^y_X+n_X\right)P_X\right)\sigma^x_X-\frac{1}{2}\tilde{\sigma}^x_X\right]\nonumber\\
&&+\int_X\tilde{\sigma}^y_X\left[\left(\partial_t+\frac{\chi+1}{2}+
n_X P_X\right)\sigma^y_X + \frac{2\omega}{\chi}  \left(2 n_X - 1 \right)P_X n_X-\frac{\omega}{\chi}\sigma^x_XP_X\sigma^x_X-\frac{1}{2}\tilde{\sigma}^y_X\right].\label{Eq31}
\end{eqnarray}
\end{widetext}
Successively integrating over the two gapped $\sigma^{x,y}$ fields by neglecting irrelevant derivative terms, we derive the effective microscopic action for the active site density $n$ alone. This procedure is detailed in the appendix \ref{AppA} and yields
\eq{Eq32}{
S=\int_X\tilde{n}_X\left[\left((\partial_t-D\nabla^2)n_X+\frac{\partial \Gamma (n_X)}{\partial n_X}\right)  -\tilde{n}_X^2\Xi(n_X)\right].}
In this functional, the information on the coupling to the $\sigma^{x,y}$ modes is encoded in the effective potential $\Gamma$ and the noise vertices $\Xi$. 

The potential has the form
\eq{Eq33}{
\Gamma (n_X)=\frac{\Delta}{2}n_X^2+\frac{u_3}{3}n_X^3+\frac{u_4}{4}n_X^4,
}
where $\Delta$ represents the gap and $u_3, u_4$ the cubic and quartic nonlinearities. The quartic one 
\eq{Eq34}{
u_4=\frac{8\omega^2}{\chi+1}
}
is always positive and ensures dynamical stability of the system, i.e. it guarantees a finite steady-state solution $n_X < + \infty$. On the other hand, the cubic nonlinearity
\eq{Eq35}{
u_3=2\chi-\frac{4\omega^2}{\chi+1}
}
experiences a negative correction due to the coherent coupling $\omega$ and becomes negative for $\omega>\sqrt{\chi(\chi+1)/2}$. The existence of the quartic coupling and the negative correction for the cubic coupling result from coherent second order conversion processes 
\eq{Conv}{|\uparrow\rangle\rightarrow\frac{\Omega}{\sqrt{2}}(|\uparrow\rangle+|\downarrow\rangle)\rightarrow\left\{\begin{array}{ll}\Omega^2|\downarrow\rangle& \text{ in }u_4\\
\Omega^2|\uparrow\rangle&\text{ in }u_3\end{array}\right.}
 and vice versa. Due to the permanent decay of the coherences, such processes are suppressed by a factor $\frac{1}{\chi+1}$.

The gap 
\eq{Eq36}{
	\Delta =1-\chi-\frac{\omega^2}{2(1+\chi)^3}
}
can be either positive or negative. In the former case ($\Delta > 0$), the decay from up-spin states exceeds the ``pumping'' processes and the system is driven towards the absorbing, state. For $\chi>1$, the gap is generally negative and the system ends up in a finite density phase, while for $\chi<1$, the strength of the coherent conversion processes $\sim \omega^2$ determines whether the system remains active or becomes inactive. The correction $\propto \omega^2$, coming from the coherent processes, is again suppressed by a factor $\frac{1}{\chi+1}$ and proportional to the fluctuations in the $\sigma^x$ channel $\sim\frac{1}{(\chi+1)^2}$, see Appendix~\ref{AppA}. The physics corresponding to the potential $\Gamma$ will be further detailed in the next Section.

The noise vertices
\eq{Eq37}{
\Xi (n_X)=  \mu_3 n_X  +  \mu_4 n_X^2
}
with couplings
\be
	\mu_3 = \ha \mand   \mu_4 = \frac{2\omega^2}{(\chi+1)^2}\left(1+\frac{32}{(\chi + 1)^4}\right)
\ee
vanish for $n_X\rightarrow0$. The linear multiplicative noise factor $\sim n_X$, which we already discussed in the previous Section, is joined here by a quadratic term $\sim n_X^2$, which is proportional to the coherent coupling $\omega^2$, stemming from second order conversion processes. The importance of the noise vertex for the phase transition will be discussed in Secs.~\ref{BiCri}, \ref{SecFO}. 

Starting from the microscopic Langevin equations, we have derived here the the effective density action \eqref{Eq32}. Its impact on the dynamics of a Rydberg atomic setting will be analyzed in the following sections.

\section{Results}
\label{sec:results}
We shall now discuss in further detail the physics of the quantum contact process, which is encoded in the effective density \eqref{Eq32}. We start by analyzing some of its general properties. Subsequently, we discuss the phase diagram, which contains active and absorbing regimes, as well as the corresponding first and second order phase transitions. In the last part of this Section, we discuss the scaling regimes corresponding to the second order phase transition and the associated critical exponents.
\subsection{The Action}
The action \eqref{Eq32} interpolates between three structurally different limits:
\begin{itemize}
\item[(A)] \emph{classical}: $\omega \to 0$, associated to a continuous phase transition in the DP universality class.
\item[(B)] \emph{quantum}: $\chi \to 0$, associated to a first-order phase transition between an absorbing and an active state.
\item[(C)] \emph{competing}: $u_3 \to 0$, featuring a bicritical point which separates between the two regimes above.
\end{itemize}
In region (A), the coupling $u_3 \approx 2 \chi > 0$ does not vanish and one can perform the transformation $n_X\rightarrow n_X/(\sqrt{2u_3})$ and $\tilde{n}_X\rightarrow\tilde{n}_X\sqrt{2u_3}$, such that the action \eqref{Eq32} becomes
\begin{eqnarray}
S&=&S_{\text{DP}}+S_4\label{Act1}\\
&=&\int_X \tilde{n}_X\left[\partial_t-D\nabla^2+\Delta+\sqrt{\frac{u_3}{2}}(n_X-\tilde{n}_X)\right]n_X\nonumber\\
&&+\int_Xn_X^2\tilde{n}_X\left[\frac{u_4}{2u_3}n_X-\mu_4\tilde{n}_X\right].\nonumber
\end{eqnarray}
The first part, $S_{\text{DP}}$, corresponds precisely to the Reggeon field theory, which is known to describe the physics of directed percolation \cite{Reggeon, Moshe1978, Grassberger1978}. It stands invariant under the transformation 
\be
	n_X\leftrightarrow -\tilde{n}_X, \ \ t\rightarrow-t,
\ee 
which is a characteristic symmetry of DP, known as ``rapidity inversion'' \cite{Janssen2005, Kamenev}.
The second term $S_4$ represents instead the modification to the classical action due to the coherent terms and scales in fact as $S_4\sim\omega^2\rightarrow0$ for $\omega\rightarrow0$. This term breaks rapidity inversion, however, for $\omega\ll\chi$ the quartic correction is negligible in a two-fold sense. First and more importantly, loop corrections to the action stemming from the integration over the cubic couplings are strongly infrared-sensitive in dimensions $d<4$ and, on large wavelengths, dominate over the quartic loop corrections. Second, the microscopic parameters $\mu_4, u_4$ are much smaller than their cubic counterparts such that even on short distances, $S_4$ can be considered an unimportant perturbation. The regime in which the dynamics is dominated by $S_{\text{DP}}$ features a DP phase transition; its extension to values $\omega > 0$ will be discussed in Sec.~\ref{secCrit}.

The second important parameter region (B) features a large and negative $u_3 < 0$, which leads to the emergence of a second, meta-stable minimum in the potential landscape. In this regime, the transition from the absorbing to the active phase is first-order and takes place at finite gap $\Delta \neq 0$. In this case, no irrelevant terms can be dropped from Eq.~\eqref{Eq32} and a different approach is required. Its discussion will be covered in Sec.~\ref{SecFO}.

The third region (C) is identified by $u_3=0$. At this point, coherent and classical processes determine the dynamics of the system on equal footing, which leads to the cancellation of the cubic coupling.
Obviously, for this point (and generally for $u_3\le0$) the above introduced transformation to directed percolation type models is not well defined; this anticipates a modified canonical power counting and a universality class different from DP. Instead, we perform the transformation $n_X\rightarrow n_X(u_4/\mu_3)^{-1/3}, \tilde{n}_X\rightarrow\tilde{n}_X(u_4/\mu_3)^{1/3}$, which yields
\begin{eqnarray}
S&=&S_{\text{QP}}-\tilde{S}_4\label{Act2}\\
&=&\int_X\tilde{n}_X\left[\partial_t-D\nabla^2+\Delta+\left(u_4\mu_3^2\right)^{\frac{1}{3}}\left(n_X^2-\tilde{n}_X\right)\right]n_X\nonumber\\
&&-\int_X\mu_4 \tilde{n}^2_Xn^2_X.\nonumber
\end{eqnarray}
For case (C), the first part $S_{\text{QP}}$ determines the long-wavelength dynamics and encodes the novel critical features of the quantum contact process. The loop corrections involving the $\mu_4$ vertex, on the other hand, are subleading and may be neglected. This will be expanded upon in Sec.~\ref{BiCri}. During the completion of this work, we became aware that an action of the form of $S_{\text{QP}}$ has been discussed in the context of generalized reaction diffusion models with a unique absorbing state in Refs.~\cite{TriDP_FT1, TriDP_FT2}, and classical models falling in this class numerically analyzed in Refs.~\cite{Lubeck2006, Grassberger2006}. In these models, as in the present case, $S_{\text{QP}}$ describes the long-wavelength physics at a critical point, which separates a continuous phase transition, corresponding to the directed percolation universality class, from a discontinuous first order phase transition. The corresponding scaling regime and dynamics has been termed ``tricritical directed percolation'' although the considered systems display only two distinct, stable thermodynamic phases. In the present work, the critical point represents the end point of a line of second order transitions, which separates an active from an inactive phase and represents thus a bicritical point. Since previous analysis of the corresponding dynamics in the literature is rare and inconclusive, we perform an independent mean-field and renormalization group analysis of $S_{\text{QP}}$, i.e. the tricritical directed percolation dynamics, in the following sections. In the present case, this regime is established by classical and coherent contact dynamics on equal footing and will term it the ``quantum contact process''.

\subsection{Mean-Field Phase Diagram}\label{PD}
\begin{figure}[t!]
	\includegraphics[width=1.05\columnwidth]{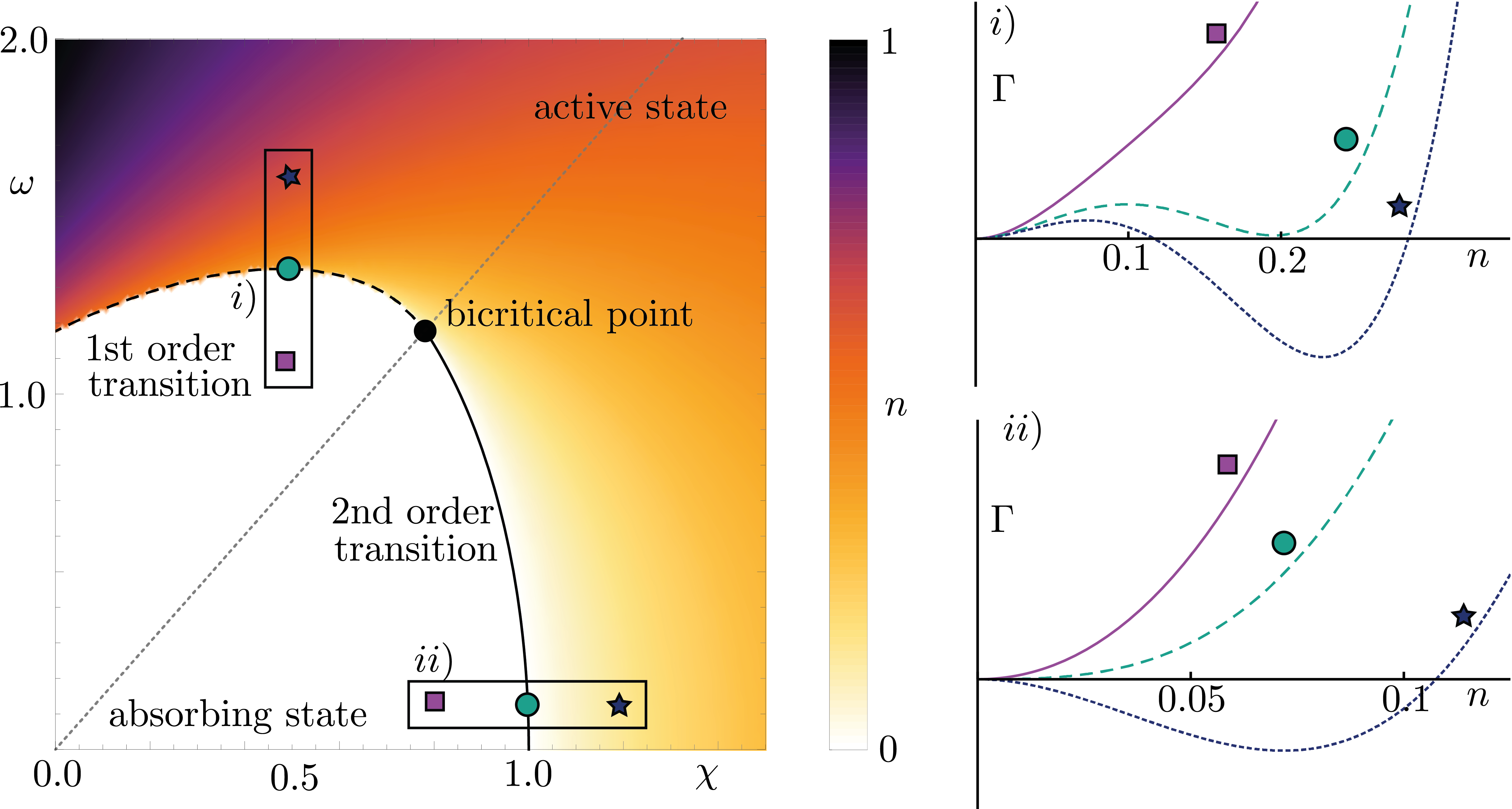}
\caption{Mean-field phase diagram of the quantum contact process. The system can undergo a phase transition from an absorbing state towards an active, non-zero density phase, which can be either continuous (solid line) or first order (dashed line). The second and first order lines meet at a bicritical point. The axes represent the rescaled classical branching rate $\chi=z\kappa/\gamma$ and the quantum branching rate $\omega=z\Omega/\gamma$ and correspond to the classical (A) and quantum (B) limits, respectively. The dotted diagonal line indicates the competing regime (C). On the right, the corresponding evolution of the potential as a function of the axes' parameters is shown for $i)$ the first order transition and $ii)$ the second order transition.
}
\label{fig:1}
\end{figure}

In the thermodynamic limit, the steady state corresponds either to the absorbing phase or to the active, finite density phase. In this Section, we discuss the mean-field phase diagram and the nature of the active-to-inactive phase transition for different parameter regimes by neglecting spatial fluctuations at the level of the action. This corresponds to restricting to a stationary, spatially-uniform ($n_X \to n$, $\tilde{n}_X \to \tilde{n}$) saddle-point approximation of the path integral, which satisfies the Euler-Lagrange equations
\be
\begin{split}
	\frac{\delta S}{\delta \tilde{n}} &= 0   \ \ \Leftrightarrow \ \   \Gamma'(n) -2 \tilde{n} \Xi (n)   = 0     ,  \\
	\frac{\delta S}{\delta n} & = 0   \ \ \Leftrightarrow \ \  \tilde{n} \lqq \Gamma''(n) - \tilde{n} \Xi'(n)   \rqq = 0,
\end{split}	
\ee
where the primes are the standard notation for differentiation with respect to the argument. A further simplification comes from the properties of the response fields: according to Eq.~\eqref{eq:response},
\be
	\tilde{n} = \avb{\tilde{n}_X} = \avb{\frac{\delta \mathbb{1}}{\delta h^n_X}} \eval{h^n = 0} = 0,
\ee
i.e., $\tilde{n}$ is the response of the identity to an external field and therefore trivially vanishes. Hence, one finds the intuitive result that the properties of the system are encoded in the potential \eqref{Eq33} (reported here for convenience)
\be
	\Gamma(n) = \frac{\Delta}{2} n^2 + \frac{u_3}{3} n^3 + \frac{u_4}{4} n^4, 
\ee
with the couplings \eqref{Eq34}-\eqref{Eq36}. The potential $\Gamma$ describes the deterministic dynamics \emph{in the absence of noise and spatio-temporal fluctuations}; in the long-time limit, this dynamics will relax towards its global minimum, whose properties thereby determine the thermodynamic phases and the in-between phase boundaries. The corresponding results are reported in the left panel Fig.\ref{fig:1}.
Recalling that $u_4$ is always positive, one can distinguish three different regimes:
\begin{itemize}
\item[(I)] For $\Delta<0$, $\Gamma$ has a single minimum at finite density ${n_X=n_{\text{MF}}=\frac{-u_3+\sqrt{u_3^2-4u_4\Delta}}{2u_4}}$. This region is thus a portion of the active phase. 
\item[(II)] For $\Delta>0, u_3>0$, there is a single minimum of the potential at $n_X=0$ and the absorbing state is the steady state of the system.
\item[(III)] For parameters $\Delta>0, u_3<0$, $\Gamma$ has one local minimum at $n_X=0$ and a second local minimum at $n_X=n_{\text{MF}}$. In the absence of noise, the system will always relax towards the global minimum of the potential, which is located at $n_X=0$ for $u_3>u_c=-3\sqrt{\frac{u_4\Delta}{2}}$ and at $n_X=n_{\text{MF}}$ for $u_3<u_c$.
\end{itemize}
The nature of the transition between the active and the absorbing phases depends on the position in parameter space. 

We start from the boundary separating (I) from (III), corresponding to the regime dominated by the quantum limit (B) discussed above: for $u_3<0$, i.e. for $\omega>\sqrt{\chi(\chi+1)}/2$, the phase transition takes place at $u_3=u_c$. As the transition line is crossed, the density jumps from zero to a finite value. Furthermore, the system remains gapped ($\Delta>0$) at the transition and keeps a finite correlation length $\xi=\sqrt{D/\Delta}<\infty$. These are hallmarks of a discontinuous, first-order phase transition.
Due to the finite correlation length, the theory remains well behaved at long wavelengths (i.e., free of infrared singularities) and the qualitative mean-field picture does not break down once fluctuations are included. The latter will only lead to perturbative corrections of the system parameters, which may become quantitatively substantial, but remain finite. An interesting situation appears in one spatial dimension, where the critical region of the neighboring bicritical point -- i.e., the region where critical fluctuations become comparable with the mean-field couplings and therefore dominate the behavior of the system -- grows to encompass part of the first-order line (see Fig.~\ref{fig:2}). This leads to strong, infrared dominated corrections to the dynamics of the first order transition. An estimation of the extension of the critical region via the calculation of the corresponding Ginzburg scale \cite{Amit1973} will be provided further below. Apart from spatial fluctuations, one has to consider the effect of the non-equilibrium noise vertices $\Xi$. Their effect is non-perturbative and leads to a shift of the transition line, which is discussed in Sec.~\ref{SecFO}.

For $u_3\ge0$, the transition takes place when the gap vanishes ($\Delta=0$), corresponding to the boundary between (I) and (II) and to the regime dominated by classical physics (B). In this case, the density varies continuously across the transition and the phase transition is of second order. Due to the vanishing gap, spatial fluctuations induce infrared divergent corrections to the dynamics and the mean-field picture is significantly modified. The relevant scaling and the corresponding regimes are discussed in Sec.~\ref{secCrit}.

Finally, a special role is played by the point $(\Delta, u_3)=(0, 0)$ (lying within (C)) at which the first and second order transition lines terminate. This represents a bicritical point, for which the physics is dominated by the coherent vertex $u_4\sim \omega^2$ alone. The corresponding scaling regime is discussed in the subsequent Section, while a renormalization group analysis is presented in Sec.~\ref{BiCri}.
\begin{figure}[t]
	\includegraphics[width=0.95\columnwidth]{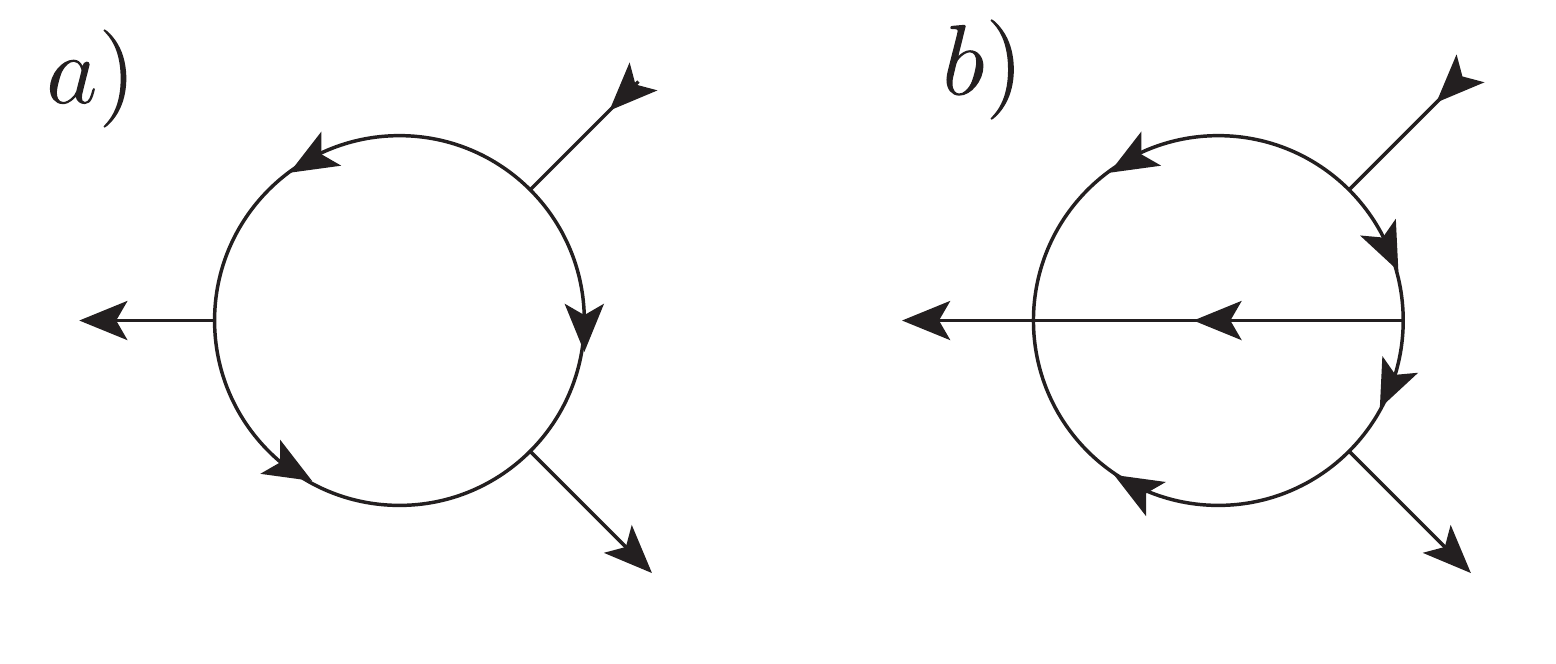}
\caption{Diagrammatic representation of the a) one-loop and b) two-loop correction of the vertex $\mu_3\tilde{n}^2n$. Ingoing lines represent density fields $n$, while outgoing lines represent response fields $\tilde{n}$. The retarded propagator \eqref{Eq42} corresponds to a directed line, where the arrow points from earlier to later times (the propagator vanishes if this order is inverted, see \eqref{eq:rsp2}).
}
\label{fig:3}
\end{figure}

\subsection{Scaling Regimes at the Second Order Transition}\label{secCrit}
\begin{figure*}
	\includegraphics[width=1.9\columnwidth]{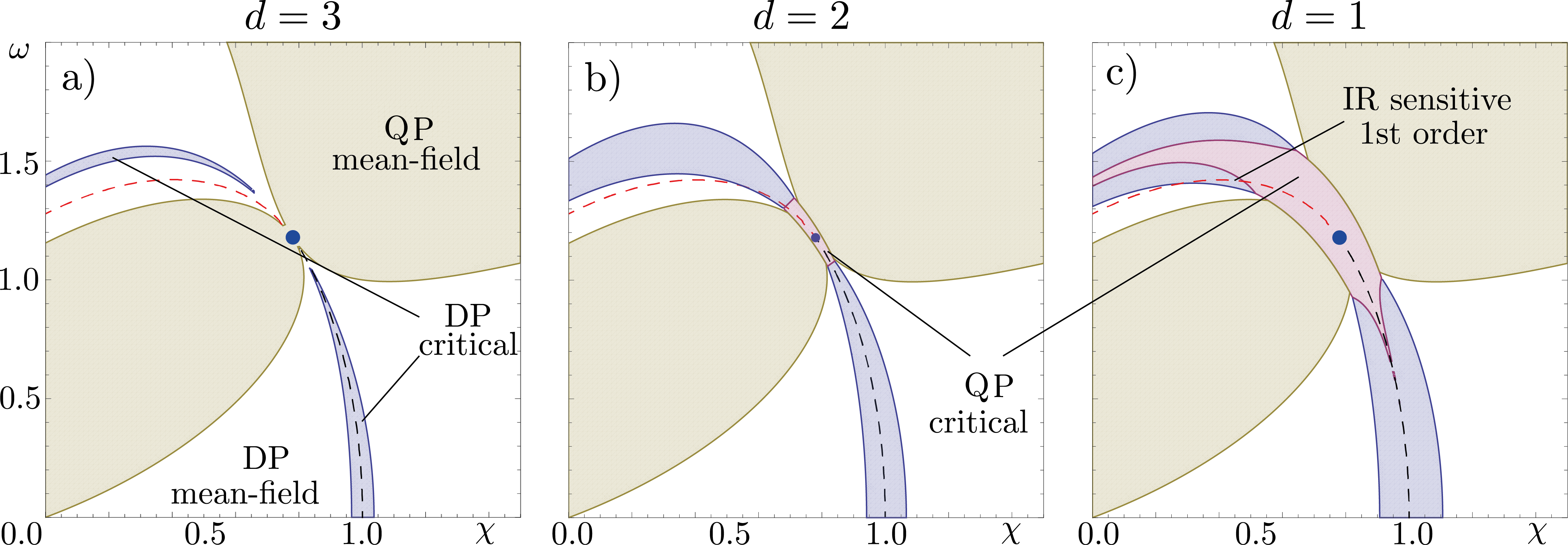}
\caption{Scaling regimes for the second order phase transition in dimensions $d=3,2,1$. In the white region, mean-field scaling behavior according to classical directed percolation (DP) is observed, while the blue region corresponds to critical scaling of the classical DP universality class below the Ginzburg scale. In the yellow regions, the dynamics is dominated by the mean-field behavior of the quantum contact process (QP). The critical behavior of the QP corresponds to the bicritical point and is found in the red region. The black (red) dashed line indicates the line of second (first) order transitions. Remarkably, in one dimension, the first order transition is located partly in the critical regime and experiences strong infrared corrections. 
}
\label{fig:2}
\end{figure*}

We present here the universal scaling behavior at the second order phase transition and identify the corresponding regimes for which it is observable. The mean-field description of the previous Section breaks down when fluctuation corrections become significantly strong. The scale at which this occurs is known as \emph{Ginzburg scale} \cite{Amit1973}; we will discuss it together with the corresponding scaling corrections. The main points highlighted in this Section are summarized in Table \ref{table:scaling} and Fig.~\ref{fig:2}.

\begin{table}[H]
\begin{center}
\begin{tabular}{|C{1.5cm} |C{1.5cm} | C{1.5cm}| C{1.5cm}|C{1.5cm}|}
  \hline
  \rule{0mm}{3ex}
Exponent & mean-field &$d=3$&$d=2$& $d=1$\\
  \hline\hline
\rule{0mm}{3ex} $\beta_{\text{QP}}$& 0.5&$ - $&$0.353$ &$0.218$\\ \rule{0mm}{3ex} 
$\beta_{\text{DP}}$&$1$&$0.81(1)$ &$0.584(4)$ &$0.2764$\\ 
\hline
\rule{0mm}{3ex} $\nu_{\text{QP}}$&0.5 &$-$ &$0.521$ &$0.545$\\ \rule{0mm}{3ex} 
$\nu_{\text{DP}}$&$0.5$ &$0.581(5)$ &$0.734(4)$ &$1.0968$\\
\hline
\rule{0mm}{3ex} $z_{\text{QP}}$&2& $-$&$1.965$ &$1.930$  \\ \rule{0mm}{3ex} 
$z_{\text{DP}}$&$2$&$1.90(1)$ &$1.76(3)$ &$1.5807$ \\
  \hline
\end{tabular}
\caption{Critical exponents for the directed percolation (DP) and the quantum contact process (QP) universality classes. The latter corresponds to the bicritical point in the phase diagram. The corresponding scaling regimes are illustrated in Fig.~\ref{fig:2}. The critical exponents for the DP universality class are exact numerical values, taken from Refs.~\cite{Jensen1992, Voigt1997, Jensen1999}, while the estimates for the exponents of the QP result from a functional renormalization group approach, presented in Sec.~\ref{BiCri}.}
\label{table:scaling}
\end{center}
\end{table}

In the following, we shall employ the standard notation \cite{Ma_book, QPT, Huang_book, PelVicari} for the critical exponents of magnetic systems, such that in a neighborhood of the critical point the order parameter (density of active sites) scales as $n \sim \Delta^\beta$ with the closing gap, the correlation length $\zeta$ diverges as $\zeta \sim \Delta^{-\nu}$ and the dispersion relation of the frequencies $\freq$ vanishes as $\freq \sim q^z$ in the limit of vanishing momenta $q \to 0$ (IR, or large-wavelength limit), with $z$ the \emph{dynamical exponent}, not to be confounded with the aforementioned coordination number of the lattice appearing e.g.~in Eqs.~\eqref{Eq8}-\eqref{Eq10}. With these conventions, the scaling dimension $d_n$ of the field $n$ can be expressed via the hyperscaling relation $d_n = -\beta / \nu$. Within the mean-field description, $\freq = iD q^2$ and $\zeta = \sqrt{D/\Delta}$. Consequently, $z_{MF} = 2$ and $\nu_{MF} = 1/2$ independently of the dimensionality.
In order to determine the order parameter exponent $\beta_{MF}$, one has to solve the deterministic (stationary) equation for the density in the absence of fluctuations
\eq{Eq38}{
\partial_t n = 0 = \frac{\delta\Gamma}{\delta n} = \left(\Delta+u_3n+u_4n^2\right)n.
}
Apart from the absorbing state solution $n=0$, one finds
\eq{Eq39}{
n=\frac{u_3}{2u_4}\left[\sqrt{1+\frac{4\Delta u_4}{u_3^2}}-1\right]=\left\{\hspace{-0.15cm}\begin{array}{ll}(\Delta/u_4)^{\frac{1}{2}} & \text{for } u_3=0\\
\Delta/u_3 & \text{for } u_4=0\end{array}\right. \hspace{-0.1cm}.\  \ \ \
}
As a consequence, the value of the order parameter exponent depends on the parameter regime. For $4\Delta u_4 \gg u_3^2$ the classical branching process dominates over the coherent process and $\beta_{MF}=1$. In this limit, the theory describes the mean-field dynamics of classical directed percolation (DP).\\
On the other hand, for $4\Delta u_4  \ll u_3^2$ the coherent processes dominate and the corresponding scaling behavior is that of a $\phi^4$-theory with a non-equilibrium noise vertex $\Xi\sim n$, which explicitly breaks the $Z_2$ symmetry $n \to - n$, $\tilde{n} \to - \tilde{n}$. We term this the quantum contact process regime (QP) and the corresponding mean-field exponent is $\beta_{MF} = \frac{1}{2}$.

The mean-field predictions remain valid as long as fluctuations remain small. In order to estimate the scale (in particular, we choose here the gap $\Delta$) at which fluctuations become sufficiently correlated to compete with the average local field, i.e. the \emph{Ginzburg scale}, we compare here the bare couplings with the one-loop perturbative corrections induced by the interaction terms, which read
\eq{Eq40}{
\Delta S=\frac{1}{2}\text{Tr}\log S^{(2)}, \text{ with } S^{(2)}_{\alpha \beta, XY}=\frac{\delta^2 S}{\delta n_{\alpha, X}\delta n_{\beta, Y}}.
}
Here, the indices $\alpha$ and $\beta$ distinguish between density ($n_{1,X} = n_X$) and response fields ($n_{2,X} = \tilde{n}_X$). A detailed computation of the loop corrections can be found in Appendix \ref{app:Loop}.
The strongest infrared divergence is associated to the most relevant non-linearity in the action and therefore produces a correction to the cubic coupling $\mu_3$, which can be represented as the Feynman diagram in Fig.~\ref{fig:3} (a). It corresponds to the frequency and momentum integral
\eq{Eq41}{
\delta \mu_3^{(1)} = u_3\mu_3^2 \int_{\freq,q}\left(G_{q,\freq}\right)^2G_{q,-\freq}\approx\gamma^{(1)}_d\frac{u_3\mu_3^2}{D^{\frac{d}{2}}}|\Delta|^{\frac{d-4}{2}},}
over the retarded Green's functions
\eq{Eq42}{
G_{q,\freq}=(-i\freq + Dq^2 + \Delta)
}
in $d$ dimensions and a dimension-dependent numerical prefactor $\gamma^{(1)}_d$, see App.~\ref{app:Loop}. The diagrammatic representation of Eq.~\eqref{Eq41} is shown in Fig.~\ref{fig:3} a). We recall that $G_{q, \freq}$ corresponds to the response function $\av{n_{q, \freq} \tilde{n}_{-q, -\freq}}$; indeed, Fourier back-transforming to time coordinates yields
\be
	G_{q,t} = \int \frac{\rmd \freq}{2\pi} \rme{-i\freq t} G_{q, \freq} = \theta(t)  \rme{- t(Dq^2 + \Delta)},
	\label{eq:rsp2}
\ee
highlighting the causal structure $G(t < 0) = 0$. $\delta \mu_3^{(1)}$ is diverging for $\Delta \rightarrow 0$ in dimensions $d<4$. The Ginzburg scale is obtained by setting $\delta \mu_3^{(1)} =\mu_3$, which defines a threshold
\be
	\Delta_G \approx \lt \frac{D^{\frac{d}{2}}}{\gamma^{(1)}_du_3 \mu_3}  \rt^{\frac{2}{d-4}} 
\ee
below which ($\abs{\Delta} < \Delta_G$) the fluctuations are strong and the system enters the critical scaling regime, and above which ($\abs{\Delta} > \Delta_G$) fluctuations are instead small and the system is approximated by the mean-field solutions.
In the critical regime, the exponents correspond to the directed percolation universality class below the critical dimension $d_c=4$ (see Table \ref{table:scaling} or Table 2 in \cite{DP_Hinrichsen}).

Approaching the bicritical point, the cubic coupling $u_3$ vanishes, alongside all one-loop corrections to the action. The leading order corrections thus consist of two-loop diagrams, of which the cubic noise correction $\delta\mu_3^{(2)}$ turns out to have the strongest infrared divergence. The diagrammatic representation of this correction is shown in Fig.~\ref{fig:3}b) and the analytical value is determined by the integral
\eq{Dq43}{
\delta\mu_3^{(2)} = \int_{q,p,\omega,\nu}\hspace{-0.5cm}G_{q,\omega}^2G_{p,\nu}G_{p,-\nu}G_{p+q,\nu-\omega}\approx\gamma^{(2)}_d\frac{u_4\mu_3^3}{D^d}\Delta^{d-3}.
}
This correction diverges for $\Delta\rightarrow0$ only in $d<3$ and the corresponding Ginzburg scale is set by $\delta\mu_3^{(2)} = \mu_3$; it reads
\be
	\Delta_G \approx \lt  \frac{D^d}{\gamma^{(2)}_d u_4 \mu_3^2}   \rt^{\frac{1}{d-3}}
\ee
and appears only in dimensions $d=2,1$. Inside the associated critical regime, the scaling behavior is determined by the bicritical point, which represents a different non-equilibrium universality class. In the following Section we set up a functional renormalization group approach and determine the relevant universal quantities.

\section{Renormalization group approach to the Bicritical point}
\label{BiCri}
In the previous Section, we have discussed the emergence of a bicritical point in the phase diagram and analyzed the associated scaling behavior. In three spatial dimensions, this point displays mean-field scaling behavior, with exponents given in Tab.~\ref{table:scaling}. In lower dimensions, however, the exponents experience strong infrared modifications below the Ginzburg scale and one finds universal corrections to the mean-field exponents.
Our estimates for the critical exponents have been determined instead via a background field functional renormalization group approach (FRG) \cite{Buchhold2016, Litim2002}, with results reported in Tab.~\ref{table:scaling}. We devote this Section to present the application of this method to our case.

\subsection{Canonical scaling dimensions}
The action describing the leading order dynamics at the bicritical point is given by $S_{\text{QP}}$ in Eq.~\eqref{Act2}. The cubic coupling $u_3$ is zero at this point and the quartic noise vertex $\sim \mu_4$ represents a subleading correction. Due to causality, all one-loop corrections of the quadratic sector vanish for $u_3=0$. On the other hand, two loop corrections involving $\mu_4$ have a less relevant infrared divergence $\sim |\Delta|^{d-2}$ than the corrections involving $\mu_3$, which diverge as $|\Delta|^{d-3}$ for $\Delta\rightarrow0$. In order to give a first estimate for the critical exponents at the bicritical point, we focus here on the leading order action $S_{\text{QP}}$.

In the absence of a quadratic noise scale $\sim T\tilde{n}_X^2$, the canonical scaling dimensions $d_n, d_{\tilde{n}}$ of the fields $n_X, \tilde{n}_X$ have been determined on the basis of the mean-field scaling behavior and the general properties of the action. As mentioned in the previous Section, the strongest infrared divergence in the absence of $u_3$ appears, below $d=3$, in the renormalization of the cubic noise vertex $\delta\mu_3^{(2)} \sim |\Delta|^{d-3}$. This sets the upper critical dimension of the bicritical point to $d_c=3$. The hyperscaling relation $\beta=-\nu d_n$, which is valid in dimensions $d\le d_c$, together with the mean-field exponents $\beta_{MF} = \nu_{MF} = 1/2$, determines the scaling dimension  $d_n = -1$ in three dimensions. This deviates from the canonical scaling of the directed percolation universality class, which is fixed instead by the rapidity inversion symmetry to $d_n=d_{\tilde{n}}=-d/2$ (i.e., $d_n = -3/2$ in $d = 3$). 
In order to determine the canonical power counting at the bicritical point, one requires the action to be invariant under the canonical scaling transformation $x\rightarrow bx, t\rightarrow b^{z}t$. Thus one finds $z=2$, $d_n+d_{\tilde{n}}=-d$ in the quadratic sector, as well as $2d_n=d_{\tilde{n}}$ in the cubic and quartic sector. This sets the canonical scaling of the fields and couplings at the bicritical point to
\eq{FRG1}{
d_n=-\frac{d}{3}, \ d_{\tilde{n}}=-\frac{2d}{3}, \  d_{\Delta}=-2, \ d_{\mu_3}=d_{u_4}=\frac{2(d-3)}{3}.
}
This result is consistent with the upper critical dimension $d_c=3$, below which the scaling of the non-linearities in Eq.~\eqref{FRG1} becomes relevant. It also reproduces the mean-field scaling $\beta=\nu=1/2$ in $d=d_c=3$.

According to the canonical scaling in Eq.~\eqref{FRG1}, there exist only two additional relevant couplings at the bicritical point. The first one is the cubic coupling $u_3$, which has $d_{u_3}=(d-6)/3$. A non-zero cubic coupling $u_3\neq0$ would therefore induce a much stronger infrared divergence than the couplings $u_4, \mu_3$ and always dominate the renormalization group flow on long wavelengths. A second relevant coupling according to the present power counting is represented by an additive noise scale, described by a term $T\tilde{n}_X^2$. In this term, $T$ acts as an effective low frequency temperature and its canonical scaling dimension is as well $d_T=(6-d)/3$ and would be a relevant perturbation if generated under RG (even if absent in the microscopic model). It is thus important to show, that such terms cannot be generated in the renormalization group flow and $T$ remains pinned exactly to its initial value $T=0$. In the following, we will give a brief argument, why this is the case for the present system based on the functional renormalization group. 

For the present system, the microscopic action has the important property $S_{\tilde{n}=0}=0= S_{n=0}$. The first equality is nothing but the causality condition and must hold for any MSRJD action. The second is a specific property of the present system, resulting from the fact that the microscopic action contains no vertex consisting solely of response fields. Moreover, it is clear that for an action with the property $S_{\tilde{n}=0}=0$ no such noise vertex can be generated on one-loop level. Since the functional renormalization group is the  one-loop exact RG evolution equation for the effective action 
\footnote{For readers that prefer Dyson-Schwinger equations (DSE) over the functional renormalization group, the argument applies in the same way. DSE can be formulated one-loop exact as well. Alternatively, the proof becomes particularly simple when accounting also for $S_{n=0}$: this implies that every vertex has at least one outgoing ($\tilde{n}$) and one ingoing ($n$) line. We look here to the possibility of generating in the RG flow a vertex with only outgoing lines. Say that at some level of the perturbative expansion there exists such a diagram. First, invert for simplicity all arrows. Second, choose any vertex and start following a directed path, always according to the arrows. Since it is impossible to terminate such a path without going out, and that is impossible since now all the external lines are ingoing, at some point a loop must form. Because of the causal structure of the theory (see Eq.~\eqref{eq:rsp2}), any closed directed loop vanishes.},
 the property $S_{\tilde{n}=0}=0$ remains exact for each single renormalization group step. It is thus invariant under renormalization.

This demonstrates the two necessary conditions in order to observe the properties of this specific bicritical point. First, one has to fine tune the coherent and classical branching $\omega, \chi$ into the scaling regime of the bicritical point, centered around the point $u_3=0$ and displayed in Fig.~\ref{fig:2}. Second, there must not be any additive noise scale $\sim T$ or any other pure noise vertex in order to ensure the condition $S_{\tilde{n}=0}=0$. 
\subsection{Functional Renormalization Group Approach}
In order to determine the critical exponents at the bicritical point, we perform a functional renormalization group (FRG) analysis of the effective action $\Gamma_{\text{eff}}$. It is important to remark that $\Gamma_{\text{eff}}$ here is not just the rescaled version of the potential $\Gamma[n]$ under RG, but the full generator of one-particle irreducible (1PI) correlation and response functions \cite{Zinn-Justin, Berges2004}. 
The FRG describes the evolution of the microscopic action $S$ towards the effective action $\Gamma_{\text{eff}}$ via the Wetterich equation \cite{Wetterich1993, Berges2002}
\eq{FRG2}{
\partial_k\Gamma_k=\frac{1}{2}\text{Tr}\ln \left[\left(\Gamma^{(2)}_k+R_k\right)^{-1}\partial_kR_k\right].
}
Here, $k$ is a running momentum scale and $\Gamma_k$ interpolates between the microscopic action $S=\Gamma_{k=\Lambda}$, where $\Lambda$ is the ultraviolet cutoff of the theory and the effective action $\Gamma_{\text{eff}}=\Gamma_{k=0}$. $\Gamma_k^{(2)}$ is the second order functional derivative of $\Gamma_k$ with respect to the fields $n, \tilde{n}$ --- analogously to $S^{(2)}$ in Eq.~\eqref{Eq40} --- and $R_k$ is an optimized momentum cutoff, which is diagonal in momentum and frequency, and has momentum-space matrix elements \cite{Litim2001}
\eq{FRG3}{
R_k(q)=D_k(k^2-q^2)\Theta(k^2-q^2).}
Here, $D_k$ is the flowing diffusion constant (i.e., the value taken by $D$ at the scale $k$).

The defining property of a critical point is the scale invariance of correlation and response functions, which is equivalent to the scale invariance of $\Gamma_{\text{eff}}$. Thus a critical point in parameter space corresponds to the scale invariance of the Wetterich equation \eqref{FRG2} in the limit $k\rightarrow0$. The present system contains at least two different critical points, namely the one corresponding to the directed percolation universality at $\omega=0$ and the bicritical point $u_3=0$. In order to ensure that the effective action flows towards the latter, we initialize Eq.~\eqref{FRG3} with $\Gamma_{k=\Lambda}=S_{\text{QP}}$ and set the cubic coupling $u_3$ to remain zero during the entire flow. Generally, a non-zero flow of $u_3$ is generated when starting from a microscopic action of the form of $S_{\text{QP}}$, which is in accordance with the fact that the bicritical point corresponds to a fine tuning of two distinct parameters, $\Delta, u_3$, both representing relevant directions. Strictly speaking, setting $u_3$ to zero during the flow thus corresponds to a microscopic starting point in the scaling regime of the bicritical point and an RG flow towards the bicritical point, which is reached in the limit $k\rightarrow0$.

For the present approach, we consider only the most relevant vertices at the bicritical point, which corresponds to a truncation of the form
\eq{FRG4}{
\Gamma_k=\int_X\tilde{n}_X\left[Z_k\partial_t+D_k\nabla^2+\Delta_k-\mu_k\tilde{n}_X+u_kn_X^2\right]n_X.
}
Within this truncation, the flow of $\Gamma_k$ is mapped onto the flow of the field independent couplings $(Z_k, D_k, \Delta_k, \mu_k, u_k)$. In the limit $k\rightarrow\Lambda$ the microscopic parameters are recovered and the ``wave function'' renormalization (i.e., the renormalization factor of the composite field $n \tilde{n}$) $Z_{\Lambda} \to 1$. As discussed in the previous Sections, the truncated action \eqref{FRG4} does not get renormalized at the one-loop level, and the corresponding FRG flow is zero according to \eqref{FRG2}. New contributions appear, however, in a two-loop computation, and indeed $\Gamma_k$ gets renormalized. Thus, for the specific dynamics at the bicritical point, the leading order corrections are of two-loop order and one has to modify the truncation \eqref{FRG4} in oder to capture this effect.

In order to do so, we approach the bicritial point not from the absorbing, but from the active phase, such that the density field $n_X\rightarrow n_X+\rho_k$ is expanded around a finite stationary value $\rho_k$. This new variable $\rho_k$ represents now a $k$-dependent background field \cite{Dashen1981, Buchhold2016}. Some of us have performed the same approach in Ref.~\cite{Buchhold2016} in order to determine the critical exponents of the directed percolation (DP) universality class. The effect of the background field is the effective inclusion of higher order loop corrections within a one-loop computation. There are, however, two major differences between the setup in \cite{Buchhold2016} and the present one. First, in \cite{Buchhold2016} the effective higher order diagrams gave a valuable correction to the leading order renormalization group flow, which led to a significant improvement in the estimates for the critical exponents. In the present setup, however, the effective higher order loop corrections represent the leading order terms in the renormalization group flow. Second, in \cite{Buchhold2016} the background field introduced an imbalance between the response and the density field during the FRG flow, which had to be compensated by an additional flowing parameter. Here, the imbalance between density and response field is present already in the microscopic action and no such parameter has to be introduced. The background field approach increases the number of flowing parameters by one and one has to determine now the flow equations for $(Z_k, D_k, \rho_k, \Delta_k, \mu_k, u_k)$. 

The standard procedure for obtaining these flow equations consists in projecting Eq.~\eqref{FRG2} onto the different (quadratic, cubic, \ldots) sectors \cite{Berges2002, Eq_vs_NonEq}. The flow of the inverse propagator for instance is determined via
\eq{FRG5}{
\partial_k G^{-1}_k(q,\omega)=\left(\frac{\delta^2}{\delta n_{q,\omega}\delta\tilde{n}_{-q,-\omega}}\partial_k\Gamma_k\right)_{n=\tilde{n}=0}
}
and the flow of the remaining couplings is determined accordingly. In order to identify a scale invariant fixed point of these equations, one has to rescale the couplings to make them dimensionless. Moreover, the wave function renormalization and diffusion constants are eliminated via the transformation
\eq{FRG6}{(n, \tilde{n}, t)\rightarrow (n Z_k^{-\frac{1}{3}}, \tilde{n}Z_k^{-\frac{2}{3}}, tZ_kD_k^{-1}),} in accordance with canonical power counting.
The rescaled couplings are
\eq{FRG7}{
\left(\begin{array}{c}\bar{\rho}_k\\[1mm] \bar{\Delta}_k\\[1mm] \bar{\mu}_k\\[1mm] \bar{u}_k\end{array}\right)=\left(\begin{array}{c}\left(\frac{Z_k}{k^d}\right)^{\frac{1}{3}}\rho_k\\[2mm] k^{-2} \frac{\Delta_k}{D_k}\\[2mm] \left(\frac{k^{d-3}}{Z_k}\right)^{\frac{2}{3}}\frac{\mu_k}{D_k}\\[2mm] \left(\frac{k^{d-3}}{Z_k}\right)^{\frac{2}{3}}\frac{u_k}{D_k}\end{array}\right).
}
In these units, the anomalous dimensions read
\begin{eqnarray}
\eta_D&=&-\frac{k\partial_kD_k}{D_k}=\frac{3C_d \bar{u}_k \bar{\mu}_k \bar{\rho}_k}{d(1+\bar{\Delta}_k)^3},\label{FRG8}\\
\eta_Z&=&-\frac{k\partial_kZ_k}{Z_k}=\frac{2(2+d-\eta_D)}{(2+d)}\eta_D,\label{FRG9}
\end{eqnarray}
where 
\be
	C_d = \frac{2 \pi^{\frac{d}{2}}}{\Gamma\lt d/2\rt}
\ee
is the surface of the $d$-dimensional unit sphere (or $d$-dimensional solid angle).
Employing $\eta_Z$ as a shorthand, the flow equations of the background field and the gap can be brought as well into a simple form
\begin{eqnarray}
k\partial_k \bar{\rho}_k&=&\left(\frac{\eta_Z}{4}\left(42+\frac{1+ \bar{\Delta}_k}{ \bar{u}_k \bar{\rho}_k^2}\right)-\frac{d+\eta_Z}{3}\right) \bar{\rho}_k,\ \ \ \ \ \\
k\partial_k \bar{\Delta}_k&=&\lt -2+\eta_D+\frac{\eta_Z}{2}(3+\frac{1}{ \bar{\Delta}_k})\rt \bar{\Delta}_k,
\end{eqnarray}
while the ones for the non-linear couplings show an increased complexity
\begin{align}
k\partial_k\bar{\mu}_k&=\lt \frac{2}{3}(d-3+\eta_Z)+\eta_D-\eta_Z(6-\mathcal{O}(\tilde{\rho}_k^2))\rt \bar{\mu}_k,\label{FRG12}\\
k\partial_k\bar{u}_k&= \lt \frac{2}{3}(d-3+\eta_Z)+\eta_D-\eta_Z(21-\mathcal{O}(\tilde{\rho}_k^2))\rt \bar{u}_k.\ \ \ \ \ \ \ \ \ 
\end{align}
Hence, the corresponding fixed point equations 
\eq{FRG14}{
k\partial_k\left(\begin{array}{c}\bar{\rho}_k\\ \bar{\Delta}_k\\ \bar{\mu}_k\\ \bar{u}_k\end{array}\right)\overset{!}{=}0
}
have to be solved numerically. The critical exponents are extracted from the numerical fixed point values via the relations,
\eq{FRG15}{
z=2-\eta_D+\eta_Z, \ \ \ \beta=\frac{d+\eta_Z}{3}\nu}
while $\nu$ corresponds to the inverse of the largest eigenvalue of the flow equations' stability matrix at the fixed point.
The results of this analysis are summarized in Tab.~\ref{table:scaling}, which provides a quantitative estimate of the critical behavior governed by the class of the bicritical point.

\section{First order transition}\label{SecFO}
In Section \ref{PD} we have seen that in regime (III) (i.e., $\Delta >0$ and $u_3 < 0$) the mean-field prescription predicts the presence of two stable stationary configurations in the dynamics (corresponding to the minima of the effective potential \eqref{Eq33}). Once fluctuations are included, however, one of them becomes metastable and eventually decays to the other one. Depending on the parameters, different minima can become stable in different regimes; the separatrices (i.e., all the points at which stability switches from one solution to the other) between these phases correspond to first-order transitions. In equilibrium, the actual stable state is the global minimum of the free energy, while the remaining local minima are metastable. The present case is however different due to the multiplicative ($\propto \sqrt{n}$) nature of the noise, which makes fluctuations much more relevant in the neighborhood of the finite-density minimum than in the one of the absorbing state. Therefore, this produces a bias towards the latter, which must be accounted for.  
The metastable dynamics takes place at a finite correlation length $\xi<\infty$ and therefore is not driven by infrared divergent spatial fluctuations. In the following, we will thus neglect them and discuss the first order phase transition in the presence of non-equilibrium noise via an optimal path approximation \cite{Kamenev}.

\subsection{Optimal path approximation}\label{SecOPA}
In the coexistence region (III), the mean-field solutions found in Sec.~\ref{PD} read $n_{0} = 0$ (absorbing) and $n_{MF} = \frac{\sqrt{u_3^2-4u_4\Delta}-u_3}{2u_4}$ (active), which solve
\be
	\Gamma'[n] = \frac{\partial \Gamma}{\partial n}[n] = 0.
\ee

In order to determine the stable phase of this model in the thermodynamic limit (in the following ``thermodynamic phase'' for short), we search for a "classical" trajectory in phase space --- i.e., a trajectory in the $(n,\tilde{n})$ space which keeps the action stationary --- which connects the two minima. This trajectory is referred to as the \emph{optimal path} and determines the preferred minimum of the potential in the presence of noise \cite{Kamenev}. The stationarity of the action is ensured by enforcing the Euler-Lagrange equations
\eq{OPA1}{
\frac{\delta S}{\delta n_X}=\frac{\delta S}{\delta \tilde{n}_X}\overset{!}{=}0.
}
The common solution $\tilde{n}_X=0$ of vanishing noise field yields the deterministic equation of motion
\be 
	\partial_t n_X = -\Gamma'[n_X]
	\label{eq:deterministic}
\ee
for $n_X$ and remains valid for small fluctuations around the minima $n_0$ and $n_{MF}$, but 
does not account for large fluctuations connecting one minimum to the other. We have therefore to look for those solutions of \eqref{OPA1} that do.

Neglecting spatial fluctuations and keeping only the dependence on time, we replace the fields $(n_X, \tilde{n}_X)\rightarrow(n_t, \tilde{n}_t)$ with spatially homogeneous and temporally fluctuating ones. The action $S$ takes the form
\eq{OPA2}{
S=V\int_t \left[\tilde{n}_t\left(\partial_t n_t + \Gamma'[n_t] \right)-\tilde{n}_t^2\Xi [n_t]\right],
}
with $V$ the volume of the system.
This leads to the dynamical saddle point equations 
\begin{eqnarray}
0&=&\frac{1}{V}\frac{\delta S}{\delta \tilde{n}_t}=\partial_tn_t + \Gamma'[n_t] - 2\tilde{n}_t \Xi [n_t],\label{OPA3}\\
0&=&\frac{1}{V}\frac{\delta S}{\delta n_t}=\left(-\partial_t+ \Gamma''[n_t] - \tilde{n}_t\Xi'[n_t]\right)\tilde{n}_t. \label{OPA4}\ \ \ \ \ 
\end{eqnarray}
Note that, by defining the effective Hamiltonian 
\be
	H[n_t, \tilde{n}_t] = \tilde{n}_t \lt  \tilde{n}_t \Xi[n_t] - \Gamma'[n_t] \rt,
\ee
these take the form of Hamilton-Jacobi equations
\be
	\partial_t n_t = \frac{\partial H}{\partial \tilde{n}_t}  \comma    \partial_t \tilde{n}_t = - \frac{\partial H}{\partial n_t},
\ee
and thus $\tfrac{dH}{dt} = 0$. Consequently, trajectories in phase space can be seen as level curves at constant $H \equiv E$. The stationary solutions $n_0$ and $n_{MF}$ have been identified for $\tilde{n} = 0$, which implies $H = 0$. The remaining trajectories at zero energy are given by either $n_t = 0$ or  
\eq{OPA5}{
\tilde{n}_t=\Xi^{-1} \Gamma'.
}
With this last choice, the equation of motion for the density field \eqref{OPA3} becomes
\eq{OPA6}{
\partial_t n_t = \Gamma',
}
corresponding to motion in an inverted potential $-\Gamma$. This implies that the stationary solutions become unstable along these trajectories, which are thus the right candidates to escape from the attraction basins of the steady states and to describe large fluctuations \cite{Kamenev}.

The exponential of the action $\rme{-S[n_t, \tilde{n}_t]}$ represents the statistical weight of a trajectory; for optimal paths we have, upon substitution of Eq.~\eqref{OPA5},
\be
	S_{OP} = V \int_{t_i}^{t_f} \tilde{n}_t \partial_t n_t \, dt = V \int_{n_i}^{n_f} \tilde{n} \, dn = V \int_{n_i}^{n_f} \frac{\Gamma'}{\Xi} \, dn,
\ee
for a generic trajectory connecting an initial field configuration $n_i = n_{t = t_i}$ with a final state configuration $n_f = n_{t = t_f}$. The remaining expression is independent of the initial and final time and one can thus choose $t_i=0$ and $t_f=t$, such that $S_{OP}$ interpolates between the initial and the current state at time $t$.
Fixing a given $n_i$ as a reference value, these rates can be used to reconstruct the probability distribution to reach $n_f$ (for details, see Appendix \ref{app:OPA}) which is proportional to
\be
	\rme{-S_{OP}(n_i,n_f)} = \rme{-V(\mal{F}(n_f) - \mal{F}(n_i))},
\ee
where the integral $\mathcal{F}(n)$ is defined (up to an irrelevant constant) by $\frac{\partial \mathcal{F}}{\partial n}=\frac{1}{\Xi}\frac{\partial\Gamma}{\partial n}$. Setting $n_i = 0$ and normalizing the distribution yields
\eq{OPA9}{
P(n_f)=Z^{-1}\ e^{-V\mathcal{F}(n_f)}, \text{ with } Z=\int dn\ e^{-V\mathcal{F}(n)}.}

We want to stress that the existence of a stationary state probability distribution of the form \eqref{OPA9} does not imply that the system is effectively in thermal equilibrium. The balance between noise and deterministic dynamics leads to an effective "free energy" $\mathcal{F}$ for the stationary distribution. However, detailed balance is not restored in the system, which can be seen by the fact that the equation of motion from the field --cannot-- be read off the function $\mathcal{F}$. Naively taking this $\mal{F}$ as describing thermal equilibrium, the equation of motion derived from Eq.~\eqref{OPA9} would read
\eq{EOMf}{
\partial_t\phi=-\frac{\partial\mathcal{F}}{\partial\phi}+\xi_t\ \ \text{ with } \langle \xi_t\xi_{t'}\rangle=\frac{\delta(t-t')}{V}.
}
This equation is not equivalent to the correct Langevin equation of motion for the dynamics of the field
\eq{EOML}{
\partial_t\phi=-\frac{\partial\Gamma}{\partial\phi}+\zeta_t\ \ \text{ with } \langle\zeta_t\zeta_{t'}\rangle=\frac{\delta(t-t')}{V}\Xi_t.
}
In this sense, the existence of an effective stationary state probability distribution does neither imply detailed balance nor can a corresponding equation of motion be derived from the functional $\mathcal{F}$ alone.

The integral $\mathcal{F}(n)$ can be determined by common integration with respect to $n$ and reads
\eq{OPA10}{
\mathcal{F}(n)=\Delta l+2u_3\tfrac{n-l}{\mu}+u_4\tfrac{n(n\mu-2)+2l}{\mu^2}.\ \ \ 
}
Here, we defined the function $l=\frac{\ln(1+\mu n)}{\mu}$ and noise ratio $\mu=\mu_4/\mu_3 = 2 \mu_4$.

In the thermodynamic limit $V\rightarrow \infty$, the volume factor in the exponent of $P(n_f)$ suppresses all field configurations except the one that minimizes $\mathcal{F}$. Thus
\eq{OPA11}{
P(n_f)\overset{V\rightarrow\infty}{\rightarrow}\delta(n_f-n_{\text{min}}),
}
where $n_{\text{min}}$ is the minimizing density field. One can thus compare $\mathcal{F}(n=0)=0$ with the value $\mathcal{F}(n=n_{\text{MF}})$ and finds the system in the active phase for $\mathcal{F}(n=n_{\text{MF}})<0$ and in the absorbing phase for $\mathcal{F}(n=n_{\text{MF}})>0$. The first order phase transition takes place at ${\mathcal{F}(n=n_{\text{MF}})=0}$. 

\begin{figure}
	\includegraphics[width=1.\columnwidth]{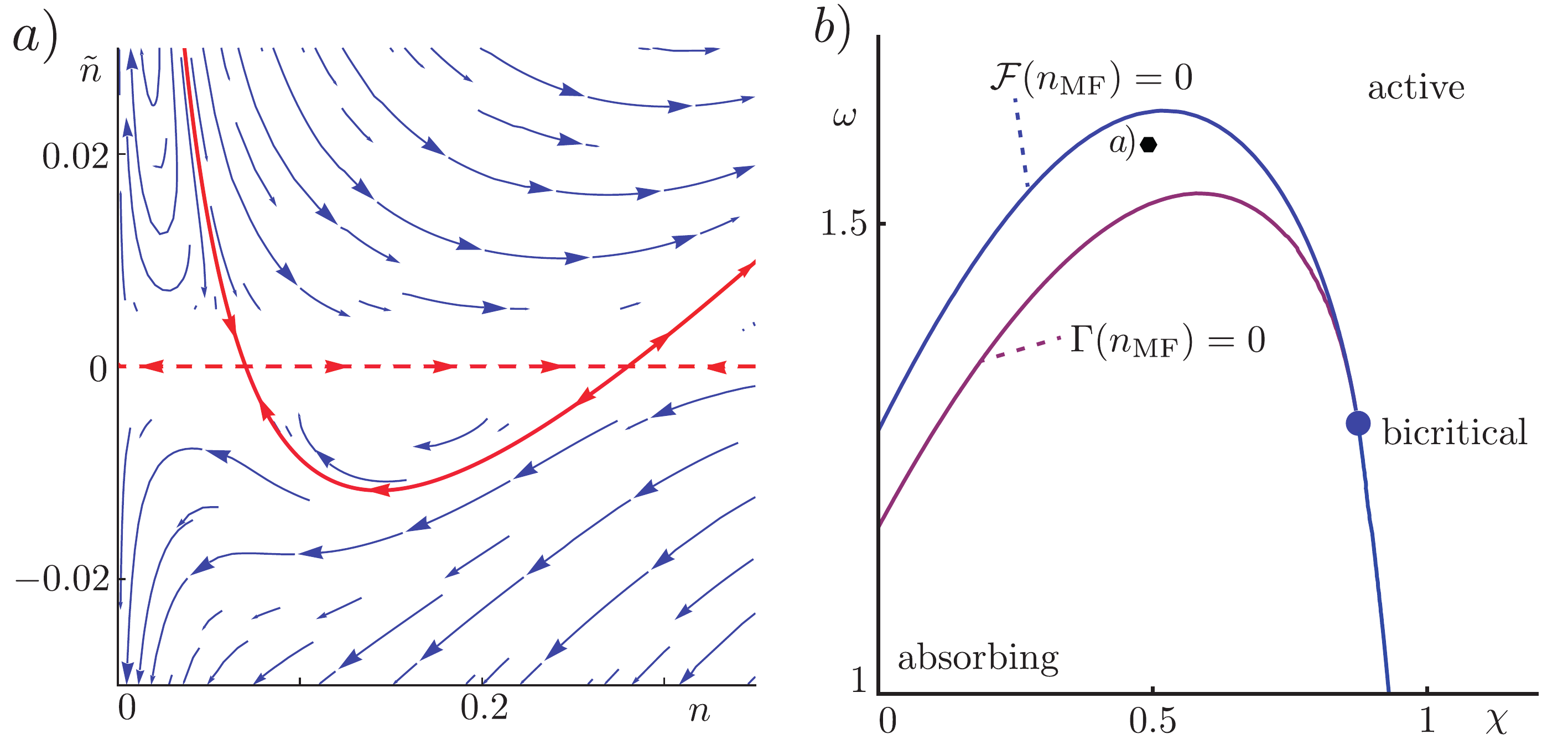}
\caption{Optimal path approximation. a) Phase space $(n, \tilde{n})$ trajectories for parameters $\omega=1.6, \chi=0.5$ (represented by the dot in parameter space in b)). The optimal path trajectories are displayed in red. One distinguishes between the deterministic, zero-noise trajectory (dashed line) and the noise dominated trajectory (solid line). b) Shift of the first order transition line from the mean-field result $\Gamma(n_{\text{MF}})=0$ (red) to the one determined by the optimal path approach $\mathcal{F}(n_{\text{MF}})=0$ (blue) in the presence of temporal, noise-induced fluctuations.
}
\label{fig:4}
\end{figure}

We compare our non-equilibrium result to the more usual case of thermal equilibrium. In the latter case, the noise kernel is simply the temperature $\Xi_t=T$ and the integration over the optimal path trajectories yields the thermal distribution according to the Boltzmann weight function
\eq{OPA12}{
P_{\text{th}}(n)=Z^{-1}_{\text{th}}e^{-\frac{V\Gamma(n)}{T}},
}
where $\Gamma(n)$ can be identified as the free energy density. In this case, the thermodynamic phase is determined by the global minimum of the potential $\Gamma$ independently of the thermal noise strength $T$. We want to stress at this point, that the mean-field value of the phase transition in the absence of noise corresponds exactly to a finite-temperature, equilibrium transition. In Fig.~\ref{fig:4} b) we draw the actual phase boundary of the first order transition in the presence of non-equilibrium noise (determined by $\mathcal{F}(n_{\text{MF}})=0$) and compare it to the thermal transition line (corresponding to $\Gamma(n_{\text{MF}})=0$) as predicted by mean-field. Crucially, the transition line is shifted to larger values of $\omega$ in the presence of a noise kernel, which prefers the absorbing state over any active field configuration and pulls the system towards an empty state.

\subsection{First order transition at finite volume}

\begin{figure}
	\includegraphics[width=1.\columnwidth]{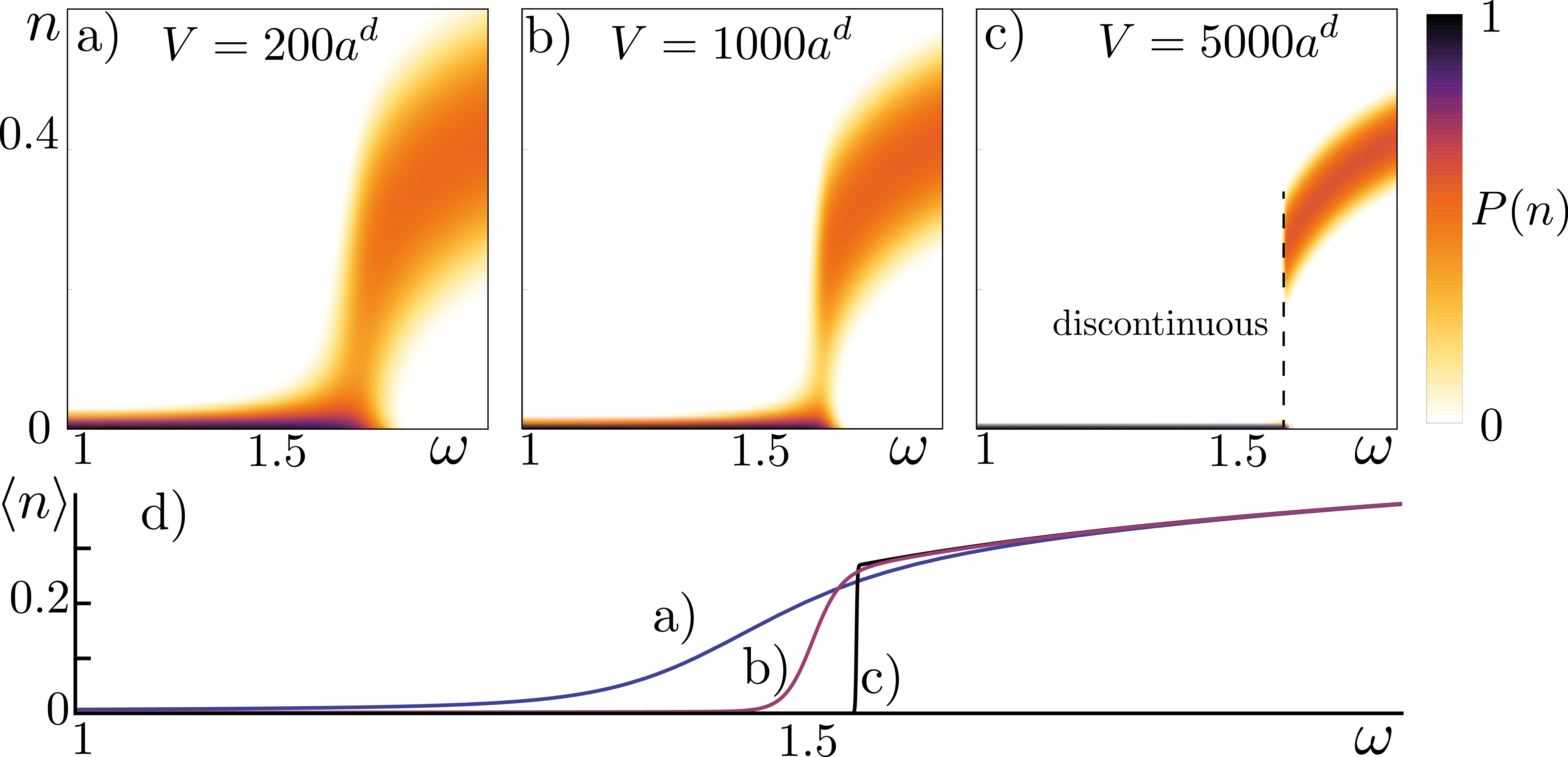}
\caption{First order phase transition at a finite system size $V<\infty$. The figures a)-c) display the density distribution function $P(n)$ in Eq.~\eqref{OPA9} as in the $(\omega, n)$ plane for fixed value of $\chi=0.4$ for different system sizes $V$ in units of the $d$-dimensional unit volume $a^d$. In the thermodynamic limit, a first order transition occurs at $\omega_c\approx1.6$. d) The density expectation value $\langle n\rangle=\int dn\ n P(n)$ is plotted as a function of $\omega$ for $\chi=0.4$. The different plots correspond to the volume $V$ in a)-c). In both rows, the discontinuity at $\omega=\omega_c$ is established for increasing system size and is significant already for moderate system sizes of $V=5000a^d$.
}
\label{fig:5}
\end{figure}

The partition function for the density distribution evolves into a discontinuous $\delta$-function only in the limit $V\rightarrow\infty$ but remains a continuous function for any finite volume $V<\infty$.  Strictly speaking, the corresponding first order phase transition occurs only in the thermodynamic limit and for all finite system sizes, macroscopic observables in the active and absorbing phase are continuously connected.

For the previously discussed second order transitions, the Ginzburg scale sets the minimal system size beyond which universal scaling behavior can be observed. For the first order transition, however, we have not determined such a scale and thus we have to detect numerically at which volumes the discontinuity at the phase transition is observable. In Fig.~\ref{fig:5}, we discuss the density distribution $P(n)$ and the behavior of the average density $\langle n\rangle$, when the system crosses the first order transition line, as a function of the system size $V$. As one can see, already for moderate system sizes of $\approx 5000$ lattice sites, the discontinuity at the transition is clearly visible. 

A comprehensive picture of the first order non-equilibrium phase transition, and a more detailed and quantitative estimation of the nucleation events and their effects on the first-order phase transition would require accounting for spatially inhomogeneous field configurations  as well ($n_t$ back to $n_X$, see e.g.~Chapter 8 of \cite{Kamenev}). In addition, in one dimension, there may be domain wall defects with a non-extensive barrier for their creation, which may act to wash out the first order transition. However, such an analysis is beyond the scope of the current work.

If we assume that every lattice site in our original description corresponds to a frozen, tightly-trapped Rydberg atom, the non-equilibrium phase transitions discussed in the present work should in principle be observable in an experimental context as well, as detailed in the next Section.

\section{Exploring the non-equilibrium phase transition with Rydberg atoms}\label{ExpSec}

In this Section we discuss the possibility of observing aspects of the non-equilibrium phase transitions discussed here in ensembles of laser excited cold atomic gases. To this end we consider a setup in which atoms are confined to a rectangular lattice with one atom per site at positions $\mathbf{r}_l$, as e.g. realized experimentally in Refs.~\cite{Ryd-lattice2, Schauss14, Labuhn2015, Nogrette2014}. The internal dynamics of the atoms is described by a two-level system where $\mid\downarrow\rangle$ is an electronic ground state and $\mid\uparrow\rangle$ is a high-lying Rydberg $S$-state \cite{Rydberg2, Ryd-QI}. Atoms are excited to the Rydberg state with a laser (Rabi frequency $\Omega$ and detuning $\Delta$). Here they interact with a van-der-Waals potential of the form $V_{lm}=C_6/|\mathbf{r}_l-\mathbf{r}_m|^6$ with $C_6$ being the so-called dispersion coefficient that parameterizes the interaction strength \cite{Beguin2013, Rydberg2}. With this modeling, the coherent dynamics of an ensemble of atoms in which Rydberg states are excited is described by the Hamiltonian
\begin{eqnarray}
  H_\mathrm{Ryd}=\Omega\sum_l \si^x_l + \Delta\sum_l \n_l + \frac{1}{2}\sum_{l \neq m} V_{lm} \n_l \n_m,
  \label{eq:RH1}
\end{eqnarray}
where we recall that $\si^x_l=\mid\uparrow\rangle_l\langle\downarrow\mid_l + \mid\downarrow\rangle_l\langle\uparrow\mid_l$ and $\n_l=\mid\uparrow\rangle_l\langle\uparrow\mid_l$ and a rotating-wave approximation has been performed (i.e., a transformation to the frame rotating with the frequency of the laser has been applied and subsequently counter-rotating terms have been neglected).

One central aspect of the non-equilibrium physics explored here is the presence of a facilitation mechanism, i.e.~an enhanced probability of creating an excitation right next to an already existing one. In the context of Rydberg lattice gases, this can be in principle achieved via the so-called anti-blockade condition \cite{Ates07, Amthor2010, Lesanovsky14, PRL-KinC, Valado2015}. In order to realize it, the laser detuning is set to cancel exactly the interaction energy between nearest neighbors, $\Delta+V_{12}=0$. In this case transitions (in a one-dimensional chain) of the kind $\mid...\downarrow\uparrow\downarrow...\rangle\rightarrow \mid...\downarrow\uparrow\uparrow...\rangle$ become resonant. For sufficiently large detuning $|\Delta|\gg|\Omega|$, off-resonant transitions such as $\mid...\downarrow\downarrow\downarrow...\rangle\rightarrow \mid...\downarrow\uparrow\downarrow...\rangle$ are instead suppressed; note, however, that for any finite (albeit large) $\Delta$, these processes are not completely absent. Therefore, the absorbing property of the ``all-down'' state is only an approximation.
With this in mind, we can now formulate an effective Hamiltonian which describes only (near-)resonant transitions:
\begin{eqnarray}
  H_\mathrm{res}=\Omega\sum_l \wh{\Lambda}_l \si_l^x + \frac{1}{2}{\sum_{lm}}^\prime V_{lm} \n_l \n_m.\label{eq:res_rydberg}
\end{eqnarray}
Here the ${\sum}^\prime$ denotes a summation excluding nearest neighbors. The operator $\wh{\Lambda}_l$ is a projector with support on the nearest neighbors of the $l$-th site. It yields $1$ for configurations that contain exactly one single excitation and $0$ otherwise. In one dimension its explicit form reads $\wh{\Lambda}_l= \n_{l-1} + \n_{l+1} - 2 \n_{l-1} \n_{l+1}$.

The Hamiltonian (\ref{eq:res_rydberg}) has a striking resemblance to the one given in Eq.~\eqref{Eq2}. Discrepancies arise in the structure of the operators $\wh{\Lambda}_l$, which however differ from the operators $\PP_l$ only through higher order terms in the local densities $\n_m$. These differences are actually irrelevant to the determination of the critical properties, as detailed later in this Section. The second discrepancy arises from the residual interaction terms which are a consequence of the power-law tail of the van-der-Waals interaction. Clearly, the energy shifts caused by them become more severe in higher dimensions $d>1$, as the distance between next-nearest neighbors decreases. Note, that in principle the importance of the residual long-range interaction can be further suppressed by employing potential shaping techniques as discussed in Ref.~\cite{Marcuzzi2015}.
\begin{figure}[t]
	\includegraphics[width=1\columnwidth]{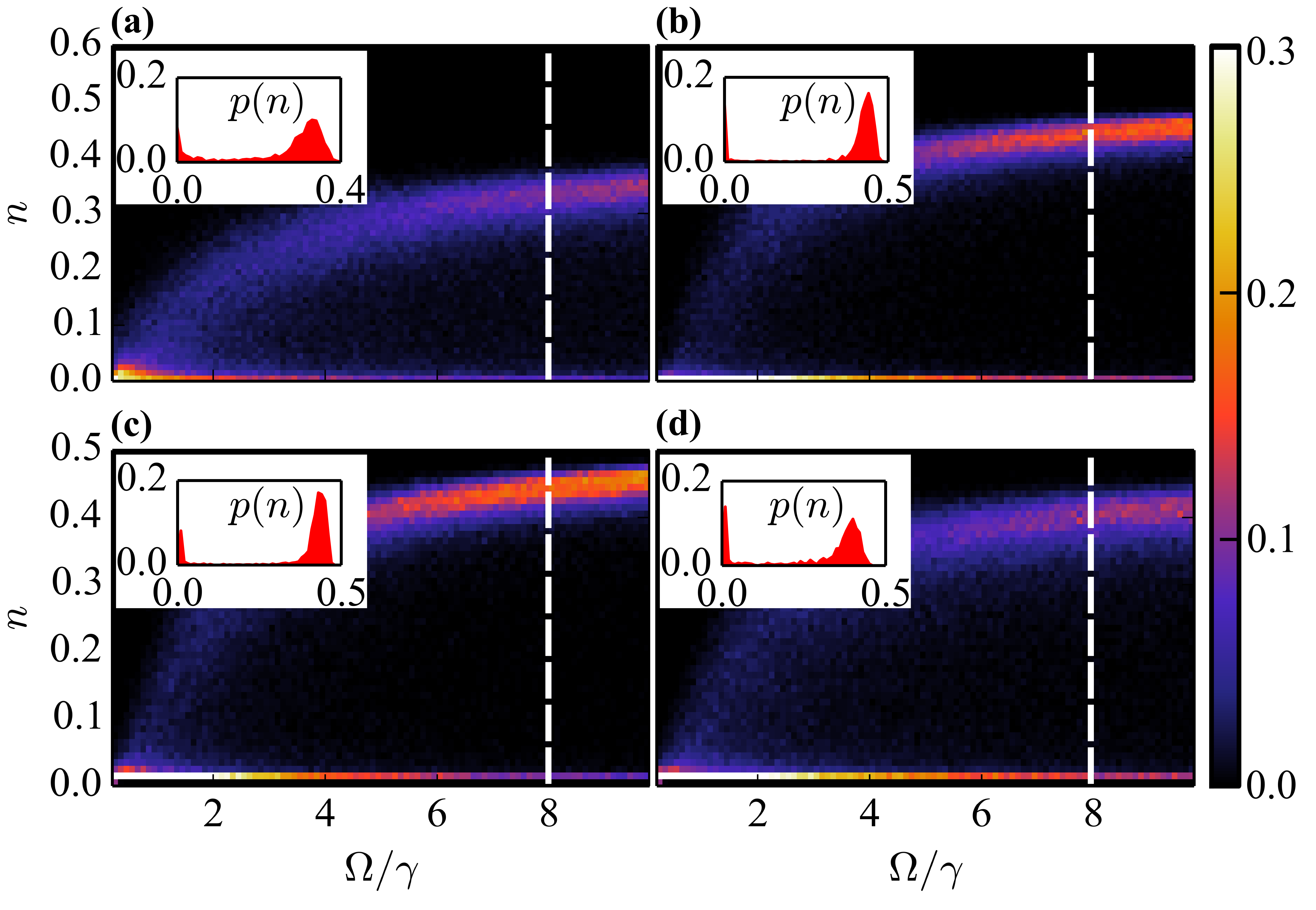}
\caption{First order transition in the quantum limit for a one-dimensional chain of $12$ sites. The main panels are density plots of the distribution $P(n)$ of the excitation density $n$ as a function of $\omega = \Omega /\gamma$. Each inset shows an instance for fixed $\Omega = 8 \gamma$ (indicated by the white dashed line) from the corresponding panel. The four panels correspond to: (a) Main model (Eq.~\eqref{Eq2} plus decay) in the quantum limit $\kappa=0$, (b) effective anti-blockaded model corresponding to Eq.~\eqref{eq:res_rydberg} (plus decay) with $V_{lm}$ set to $0$, (c) Rydberg chain (Eq.~\eqref{eq:RH1} plus decay) with nearest-neighbor interactions only, (d) Rydberg chain with full van-der-Waals tails. All plots display a crossover from an (almost) absorbing state at small $\Omega$ to a state with finite excitation density. For intermediate values of $\Omega$ the counting statistics in the insets feature a bimodal shape which can be regarded as a signature of the anticipated first order phase transition. 
All results are obtained via Quantum Jump Monte Carlo calculations \cite{Q_jump} with averages performed over $1000$ runs. The simulation times are $\gamma t=4$ (a) and $\gamma t=6$ (b-d).  For the computations for the Rydberg systems [panels (c,d)], the remaining parameters have been set to $\Delta=10\Omega=-V$. Note that the colorbar is bounded by $0.3$, despite the peak around the absorbing state exceeding this threshold (for lower values of $\Omega$). This was done in order to improve visibility of the finite-density features.}
\label{fig:rydberg}
\end{figure}

Before addressing further the influence of the differences in the Hamiltonians (\ref{eq:res_rydberg}) and \eqref{Eq2} on the features of the phase diagram, let us first discuss how the required dissipative processes are realized within a Rydberg setting. Spontaneous decay, e.g. the process described by Eq.~\eqref{Eq3}, is to a good degree of approximation realized naturally by the spontaneous emission of photons and a subsequent de-excitation of a Rydberg atom. In practice such events might proceed through a cascade, but within the two-level approximation one may model it by Eq.~\eqref{Eq3}, where $\gamma$ is the radiative decay rate.

Processes similar to the classical branching and coagulation described by Eqs.~\eqref{Eq4} and \eqref{Eq5} can be introduced by exciting Rydberg atoms with a laser source subject to strong phase noise, with a dephasing rate $\Gamma$, and a spatial correlation length that is shorter than the inter-atomic distance. The derivation of such effective classical branching has been discussed extensively in Refs.~\cite{Lesanovsky14, PRL-KinC, Marcuzzi2015} and experimentally confirmed in Ref.~\cite{Valado2015}. Here, we sketch a slightly different derivation of the corresponding dynamics as this allows us to include both coherent as well as incoherent branching in the equations of motion. To this end, we introduce for each atomic position a (bosonic) laser mode $\wh{a}_l$. Following the derivation leading to Eq.~\eqref{eq:res_rydberg}, but without substituting $\wh{a}_l$ by the classical Rabi frequency, we obtain the atom-light interaction Hamiltonian
\begin{eqnarray}
  H_\mathrm{AL}=g\sum_l\wh{\Lambda}_l(\wh{a}_l\si^+_l+\wh{a}^+_l\si_l),
\end{eqnarray}
with coupling $g$. The dynamics of the atom-light density matrix $\bar{\rho}$ is now given (in the interaction picture) by
\begin{eqnarray}
  \dot{\bar{\rho}}&=&-i\lqq   \sum_l g \wh{\Lambda}_l (\si^+_l \wh{a}_l+\si^-_l \wh{a}_l^\dagger),\bar{\rho}   \rqq \nonumber\\
  &&+ \Phi \sum_l\left(  \wh{a}_l^\dagger \wh{a}_l \,\bar{\rho}\, \wh{a}_l^\dagger \wh{a}_l - \frac{1}{2}\left\{ \lt \wh{a}^\dagger_l \wh{a}_l \rt^2 ,\bar{\rho}\right\}\right).
\end{eqnarray}
The final Lindblad dissipator, with rate $\Phi$, describes the laser phase noise \cite{Walls85}.

One can now obtain an effective equation of motion for the atoms by first performing second order perturbation theory in the small parameter $g/\Phi$ along the lines of Ref.~\cite{Marcuzzi2015} (see also \cite{Degenfeld14, Marcuzzi14}). Taking subsequently the expectation value of the light field, i.e. disregarding the back action of the atoms onto the light, the effective master equation for the reduced atomic density matrix reads
\begin{eqnarray*}
  \dot{\rho} &=& \sum_l\frac{4 g^2 \langle \wa^\dagger_l \wa_l \rangle}{\Phi} \left( \wh{\Lambda}_l \si^+_l\rho \wh{\Lambda}_l \si^-_l-\frac{1}{2}\left\{ \wh{\Lambda}^2_l \si^-_l \si^+_l , \rho  \right\}\right)\\
  &+&\sum_l\frac{4 g^2 \langle \wa^\dagger_l \wa_l+1 \rangle}{\Phi} \left( \wh{\Lambda}_l \si^-_l\rho \wh{\Lambda}_l \si^+_l - \frac{1}{2} \left\{\wh{\Lambda}^2_l \si^+_l \si^-_l,\rho \right\}\right).
\end{eqnarray*}
For sufficiently high light intensity, i.e. the photon occupation obeying $\langle \wa^\dagger_l \wa_l \rangle \gg 1$, and a homogenous system, $\langle \wa^\dagger_l \wa_l \rangle = \langle \wa^\dagger_m \wa_m \rangle$ one can define the rate $\kappa=4 g^2 \langle \wa^\dagger_l \wa_l \rangle/\Phi$. The resulting dissipative dynamics thus has the form of Eqs.~\eqref{Eq4} and \eqref{Eq5} where $\PP_l$ is replaced by $\wh{\Lambda}_l$.

Adding all contributions together we find that the dynamics of the Rydberg lattice gas under anti-blockade conditions is approximately described by the master equation
\begin{eqnarray}
  \dot{\rho}&=&-i\left[\Omega\sum_l \wh{\Lambda}_l \si_l^x + \frac{1}{2}{\sum_{km}}^\prime V_{km} \n_l \n_m, \, \rho\right]\label{eq:Rydberg_model}\\
  &&+\sum_l\kappa\left( \wh{\Lambda}_l \si^+_l\rho \wh{\Lambda}_l \si^-_l-\frac{1}{2}\left\{\wh{\Lambda}^2_l \si^-_l \si^+_l , \, \rho \right\}\right)\nonumber\\
  &&+\sum_l\kappa \left( \wh{\Lambda}_l \si^-_l\rho \wh{\Lambda}_l \si^+_l-\frac{1}{2}\left\{\wh{\Lambda}^2_l \si^+_l\si^-_l  , \, \rho \right\}\right)\nonumber\\
  &&+\sum_l\gamma \left( \si^-_l\rho \si^+_l-\frac{1}{2}\left\{ \si^+_l\si^-_l  , \, \rho  \right\}\right).\nonumber
\end{eqnarray}

The projector $\wh{\Lambda}_l$ enables the excitation of a ground state atom if and only if a single neighboring lattice site is excited. On the other hand, the operator $\PP_l$ in \eqref{Eq2} is ``proportional'' to the number of neighboring lattice sites that are excited. The difference between the physical projector $\wh{\Lambda}_l$ and the operator $\PP_l$ is thus expected to become irrelevant in the limit of small densities, when multiple occupancies in the proximity of a lattice site $l$ are unlikely. This should be the case in the entire absorbing state phase and for the active phase sufficiently close to the transition line. As long as one stays in this parameter regime, the substitution $\wh{\Lambda}_l \rightarrow  \PP_l$ in the master equation is justified. More rigorously, in terms of continuous density fields, the difference 
\eq{Diff}{
\Pi_X - \Lambda_X = 2n_X^2+2D\left(n_X\nabla^2n_X-(\nabla n_X)^2\right),
}
or, in higher dimensions,
\be
\begin{split}
	\Pi_X - \Lambda_X & = zn_X \lqq 1 - (1-n_X)^{z-1} \rqq + \\
	&+ D \lqq 1 - (1-n_X)^{z-2} \lt 1 - z n_X \rt \rqq \lt \nabla^2 n_X \rt + \\
	  & - 2D (1-n_X)^{z-3} (1-zn_X) \abs{\nabla n_X}^2  + \\
	  & + (\text{higher order derivatives})  ,
\end{split}
\ee
only includes shifts to $u_4$ and $\mu_4$ in the density action \eqref{Eq32}, and therefore affects neither the qualitative features of the phase diagram nor the universal properties.

Finally, we perform a numerical simulation of the many-body dynamics in order to gain some insight as to whether the predicted phases structure is also present for Rydberg gases. The classical limit $\Omega=0$ was explored in Ref.~\cite{Marcuzzi2015} and indeed signatures of the expected second order phase transition, falling in the DP universality class, have been identified. We will thus focus on the opposite limit in which there is solely quantum branching and coagulation ($\kappa=0$). This situation is far more challenging to treat numerically and only rather small systems can be studied. In Fig.~\ref{fig:rydberg} we show the histogram of the excitation count as a function of $\Omega/\gamma$ for the model studied in the previous Sections, the Rydberg lattice gas with and without van-der-Waals tails as well as for the effective Rydberg model (\ref{eq:Rydberg_model}). All simulations are performed on a one-dimensional chain with periodic boundary conditions and all of them show the onset of the expected first order phase transition which becomes manifest in the bimodal shape of the histogram.

Thus we can expect the Rydberg system to reproduce the physics in the classical and the quantum limit at least in one dimension. In an experiment one might therefore be able to study the competition between quantum and classical fluctuations, and to ultimately probe the physics at the bicritical point shown in the phase diagram (Fig.~\ref{fig:1}).

\section{Conclusion}
\label{sec:concl}
We have introduced a model of driven-dissipative two-level systems with coherent and dissipative branching and coagulation dynamics, which features a unique absorbing state throughout the entire parameter regime. By mapping the dissipative Heisenberg-Langevin equations to a non-equilibrium path integral for the density of the excited atomic levels, we have shown that this model undergoes a phase transition from the absorbing state towards an active, finite excitation density state for sufficiently strong branching rates. In the classical limit, i.e. in the limit of weak coherent branching, the system corresponds to a classical contact process and the absorbing-state phase transition belongs to the universality class of directed percolation. On the other hand, in the quantum limit, i.e. in the limit of vanishing incoherent branching, the phase transition is drastically modified and becomes a discontinuous non-equilibrium first order transition. These two regimes are separated by a bicritical point, which features a continuous absorbing state phase transition, which resembles the tricritical DP class. The dynamics at this point represents the quantum analog of the classical contact process. Performing a functional renormalization group analysis, we have analyzed the critical scaling behavior and characterized the universality class of this quantum contact process below its upper critical dimension $d_c=3$. By showing that the critical scaling regime of the bicritical point is extended in parameter space, we have demonstrated that the quantum contact universality class can be explored experimentally for reasonably large system sizes and with moderate parameter fine tuning. The experimental realization of the quantum contact process with ensembles of laser excited Rydberg atoms opens the door for the exploration of novel quantum and classical non-equilibrium phase transitions in the framework of current cold atom experiments.

\acknowledgments{\emph{Acknowledgments. ---} B.E., M.M.~and I.L.~wish to express their gratitude for the insightful discussions with J.P.~Garrahan and for access to the University of Nottingham High Performance Computing Facility. I.L.~acknowledges that the research leading to these results has received funding from the European Research Council under the European Union's Seventh Framework Programme (FP/2007-2013) / ERC Grant Agreement n. 335266 (ESCQUMA). Further funding was received through the H2020-FETPROACT-2014 grant No.~640378 (RYSQ) and from EPSRC Grant no.\ EP/M014266/1. M.B.~and S.D.~acknowledge funding by the German Research Foundation (DFG) through the Institutional Strategy of the University of Cologne within the German Excellence Initiative (ZUK 81), and by the European Research Council via ERC Grant Agreement n.~647434 (DOQS).

\bibliography{biblio1}

\appendix
\begin{widetext}
\section{General remarks on the Heisenberg-Langevin equations}\label{AppLast}
The quantum master equation \eqref{Eq1} is amenable to exact numerical treatment only for very small system sizes (results of quantum-jump Monte Carlo simulations can be found in Sec.~\ref{ExpSec}). In order to determine the properties of the system in the thermodynamic limit, we undertake in the following an analytical approach aiming at constructing a non-equilibrium path integral. As a first step, we derive the Heisenberg-Langevin equations \cite{Scully} for the one-spin observables of our system. For an arbitrary operator $O$, these are obtained by adding a quantum noise term $\xi^O$ -- whose properties are briefly discussed further below -- to the conjugate master equation $\partial_t O = \mal{S}^\ast O$, where $\mal{S}^\ast$ is the adjoint operator to $\mal{S}$, i.e. it satisfies 
\be
	\trace{O \lt \mal{S} \rho \rt} = \trace{\lt \mal{S}^\ast O  \rt \rho} \quad \forall \, O, \rho,
\ee
and reads
\be
	\mal{S}^\ast O = i[H,O] + \sum_l\mathcal{L}^{(d) \ast}_{l}  O + \sum_l\mathcal{L}^{(b)\ast}_{l} O + \sum_l\mathcal{L}^{(c)\ast}_{l}O.
\ee
The adjoint Liouvillians $\mal{L}^\ast$ are defined in the same way and it is not difficult to see from Eq.~\eqref{eq:struct} that
\be
	\mathcal{L}^\ast O = \sum_m  \lqq  L_m^\dag O L_m - \ha \acomm{L_m^\dag L_m}{O}
	\rqq.
	\label{eq:struct2}
\ee
Including the quantum noise term, the evolution equation for $O$ thus reads
\be
	\partial_t O = i[H,O] + \sum_l\mathcal{L}^{(d) \ast}_{l}  O + \sum_l\mathcal{L}^{(b)\ast}_{l} O + \sum_l\mathcal{L}^{(c)\ast}_{l}O + \xi^O.
	\label{eq:full_O}
\ee
From a physical point of view, the origin of the noise lies in the coupling to the environment which is producing dissipation upon the system; because of this, averaging over it ideally corresponds to averaging over the action of the bath degrees of freedom and constitutes a distinct operation from taking the quantum expectation value $\av{\cdot} = \trace{(\cdot) \rho}$. We shall thereby employ a different notation $\avb{\cdot}$ to indicate it. In order to highlight the significance of the noise term, we first remark that the evolution under $\mal{S}^\ast$ alone (which we denote for brevity by $O^\ast (t) \equiv \rme{\mal{S}^\ast t} O$) does not generally satisfy $\lt O^2 \rt^\ast (t) = O^\ast (t)  \,  O^\ast (t)  $, due to its non-unitary character. The noise is introduced to ensure that this condition is met again once the average is taken, i.e.,
\be
	\avb{ O^2 (t) } = \avb{O(t) O(t)},
\ee
where by $O(t)$ we mean here the operator $O$ evolved according to Eq.~\eqref{eq:full_O}.  
In general, $\xi^O$ is an $O$-dependent, operator-valued random variable whose moments are defined via the consistency relation above. Its average must identically vanish ($\avb{\xi^O} \equiv 0$), which implies that $\avb{O(t)} = O^\ast (t)$. In other words, $\avb{O(t)}$ as an operator evolves under $\mal{S}^\ast$. Note that this must be the case in order to guarantee that the correct state evolution is recovered:
\be
\begin{split}
	\av{  \avb{O(t)} }  =  \trace{  \avb{O(t)} \rho(0)} =  \\
	= \trace{\avb{O(0)}  \rho(t)} = \trace{O  \rho(t)},
\end{split}
\ee
where the last equality comes from the fact that the initial conditions of Eq.~\eqref{eq:full_O} are fixed. Now, since $\rho(t) = \rme{\mal{S}t} \rho(0)$, we have $\avb{O(t)} = \rme{\mal{S}^\ast t} \avb{O(0)}$. In order to determine higher-order correlations of $\xi^O$, one way is to enforce that canonical (anti-)commutation relations are preserved under time evolution. 

In this work, we will follow a different path and derive them instead from the coupling of the system to an auxiliary bath of harmonic oscillators. \changer{This is equivalent to the derivation of the Heisenberg-Langevin equations directly from the microscopic system-bath coupling. We will, however, choose a simplified bath compared to the microscopic one, which produces the same noise operators but is much more convenient and instructive to deal with.}
In this approach, which preserves the commutation relations of all operators and is therefore physically consistent, the noise correlations turn out to be simply the correlation functions of the bath\changer{, which is in agreement with physical intuition.}
\section{Derivation of the density-only action}\label{AppA}
Starting from Eq.~\eqref{Eq31}, we realize that the action is at most quadratic in $\sigma^y, \tilde{\sigma}^y$. Hence, the corresponding functional integration is Gaussian and can be carried out exactly. However, to simplify it further, we recall that, in the quadratic sector, the mass $(\chi + 1)/2$ is always $> 1/2$; therefore, the propagator is strongly gapped in the entire parameter regime and fluctuations $\sim \partial_t, n, \tilde{n}$ are negligibly small. Eliminating these terms and grouping at the end the remaining quadratic and linear ones yields the action
\begin{eqnarray}
S&=&\int_X\tilde{n}_X\left[\left(\partial_t-D\nabla^2+1-\chi\right)n_X + 2 n_XP_Xn_X-\frac{1}{2}\left(\tilde{n}_Xn_X-\tilde{\sigma}^x_X\sigma^x_X\right)\right]\nonumber\\
&&+\int_X\tilde{\sigma}^x_X\left[\left(\partial_t+\frac{\chi+1}{2} + n_X P_X\right)\sigma^x_X-\frac{1}{2}\tilde{\sigma}^x_X\right]\nonumber\\
&&+\int_X\tilde{\sigma}^y_X\left[\frac{\chi+1}{2}
\sigma^y_X-\frac{1}{2}\tilde{\sigma}^y_X\right]\nonumber\\
&&+\int_X \sigma^y_X\left(-\frac{\omega}{\chi}\tilde{n}_XP_Xn_X
 + \frac{\omega}{\chi}\tilde{\sigma^x_X}P_X\sigma^x_X
\right)+\tilde{\sigma}^y_X\left(\frac{2\omega}{\chi}(2n_X-1)P_Xn_X-\frac{\omega}{\chi}\sigma^x_XP_X\sigma^x_X
\right).\label{A1}
\end{eqnarray}
We recall that $P_X = (D \nabla^2 + \chi)$. The $\sigma^y, \tilde{\sigma}^y$ fields can now straightforwardly be integrated out. This replaces the last two lines in the expression above with
\be
	- \int_X \frac{2\omega^2}{\chi^2 (\chi + 1)}  \left\{  \frac{1}{\chi+1} \lqq  \tilde{\sigma}^x_X P_X \sigma^x_X  -\tilde{n}_X P_X n_X  \rqq^2 + \lqq  (4n_X - 2)P_X n_X - \sigma^x_X P_X \sigma^x_X      \rqq   \lqq  \tilde{\sigma}^x_X P_X \sigma^x_X  -\tilde{n}_X P_X n_X  \rqq   \right\} .
\ee
Collecting now the terms according to their order in the $\sigma^x$, $\tilde{\sigma}^x$ fields we find
\be
\begin{split}
	S = &\int_X\tilde{n}_X  \left[  \left(  \partial_t - D\nabla^2 + 1 - \chi \right)n_X+2 \left(  n_X P_X n_X -\frac{2\omega^2}{\chi^2 (\chi+1)} (P_X n_X)^2  \right)   -\frac{1}{2}\tilde{n}_Xn_X     \right]  \\
	 & - \int_X \lqq \frac{2\omega^2}{\chi^2 (\chi+1)^2}\tilde{n}_X^2 (P_X n_X)^2   -\frac{8\omega^2}{\chi^2 (\chi+1)} \tilde{n}_X n_X (P_X n_X)^2 \rqq \\
	 & +\int_X\tilde{\sigma}^x_X  \left[ \left(\partial_t+\frac{\chi+1}{2} + n_X P_X  + \frac{4\omega^2}{\chi^2 (\chi+1)^2} \tilde{n}_X P_X n_X P_X - \frac{4\omega^2}{\chi^2 (\chi+1)} (2n_X - 1) P_X n_X P_X \right)\sigma^x_X-\frac{1}{2}\tilde{\sigma}^x_X\right] \\
	 & - \int_X  \lqq \frac{2 \omega^2}{\chi^2 (\chi+1)} \sigma^x_X P_X \sigma^x_X \tilde{n}_X P_X n_X  \rqq \\
	 & + \int_X  \lqq     \frac{2\omega^2}{\chi^2 (\chi+1)} \tilde{\sigma}^x_X \sigma^x_X       (P_X \sigma_X)^2 -   \frac{2\omega^2}{\chi^2 (\chi+1)^2} \lt \tilde{\sigma}^x_X P_X \sigma_X \rt^2   \rqq 
\end{split}	
\ee
At this level we perform two manipulations. First,the elimination of negligible fluctuations in the quadratic $\sigma^x, \tilde{\sigma}^x$ sector, exploiting the presence of a gap $(\chi+1)/2 > 1/2$. Second, we neglect all spatial fluctuations in cubic and quartic nonlinearities since the corresponding terms are irrelevant in the renormalization group sense (we recall that the dynamic exponent here is $z = 2$). This corresponds to substituting all $P_X \to \chi$. This yields
\begin{eqnarray}
S&=&\int_X\tilde{n}_X\left[\left(\partial_t-D\nabla^2+1-\chi\right)n_X+2\left(\chi-\frac{2\omega^2}{\chi+1}\right)n_X^2 -\frac{1}{2}\tilde{n}_Xn_X\right]-\int_X\left[\frac{2\omega^2}{(\chi+1)^2}\tilde{n}_X^2n_X^2-\frac{8\omega^2}{\chi+1}n^3_X\tilde{n}_X
\right]
\nonumber\\
&&+\int_X\tilde{\sigma}^x_X\left[\frac{\chi+1}{2}\sigma^x_X-\frac{1}{2}\tilde{\sigma}^x_X\right]\nonumber\\
&&-\int_X\left[ \frac{2\omega^2}{(\chi+1)^2}\left(\tilde{\sigma}^x_X \sigma^x_X
\right)^2+\frac{2\omega^2}{\chi+1}\left( \tilde{n}_Xn_X\sigma^x_X\sigma^x_X
- 2\tilde{\sigma^x_X}(\sigma^x_X)^3
\right)\right].
\label{eq:A4}
\end{eqnarray}
The only relevant coupling of the density with the $\sigma^x$ sector is in the $(\sigma^x)^2$ component, which is not gapped due to causality. Therefore this coupling has to be taken seriously. The quadratic part in this sector can be expressed as
\be
	 \matb{c} \sigma^x  \\ \tilde{\sigma}^x  \mate^\intercal   \ast  G_x^{-1}  \ast \matb{c} \sigma^x  \\ \tilde{\sigma}^x  \mate   \quad \quad \text{with}  \quad \quad   \lt G_{x}^{-1} \rt_{XY} = \delta (X-Y) \matb{cc}  -\frac{2\omega^2}{\chi + 1} \tilde{n}_X n_X   &   \frac{\chi + 1}{4}   \\[2mm]   \frac{\chi + 1}{4}    &  -\ha    \mate
\ee
the inverse Green's function.
 Since the $\sigma^x$ field is strongly gapped, it is a good approximation to integrate it out in a quadratic approach. 
We remark here that by doing this the terms generated by $-\Omega \si_l^y \wh{s}_l$ in \eqref{Eq9} (the ones $\propto \tilde{\sigma}^x$ in the third line of \eqref{eq:A4}) disappear, making the original sign (and hence the typo in \cite{Marcuzzi2016}) irrelevant.
The remaining Gaussian path-integral over $\sigma^x$, $\tilde{\sigma}^x$ produces a factor $\lt \det{G_x^{-1}} \rt^{-1/2}$ which can be exponentiated and included in the action as a correction  $\Delta S= \tfrac{1}{2}\log\det(G_x^{-1}) = \tfrac{1}{2} \trace{\log(G_x^{-1})}$. Apart from a field-independent part, which we disregard, this correction reads
\be
	\Delta S = \ha   \int_X  \log \lqq  1 - \frac{16 \omega^2}{(\chi + 1)^3}  \tilde{n}_X n_X  \rqq  =  -  \frac{8 \omega^2}{(\chi + 1)^3}  \tilde{n}_X n_X  - \frac{64 \omega^4}{(\chi + 1)^6} (\tilde{n}_X n_X)^2  + \ldots,
\ee	
where the omitted terms in the Taylor expansion on the r.h.s. are irrelevant in a RG sense.
Collecting all terms, the action is now cast in the form
\begin{eqnarray}
S&=&\int_X\tilde{n}_X\left[\left(\partial_t-D\nabla^2+1-\chi-\frac{8 \omega^2}{(1+\chi)^3}\right)n_X+2\left(\chi-\frac{2\omega^2}{\chi+1}\right)n_X^2  + \frac{8\omega^2}{\chi+1}n^3_X  \right]\nonumber\\
&&-\int_X \tilde{n}_X^2 \left[ \frac{1}{2} n_X +  \left(\frac{2\omega^2}{(\chi+1)^2}+\frac{64 \omega^4}{(\chi + 1)^6}\right) n_X^2 
\right].\label{A3}
\end{eqnarray}
This is the density action \eqref{Eq32} for the quantum contact process, with the parameters corresponding to Eqs.~\eqref{Eq33}-\eqref{Eq37}.

\section{First non-trivial contributions in the loop expansion}
\label{app:Loop}

We briefly report here the calculation of the one- and two-loop corrections represented by the Feynman diagrams in Fig.~\ref{fig:3}. Throughout this section we assume $\Delta > 0$, so that we are adding fluctuations around the absorbing solution. The one-loop contribution $\delta \mu_3^{(1)}$ in panel (a) comes with a combinatoric factor $8$, i.e., $\delta \mu_3^{(1)} = 8 I_3^{(1)}$ with
\be
	I_3^{(1)} = u_3 (\mu_3)^2 \int \frac{\rmd^d q}{(2\pi)^d} \frac{\rmd \freq}{2\pi} (G_{q, \freq})^2 G_{q, -\freq}
\ee
the integral in Eq.~\eqref{Eq41} and $G_{q, \freq}$ as in Eq.~\eqref{Eq42}, which, once the substitution is performed, reads
\be
	i_3 \equiv \frac{I_3^{(1)}}{u_3 \mu_3^2} = \int \frac{\rmd^d q}{(2\pi)^d} \frac{\rmd \freq}{2\pi} \lqq  - i\freq + Dq^2 + \Delta  \rqq^{-2}   \lqq   i\freq + Dq^2 + \Delta  \rqq^{-1} .
\ee
The frequency integration here is particularly simple, as it can be performed in the complex plane, where the structure of the poles of the integrand is apparent. In the upper half plane, e.g., the only one is $\freq = i (Dq^2 + \Delta)$. We thus find
\be
	i_3^{(1)} = \int \frac{\rmd^d q}{(2\pi)^d} \frac{1}{ 4 \lqq  Dq^2 + \Delta  \rqq^2},
\ee
which can be then exponentiated to yield
\be
	i_3^{(1)} = \frac{1}{4} \int \frac{\rmd^d q}{(2\pi)^d} \int_0^\infty \rmd T  \, T \rme{- T(Dq^2 + \Delta)} = \frac{1}{4}  \int_0^\infty \rmd T \,  T  \lt 4 \pi D T  \rt^{-d/2}   \rme{- T \Delta}   =  \frac{\Delta^{\frac{d-4}{2}} \Gamma_E \lt 2 - \frac{d}{2}  \rt}{4 (4\pi D)^{d/2} }   ,
\ee
where $\Gamma_E$ denotes the Euler Gamma function, implying 
\be
	\delta \mu_3^{(1)} = 2 u_3 \mu_3^2  \frac{\Delta^{\frac{d-4}{2}} \Gamma_E \lt 2 - \frac{d}{2}  \rt}{ (4\pi D)^{d/2} }.
\ee

The two-loop contribution in panel (b) comes with a combinatoric factor $2^3 \times 3! = 48$. We take this factor out, so that $\delta \mu_3^{(2)} = 48 I_3^{(2)} = 48 (\mu_3)^3 u_4 \, i_3^{(2)}$ with
\be 
\begin{split}
	i_3^{(2)} &= \int \frac{\rmd^d q \rmd^d p \, \rmd \freq \, \rmd \freqd}{(2\pi)^{2d+2}} (G_{q, \freq})^2 G_{p, \freqd} G_{p, -\freqd} G_{p-q, \freqd - \freq} = \\
	  & \int \frac{\rmd^d q \rmd^d p \, \rmd \freq \, \rmd \freqd}{(2\pi)^{2d+2}} \lqq  -i\freq + Dq^2 + \Delta  \rqq^{-2} \lqq  -i\freqd + Dp^2 + \Delta  \rqq^{-1} \lqq  i\freqd + Dp^2 + \Delta  \rqq^{-1} \lqq  -i(\freqd - \freq) + D(\vec{p} - \vec{q})^2 + \Delta  \rqq^{-1}.
\end{split}
\ee
The integration over the frequencies yields now 
\be
\begin{split}
	i_3^{(2)} &= \int \frac{\rmd^d q \rmd^d p \, \rmd \freq }{(2\pi)^{2d+1}} \lqq  -i\freq + Dq^2 + \Delta  \rqq^{-2} \lqq  2( Dp^2 + \Delta )  \rqq^{-1}  \lqq  i\freq + D p^2 + D(\vec{p} - \vec{q})^2 + 2\Delta   \rqq^{-1}  =  \\
	& \int \frac{\rmd^d q \rmd^d p  }{(2\pi)^{2d}}   \lqq  D p^2 + D(\vec{p} - \vec{q})^2 + Dq^2 + 3\Delta   \rqq^{-2} \lqq  2(Dp^2 + \Delta)   \rqq^{-1}.
\end{split}
\ee
After a translation $\vec{q} \to \vec{q} + \vec{p}/2$ we have
\be
\begin{split}
	i_3^{(2)} & = \int \frac{\rmd^d q \rmd^d p  }{(2\pi)^{2d}}   \lqq  \frac{3}{2} D p^2 + 2Dq^2 + 3\Delta   \rqq^{-2} \lqq  2(Dp^2 + \Delta)   \rqq^{-1} = \\
	& = \frac{1}{18}  \frac{\Gamma(3)}{\Gamma(2) \Gamma(1)} \int_0^1 \rmd x  \int \frac{\rmd^d q \rmd^d p  }{(2\pi)^{2d}}  x  \lqq  x\lt  \ha Dp^2 + \frac{2}{3} Dq^2 + \Delta  \rt + (1-x)  \lt Dp^2 + \Delta  \rt   \rqq^{-3} = \\
	&= \frac{1}{9} \int_0^1 \rmd x \, x \int_0^\infty \rmd T \, T^2  \int \frac{\rmd^d q \rmd^d p  }{(2\pi)^{2d}} \rme{-T \lqq \lt 1 - \ha x  \rt Dp^2 + \frac{2}{3} xDq^2 + \Delta   \rqq} = \\
	& = \frac{1}{9} \int_0^1 \rmd x \, x \int_0^\infty \rmd T \, T^2 \rme{-T\Delta} \lt  4 \pi D T  \rt^{-d} \lqq  \lt 1 - \ha x \rt \frac{2}{3} x \rqq^{-\frac{d}{2}} = \\
	& =  \frac{\Gamma_E(3-d)}{9 (4\pi D)^d} \Delta^{d-3}  \lt \frac{2}{3}  \rt^{-\frac{d}{2}} \Theta_d
\end{split}	
\ee
with 
\be
	\Theta_d = \int_0^1 \rmd x \, x^{1-\frac{d}{2}}    \lt 1 - \ha x \rt^{-\frac{d}{2}} = \sysb{lcc} \frac{\pi - 2}{\sqrt{2}} \approx 0.8072 & \ \text{for} \ & d = 1,  \\[2mm]
	\ln 4 \approx 1.3863  & \ \text{for} \ & d = 2, \\[2mm]
	2 \sqrt{2} \approx 2.8284 & \ \text{for} \ & d = 3.   \syse 
\ee
Collecting all factors, we finally find
\be
	\delta \mu_3^{(2)} = 16 \lt \frac{3}{32 \pi^2}  \rt^{\frac{d}{2}} \Theta_d \,\Gamma_E(3 - d)  \, (\mu_3)^3 u_4 D^{-d} \Delta^{d-3}.
\ee

\section{Optimal path approximation}
\label{app:OPA}

We comment here on the meaning of the optimal path approximation and how this can yield the probability distribution of the density field $n$. We recall that we focus entirely on infinite-time trajectories. This implies that only paths which start or end at a stationary point (or both) are to be accounted. Therefore, we separate the space of values of the density field into a stationary subset (the solutions of $\Gamma'[n] = 0$)
\be
	\mal{S} = \left\{  0, n_{U} = -\frac{u_3 + \sqrt{u_3^2 - 4\Delta u_4}}{2u_4}  , n_{MF} = \frac{-u_3 + \sqrt{u_3^2 - 4\Delta u_4}}{2u_4}  \right\}
\ee
(with $n_U$ the unstable local maximum of $\Gamma$) and the remaining set of transient values $\mal{T} = \R^+ / \mal{S}$. We further divide it into the two basins of attraction $\mal{T}_0 = \mal{T} \cap \left\{ n < n_U  \right\} $ and $\mal{T}_{MF} = \mal{T} \cap \left\{ n > n_U  \right\} $. In the optimal path approximation, the statistical weights $\rme{-S}$ can be now interpreted as rates at which the system can switch, over extensively long times, from very close to a stationary point ($\in \mal{S}$) to any other value of $n$ or vice versa. In this way, the dynamics reduces to an effective stochastic process (in discrete time) between different values of $n$. We shall demonstrate here that this process satisfies detailed balance, which will allow us to extract the probability distribution $p(n)$. Intuitively, one could argue that, since thermal fluctuations would produce a term $\propto T \tilde{n}^2$ in the action, $\Xi[n]$ can be regarded as an effective, $n$-dependent temperature. For constant $\Xi = T$, one would have the ordinary Boltzmann weights $\sim \rme{- \Gamma[n]/T} = \rme{\int^n \Gamma'[n]/T}$ in terms of the energy functional $\Gamma$. Thus, replacing $T$ with $\Xi[n]$, one gets the expression $\rme{\int^n \Gamma' / \Xi}$ introduced in the main text, which can be interpreted as an effective thermal process with temperature $T_{eff} = 1$ and energy functional $\int^n \Gamma'/\Xi$. In the following, we reformulate this picture on more solid grounds. For simplicity, we will omit here the volume factor $V$, since it does not affect the discussion.

\begin{figure}[t]
	\includegraphics[width=0.6\columnwidth]{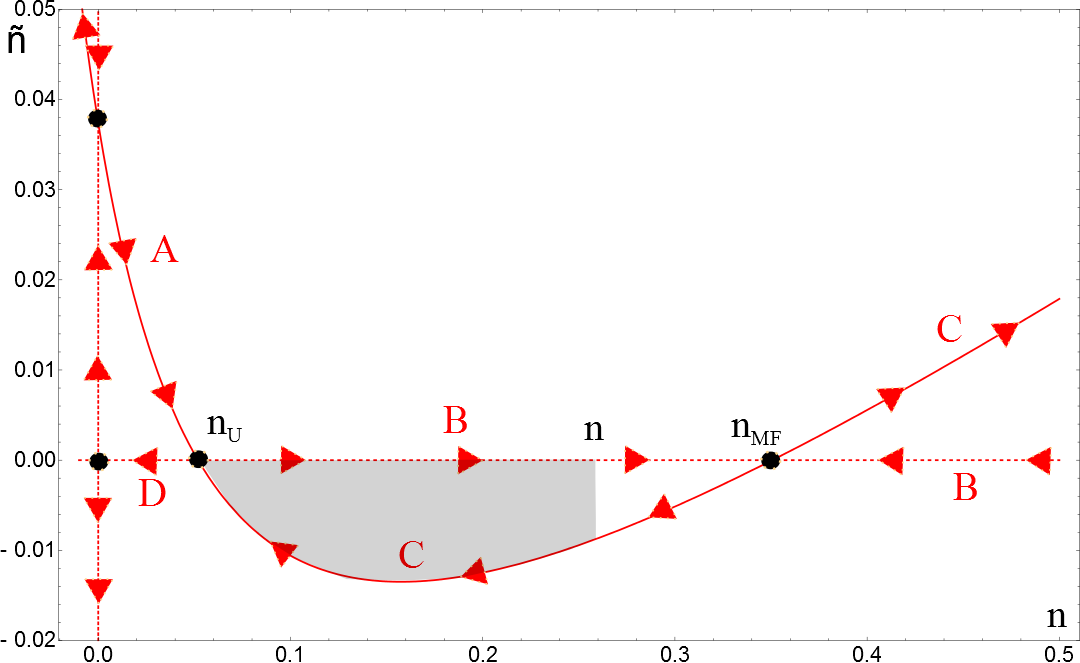}
\caption{Optimal trajectories at $H[n, \tilde{n}] = 0$ in the ($n,\tilde{n}$) phase space for $\chi = 0.05$ and $\omega = 0.3671$. Black circles identify the stationary points, whereas the arrows indicate the direction the dynamics proceeds towards along each path. The dashed lines correspond to the deterministic solution $\tilde{n} = 0$ (divided in paths B and D up to $n = n_{MF}$) and the irrelevant one $n = 0$. The solid line is instead the escape solution $\tilde{n} = \Gamma'[n] / \Xi[n]$ and includes paths A and C. The greyed area corresponds to the action $S(n,n_U)$ which defines the rate $w (n \to n_U) = \rme{-S(n,n_U)}$.}
\label{fig:paths}
\end{figure}
In Fig.~\ref{fig:paths} we display the paths the dynamics can take to connect the various configurations. Within our scheme, the allowed processes can be summarized as in Fig.~\ref{fig:stoch}.
\begin{figure}
\centering
\begin{tikzpicture}
\node[draw,minimum size=1cm,circle] (B) at (2,0) {$0$};
\node[draw,minimum size=1.3cm,circle] (A) at (0,3.4) {$\mal{T}_0$};
\node[draw,minimum size=1cm,circle] (C) at (4,3.4) {$n_U$};
\node[draw,minimum size=1.3cm,circle] (E) at (8,3.4) {$\mal{T}_{MF}$};
\node[draw,minimum size=1cm,circle] (D) at (6,0) {$n_{MF}$};
\draw[-latex] (A) to[bend right=10] node[below,rotate=300] {$D$} (B);
\draw[-latex] (B) to[bend right=10] node[above,rotate=-60] {$A$} (A);
\draw[-latex] (A) to[bend right=10] node[below] {$A$} (C);
\draw[-latex] (C) to[bend right=10] node[above,rotate=0] {$D$} (A);
\draw[-latex] (B) to[bend right=10] node[below,rotate=60] {$A$} (C);
\draw[-latex] (C) to[bend right=10] node[above,rotate=60] {$D$} (B);
\draw[-latex] (D) to[bend right=10] node[above,rotate=-60] {$C$} (C);
\draw[-latex] (C) to[bend right=10] node[below,rotate=-60] {$B$} (D);
\draw[-latex] (D) to[bend right=10] node[below,rotate=60] {$C$} (E);
\draw[-latex] (E) to[bend right=10] node[above,rotate=60] {$B$} (D);
\draw[-latex] (C) to[bend right=10] node[below] {$B$} (E);
\draw[-latex] (E) to[bend right=10] node[above] {$C$} (C);
\end{tikzpicture}
\caption{Stochastic process which mimics the dynamics of large fluctuations in the system. Small nodes in this graph indicate stationary $n$ field configurations, whereas larger ones denote sets of transient values, as defined in the text. Arrows indicate allowed transitions, with the letters referring to the trajectory being followed.}
\label{fig:stoch}
\end{figure}
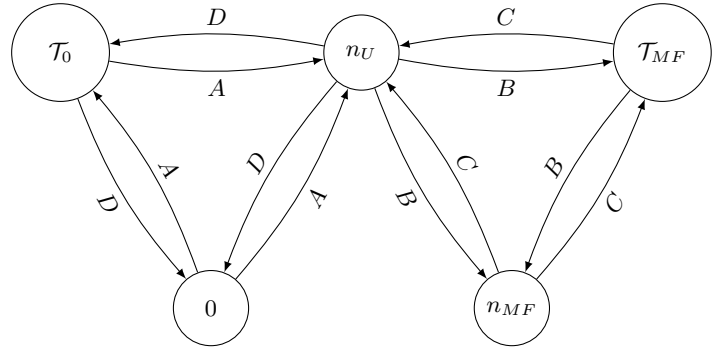
Each allowed transition follows a given path in phase space, as indicated by the arrow labels. One could also include for completeness the asymptotic states at $n = \pm \infty$, but the rates for getting there vanish, as the corresponding actions diverge. The paths can be divided in \emph{deterministic} ($B$ and $D$), which describe fluctuationless relaxation to the minima of the effective potential $\Gamma$, and \emph{escape trajectories} ($C$ and $D$), which allow the latter to be left. On the former, $\tilde{n}_B = \tilde{n}_D = 0$, whereas on the latter $\tilde{n}_A = \tilde{n}_C = \Gamma'[n] / \Xi[n]$, as reported in the main text. For example, a process going from $n \in \mal{T}_{MF}$ to $n_U$, as in Fig.~\ref{fig:paths}, the rate reads, up to a normalization constant which can be fixed at the very end,
\be
	w(n \to n_U) = \rme{- S(n, n_U)} = \exp\left\{-\int_n^{n_U} \tilde{n}_C \, dn \right\},
\ee
corresponding to the exponential of the shaded area. The inverse process occurs with a trivial rate
\be
	w(n_U \to n) = \rme{-S(n_U,n)} = \exp\left\{  - \int_{n_U}^n \tilde{n}_B \, dn  \right\} = 1,
\ee
since $\tilde{n}_B = 0$. For later convenience, we introduce also the functions 
\be
	w(n_1, n_2) = \rme{-S(n_1,n_2)},
\ee
which would correspond to the statistical weights of trajectories going from $n_1$ to $n_2$ along an optimal path in a \emph{finite} time, but for our purposes merely represent an auxiliary definition. Of course, for any allowed transition $n_1 \to n_2$ among the ones sketched in Fig.~\ref{fig:stoch}, $w(n_1, n_2) = w(n_1 \to n_2)$ holds. We emphasize that these functions are path-dependent. However, the paths are uniquely determined by the direction (e.g., for the transition between $n$ and $n_U$ in Fig.~\ref{fig:paths}, $B$ is chosen for going from left to right and $C$ otherwise). Hence, as long as the directionality is maintained, one can directly exploit the usual integral decomposition. In other words, $\forall n \in \lqq  n_1, n_2 \rqq$ one has
\be
	S(n_1,n_2) = S(n_1,n) + S(n,n_2) \quad \text{and also} \quad w(n_1,n_2) = w(n_1,n) w(n,n_2).
\ee
Furthermore, for any (not necessarily ordered) pair $n_1$, $n_2$, the product $w(n_1, n_2) w(n_2, n_1)$ always corresponds to the exponential of minus the geometric area enclosed between the deterministic and escape curves in the interval of extrema $n_1$ and $n_2$. It is important to remark that this is the geometric area, not the signed one arising from Riemann integration (e.g., the shaded area in Fig.~\ref{fig:paths} is taken to be positive).

To prove that detailed balance holds, we wish to verify Kolmogorov's criterion, i.e., the fact that for every finite closed sequence of (allowed) jumps
\be 
	n^{(1)} \to n^{(2)} \to n^{(3)} \to \ldots \to n^{(m-1)} \to n^{(m)} \to n^{(1)}
\ee 
the product of the rates along the loop $\prod_j w(n^{(j)} \to n^{(j+1)})$ is equal to its time-reversed counterpart $\prod_j w(n^{(j+1)} \to n^{(j)})$. First, we permute the $m$ values of $n$ to an increasing sequence $n_1 \leq n_2 \leq n_3 \leq \ldots \leq n_m$. For instance, if we consider, with the same parameters as in Fig.~\ref{fig:paths}, the $m=6$ loop 
\be
	n_U \approx 0.052 \to 0.01 \to 0 \to 0.03 \to 0 \to 0.04 \to n_U 
	\label{eq:example1}
\ee
the ordered sequence would be
\be
	n_1 = 0, \quad n_2 = 0, \quad n_3 = 0.01, \quad n_4 = 0.03 , \quad n_5 = 0.04, \quad n_6 = n_U .
	\label{eq:example2}
\ee
Correspondingly, we introduce the intervals $I_j = \left[ n_j , n_{j+1}  \right)$, $j = 1, \ldots , m-1$ and the areas $A_j$ enclosed between the deterministic and escape paths on each of them, so that $w(n_{j}, n_{j+1}) w(n_{j+1}, n_j) = \exp(-A_j)$. The generic stochastic jump $n^{(j)} \to n^{(j+1)}$ will correspond in the ordered sequence to a jump $n_a \to n_b$ for some given $a$ and $b$. We now decompose this rate on all the covered intervals: 
\be
	w(n^{(j)} \to n^{(j+1)}) = w(n_a \to n_b) = w(n_a,n_b) = \sysb{lcc} \prodl{k=a}{b} w(n_k, n_{k+1})  && \text{if } b > a  \\[2mm]
	\prodl{k=a}{b} w(n_k, n_{k-1})  && \text{if } b < a
	  \syse
\ee
In order for the sequence in configuration space to be closed, i.e. to form a loop, every interval $I_j$ must be covered an equal number of times $M_j$ forward and backward. For example, the interval $I_3 = \lqq 0.01, 0.03  \rt$ in the example \eqref{eq:example2} is covered twice forward (in the jumps $0 \to 0.03$ and $0 \to 0.04$) and twice backward (in the jumps $n_U \to 0.01$ and $0.03 \to 0$), so that $M_3 = 2$. Therefore, we have reduced the product of the rates along the loop to
\be
	\prodl{j=1}{m} w(n^{(j)} \to n^{(j+1)}) = \prodl{k=1}{m-1} \lqq w(n_k, n_{k+1}) w(n_{k+1}, n_k)   \rqq^{M_k} = \exp\left\{ - \suml{k=1}{m-1} M_j A_j  \right\},
\ee
where we took $n^{(m+1)} \equiv n^{(1)}$ for brevity. Therefore, every interval contributes $A_j$ to the action every time it is covered. The extension of the intervals and their multiplicities $M_j$ are thereby the only elements relevant for determining the overall rate. But, time-reversing the loop changes neither. Hence, 
\be
	\prodl{j=1}{m} w(n^{(j)} \to n^{(j+1)}) = \prodl{j=1}{m} w(n^{(j+1)} \to n^{(j)})
\ee
and the process satisfies detailed balance.

Exploiting this property, we can now look for the stationary distribution $P(n)$. Taking $P(0)$ as a reference, we can write
\be
	P(n) = P(0) \frac{w(0,n)}{w(n,0)} =  P(0) \exp \left\{   -S(0,n) + S(n,0)     \right\}.
\ee
It is not difficult to see that the deterministic paths yield no contribution and 
\be
	S(0,n) = \theta(n_U - n) \int_0^{n} \frac{\Gamma'}{\Xi}[n] \, dn + \theta(n - n_U) \int_0^{n_U} \frac{\Gamma'}{\Xi}[n] \, dn + \theta(n-n_{MF}) \int_{n_{MF}}^{n)} \frac{\Gamma'}{\Xi}[n] \, dn,
\ee
while 
\be
	S(n,0) = \theta(n-n_U) \lqq  \theta(n_{MF}-n) \int_{n}^{n_U} \frac{\Gamma'}{\Xi}[n] \, dn + \theta(n - n_{MF}) \int_{n_{MF}}^{n_U} \frac{\Gamma'}{\Xi}[n] \, dn \rqq.
\ee
Summing up the two contributions simply yields
\be
	P(n) = P(0) \exp\left\{  \int_0^n \frac{\Gamma'}{\Xi}[n] \, dn   \right\}
\ee
which, once the volume factor $V$ is reinstated and the normalization made explicit, exactly corresponds to Eq.~\eqref{OPA9} in the main text.


\section{Branching and coagulation noise}
\label{app:B}

\subsubsection{Branching noise}
The Heisenberg equations for branching are
\begin{subequations}
\begin{align}
\partial_t\si^- &= i[H_b + H_{b,\auxn},\si^-]  =  \nol
& =   i\sum_k \alpha_k \lqq  \bb_k^\dag \n_{\auxn} \si^z - \si^-  \si^-_\auxn \bb_{k,\auxn} -  \bb_{k,\auxn}^\dag \si^+_\auxn \si^-      \rqq,\label{Eq14-2}\\
\partial_t \n   &= i[H_b + H_{b,\auxn},\n]  = i \sum_k\alpha_k \n_\auxn \lt  \si^- \bb_k  - \bb_k^{\dagger} \si^+  \rt ,\label{Eq15-2}\\
\partial_t \bb_k  & =  i[H_b + H_{b,\auxn}, \bb_k] = -i \alpha_k \n_\auxn \si^+ - i\nu_k \bb_k.  \label{Eq16-2}
\end{align}
\end{subequations}
The equations for the variables on the neighboring (nn) site have the same structure with the indices exchanged between ``nn'' and ``non-nn'' operators.
Substituting the integral version
\be
	\bb_k(t) = \rme{-i\nu_k t} \bb_k(0) - i\alpha_k \int_0^t \rmd \tau \si^+(\tau) \n_\auxn(\tau) \rme{-i\nu_k (t-\tau)}
\ee
of the last equation in the remaining two yields, after a Born-Markov approximation on the ``deterministic'' part
\begin{equation}
	\partial_t\si^- = -\frac{\kappa}{2}  \si^-   + \x^-_b  
\end{equation}
with
\be
\begin{split}
	\x^-_b (t) &=  i\sum_k \alpha_k \lqq  \bb_k^\dag(0) \rme{i\nu_k t}  \n_{\auxn}  \si^z - \si^-  \si^-_\auxn \bb_{k,\auxn}(0) \rme{-i\nu_k t}  +  \right. \\
	& - \left.  \bb_{k,\auxn}^\dag(0) \rme{i\nu_k t} \si^+_\auxn \si^-  \rqq.
\end{split}
\ee
Again, $\avb{\x_b^-(t)} = 0$ and only contractions of the form $\avb{\bb \bb^\dag}$ (with equal indices) will yield a non-vanishing result, implying, in the Born-Markov approximation, 
\be
\begin{split}
	\avb{\x_b^- (t) \, \x_b^-(t')}& = \avb{\x_b^+ (t) \, \x_b^+ (t')} = 0, \\
	\avb{\x_b^- (t) \, \x_b^+(t')} & = \kappa \delta(t-t') (1 - \n) (1 - \n_\auxn), \\
	\avb{\x_b^+ (t) \, \x_b^-(t')} & = \kappa \delta(t-t') \lt \n_\auxn  + \n (1-\n_\auxn)   \rt 
\end{split}
\ee
or, equivalently,
\be
\begin{split}
	\avb{\x_b^x (t) \, \x_b^x (t')} & = \avb{\x_b^y (t) \, \x_b^y(t')} =  \kappa \delta(t-t'), \\
	\avb{\x_b^x (t) \, \x_b^y (t')} & = - i \kappa\delta(t-t') \lt 1 - 2\n -2\n_\auxn + 2 \n\n_\auxn \rt, \\
	\avb{\x_b^y (t) \, \x_b^x(t')} & = i \kappa\delta(t-t') \lt 1 - 2\n -2\n_\auxn + 2 \n\n_\auxn \rt .
\end{split}
\ee
Similarly, one can work out the equation for the density
\be
	\partial_t \n = \kappa (1-\n) \n_\auxn + \x^n_b
\ee
with
\be
	\x^n_b = i \sum_k \alpha_k \n_\auxn \lt  \si^- \bb_k (0) \rme{-i\nu_k t} - \rme{i\nu_k t} \bb_k^\dag (0) \si^+   \rt .
\ee
The remaining variances thus read
\be
\begin{split}
	\avb{\x_b^n (t) \, \x_b^n (t')} & = \kappa \delta(t - t')  \n_\auxn (1- \n), \\
	\avb{\x_b^n (t) \, \x_b^- (t')} & = -\kappa \delta (t-t') \n_\auxn \si^- , \\
	\avb{\x_b^n (t) \, \x_b^+ (t')} & = \avb{\x_b^- (t) \, \x_b^n (t')} = 0 , \\
	\avb{\x_b^+ (t ) \, \x_b^n (t')} & = -\kappa \delta (t-t') \n_\auxn \si^+ 
\end{split}
\ee
and, in the $(x,y)$ basis,
\be\begin{split}
	\avb{\x_b^x (t) \, \x_b^n (t')} & = -\kappa \delta (t-t') \n_\auxn \si^+ ,\\
	\avb{\x_b^y (t) \, \x_b^n (t')} & = i\kappa \delta (t-t') \n_\auxn \si^+, \\
	\avb{\x_b^n (t) \, \x_b^x (t')} & = -\kappa \delta (t-t') \n_\auxn \si^- , \\
	\avb{\x_b^n (t) \, \x_b^y (t')} & = -i\kappa \delta (t-t') \n_\auxn \si^- .
\end{split}
\ee
Additionally, cross-correlations between neighbors develop, i.e.,
\be
\begin{split}
	\avb{\x_b^- (t) \, \x_{b,\auxn}^- (t')}& = \kappa \delta(t-t') \si^- \si^-_\auxn, \\
	\avb{\x_b^+ (t) \, \x_{b,\auxn}^+ (t')}& = \kappa \delta(t-t') \si^+ \si^+_\auxn, \\
	\avb{\x_b^+ (t) \, \x_{b,\auxn}^- (t')}& = \avb{\x_b^- (t) \, \x_{b,\auxn}^+ (t')} = 0 , \\
	\avb{\x_b^n (t) \, \x_{b,\auxn}^n (t')}& = 0, \\
	\avb{\x_b^n (t) \, \x_{b,\auxn}^+ (t')}& = - \kappa \delta(t-t') \si^+_\auxn (1-\n) ,\\
	\avb{\x_b^- (t) \, \x_{b,\auxn}^n (t')}& = - \kappa \delta(t-t') \si^- (1-\n_\auxn) ,\\
	\avb{\x_b^n (t) \, \x_{b,\auxn}^- (t')}& =  \avb{\x_b^+ (t) \, \x_{b,\auxn}^n (t')} = 0 \\
\end{split}
\ee
which, in the $(x,y)$ basis, read
\be
\begin{split}
	\avb{\x_b^x (t) \, \x_{b,\auxn}^x (t')}& = \frac{\kappa}{2} \delta(t-t') \lt \si^x \si^x_\auxn - \si^y \si^y_\auxn \rt, \\
	\avb{\x_b^y (t) \, \x_{b,\auxn}^y (t')}& = \frac{\kappa}{2} \delta(t-t') \lt \si^y \si^y_\auxn - \si^x \si^x_\auxn \rt, \\
	\avb{\x_b^x (t) \, \x_{b,\auxn}^y (t')}& = \frac{\kappa}{2} \delta(t-t') \lt \si^x \si^y_\auxn + \si^y \si^x_\auxn \rt, \\
	\avb{\x_b^n (t) \, \x_{b,\auxn}^x (t')} & = - \kappa \delta(t-t') \si^+_\auxn (1-\n), \\
	\avb{\x_b^n (t) \, \x_{b,\auxn}^y (t')}& = i\kappa \delta(t-t') \si^+_\auxn (1-\n)   .
\end{split}
\ee
As for decay, we can introduce the vectorial notation
\be 
	\bm{\x}^{\dagger}_{b,l,m}(t)=( \x^x_{b,l}(t), \x^y_{b,l}(t) , \x^n_{b,l}(t), \x^x_{b,m}(t), \x^y_{b,m}(t) , \x^n_{b,m}(t)),
\ee
with $m$ denoting a neighbor of $l$, and write
\be
\begin{split}
	\avb{\bm{\x}_{b,l,m}(t) \bm{\x}^{\dagger}_{b,l',m'}(t')} = \kappa \delta(t - t') \delta_{l,l'} \delta_{m,m'} \times \\[2mm]
	 \times \matb[c]{c|c|c|c|c|c}   1 & -i (2\p_l \p_m - 1) & -  \n_m \si^+_l & \frac{\si^x_l \si^x_m - \si^y_l \si^y_m}{2}  & \frac{\si^x_l \si^y_m  + \si^y_l \si^x_m}{2} & -\si^-_l (1-\n_m)   \\[0.5mm] \hline & & & & &  \\[0.1mm]
	i (2\p_l \p_m - 1)  &  1  &  i \n_m \si^+_l  &  \frac{\si^x_l \si^y_m  + \si^y_l \si^x_m }{2}  &   \frac{\si^y_l \si^y_m  - \si^x_l \si^x_m}{2}   & -  i \si^-_l (1-\n_m)  \\[0.5mm] \hline & & & & &  \\[0.1mm]
	- \n_m \si^-_l  &  -i \n_m \si_l^-  &   \n_m (1 - \n_l)  &  - \si_m^+ (1- \n_l)  & i \si^+_m (1 - \n_l) & 0 \\[0.5mm] \hline & & & & &  \\[0.1mm] 
	 \frac{\si^x_l \si^x_m - \si^y_l \si^y_m}{2}   &   \frac{\si^x_l \si^y_m  + \si^y_l \si^x_m}{2}   &   -  \si_m^- (1 - \n_l)  &   1  &   -i (2\p_l \p_m - 1) & -  \n_l \si^+_m \\[0.5mm] \hline & & & & &  \\[0.1mm]
	 \frac{\si^x_l \si^y_m  + \si^y_l \si^x_m}{2} &   \frac{\si^y_l \si^y_m  - \si^x_l \si^x_m}{2}   &   -  i \si^-_m (1-\n_l)   &  i (2\p_l\p_m - 1)  &  1 &  i\n_l \si^+_m  \\[0.5mm] \hline & & & & &  \\[0.1mm] 
	 -\si_l^+ (1-\n_m)  &  i\si_l^+ (1-\n_m) &   0  &   - \n_l \si^-_m  &   -i\n_l \si^-_m   &  \n_l (1-\n_m) 
	    \mate
\end{split}
\ee
with the shorthand $\p = 1 - \n = \proj{\dar}$. Keeping only the ``leading'' terms in the density operators, the matrix above reduces to
\be
	\matb[c]{c|c|c|c|c|c}   1 & -i  &  -  \n_m \si^+_l & \frac{\si^x_l \si^x_m - \si^y_l \si^y_m}{2}  & \frac{\si^x_l \si^y_m  + \si^y_l \si^x_m}{2} & -\si^-_l    \\[0.5mm] \hline & & & & &  \\[0.1mm]
	i   &  1  &  i \n_m \si^+_l  &  \frac{\si^x_l \si^y_m  + \si^y_l \si^x_m }{2}  &   \frac{\si^y_l \si^y_m  - \si^x_l \si^x_m}{2}   & -  i \si^-_l   \\[0.5mm] \hline & & & & &  \\[0.1mm]
	- \n_m \si^-_l  &  -i \n_m \si_l^-  &   \n_m   &  - \si_m^+   & i \si^+_m  & 0 \\[0.5mm] \hline & & & & &  \\[0.1mm] 
	 \frac{\si^x_l \si^x_m - \si^y_l \si^y_m}{2}   &   \frac{\si^x_l \si^y_m  + \si^y_l \si^x_m}{2}   &   -  \si_m^-   &   1  &   -i  & -  \n_l \si^+_m \\[0.5mm] \hline & & & & &  \\[0.1mm]
	 \frac{\si^x_l \si^y_m  + \si^y_l \si^x_m}{2} &   \frac{\si^y_l \si^y_m  - \si^x_l \si^x_m}{2}   &   -  i \si^-_m    &  i   &  1 &  i\n_l \si^+_m  \\[0.5mm] \hline & & & & &  \\[0.1mm] 
	 -\si_l^+   &  i\si_l^+  &   0  &   - \n_l \si^-_m  &   -i\n_l \si^-_m   &  \n_l  
	    \mate.
\ee
There are no elements which are leading with respect to those of the decay matrix. However, keeping only those of the same order in the density, we still have a non-trivial matrix
\be
	\matb[c]{c|c|c|c|c|c}   1 & -i  & 0 & \frac{\si^x_l \si^x_m - \si^y_l \si^y_m}{2}  & \frac{\si^x_l \si^y_m  + \si^y_l \si^x_m}{2} & -\si^-_l    \\[0.5mm] \hline & & & & &  \\[0.1mm]
	i   &  1  &  0  &  \frac{\si^x_l \si^y_m  + \si^y_l \si^x_m }{2}  &   \frac{\si^y_l \si^y_m  - \si^x_l \si^x_m}{2}   & -  i \si^-_l   \\[0.5mm] \hline & & & & &  \\[0.1mm]
	0 &  0  &   \n_m   &  - \si_m^+   & i \si^+_m  & 0 \\[0.5mm] \hline & & & & &  \\[0.1mm] 
	 \frac{\si^x_l \si^x_m - \si^y_l \si^y_m}{2}   &   \frac{\si^x_l \si^y_m  + \si^y_l \si^x_m}{2}   &   -  \si_m^-   &   1  &   -i  & 0 \\[0.5mm] \hline & & & & &  \\[0.1mm]
	 \frac{\si^x_l \si^y_m  + \si^y_l \si^x_m}{2} &   \frac{\si^y_l \si^y_m  - \si^x_l \si^x_m}{2}   &   -  i \si^-_m    &  i   &  1 &  0  \\[0.5mm] \hline & & & & &  \\[0.1mm] 
	 -\si_l^+   &  i\si_l^+  &   0  &  0  &   0  &  \n_l  
	    \mate.
\ee

\subsubsection{Coagulation noise}
We repeat here the same derivation employed for branching, adapted at the case of coagulation processes. The Heisenberg equations of motion in this case are
\begin{subequations}
\begin{align}
\partial_t\si^- &= i[H_c + H_{c,\auxn},\si^-]  =  \nol
& =   i\sum_k \alpha_k \lqq   \n_{\auxn} \si^z \cc_k    -    \si^-  \si^+_\auxn \cc_{k,\auxn} -  \cc_{k,\auxn}^\dag \si^-_\auxn \si^-      \rqq,\label{Eq14-2}\\
\partial_t \n   &= i[H_c + H_{c,\auxn},\n]  = i \sum_k\alpha_k \n_\auxn \lt  \cc_k^{\dagger} \si^-  - \si^+ \cc_k  \rt ,\label{Eq15-2}\\
\partial_t \cc_k  & =  i[H_c + H_{c,\auxn}, \cc_k] = -i \alpha_k \n_\auxn \si^- - i\nu_k \cc_k.  \label{Eq16-2}
\end{align}
\end{subequations}
The equations for the variables on the neighboring (nn) site have the same structure with the indices exchanged between ``nn'' and ``non-nn'' operators.
Substituting the integral version
\be
	\cc_k(t) = \rme{-i\nu_k t} \cc_k(0) - i\alpha_k \int_0^t \rmd \tau \si^-(\tau) \n_\auxn(\tau) \rme{-i\nu_k (t-\tau)}
\ee
of the last equation in the remaining two yields, after a Born-Markov approximation on the ``deterministic'' part
\begin{equation}
	\partial_t\si^- = -\kappa  \n_\auxn \si^-   + \x^-_c  
\end{equation}
with
\be
\begin{split}
	\x^-_c (t) &=  i\sum_k \alpha_k \lqq  \rme{-i\nu_k t}  \n_{\auxn}  \si^z  \cc_k(0) - \si^-  \si^+_\auxn \cc_{k,\auxn}(0) \rme{-i\nu_k t}  +  \right. \\
	& - \left.  \cc_{k,\auxn}^\dag(0) \rme{i\nu_k t} \si^-_\auxn \si^-  \rqq.
\end{split}
\ee
Again, $\avb{\x_c^-(t)} = 0$ and only contractions of the form $\avb{\cc \cc^\dag}$ (with equal indices) will yield a non-vanishing result, implying, in the Born-Markov approximation, 
\be
\begin{split}
	\avb{\x_c^- (t) \, \x_c^-(t')}& = \avb{\x_c^+ (t) \, \x_c^+ (t')} = 0, \\
	\avb{\x_c^- (t) \, \x_c^+(t')} & = \kappa \delta(t-t') (2 - \n) \n_\auxn, \\
	\avb{\x_c^+ (t) \, \x_c^-(t')} & = \kappa \delta(t-t')  \n_\auxn \n    
\end{split}
\ee
or, equivalently,
\be
\begin{split}
	\avb{\x_c^x (t) \, \x_c^x (t')} & = \avb{\x_c^y (t) \, \x_c^y(t')} =  2\kappa \n_\auxn \delta(t-t'), \\
	\avb{\x_c^x (t) \, \x_c^y (t')} & = - 2 i \kappa\delta(t-t') \lt 1 - \n \rt \n_\auxn , \\
	\avb{\x_c^y (t) \, \x_c^x(t')} & = 2 i \kappa\delta(t-t') \lt 1 - \n \rt \n_\auxn .
\end{split}
\ee
Similarly, one can work out the equation for the density
\be
	\partial_t \n = -\kappa \n \n_\auxn + \x^n_c
\ee
with
\be
	\x^n_c = i \sum_k \alpha_k \n_\auxn \lt   \rme{i\nu_k t} \cc_k^\dag (0) \si^-   -  \si^+ \cc_k (0) \rme{-i\nu_k t}     \rt .
\ee
The remaining variances thus read
\be
\begin{split}
	\avb{\x_c^n (t) \, \x_c^n (t')} & = \kappa \delta(t - t')  \n_\auxn  \n, \\
	\avb{\x_c^n (t) \, \x_c^- (t')} & =  \avb{\x_c^+ (t) \, \x_c^n (t')} = 0, \\
	\avb{\x_c^n (t) \, \x_c^+ (t')} & = \kappa \delta(t-t')  \n_\auxn \si^+ , \\
	\avb{\x_c^- (t ) \, \x_c^n (t')} & = \kappa \delta (t-t') \n_\auxn \si^- 
\end{split}
\ee
and, in the $(x,y)$ basis,
\be\begin{split}
	\avb{\x_c^x (t) \, \x_c^n (t')} & = \kappa \delta (t-t') \n_\auxn \si^- ,\\
	\avb{\x_c^y (t) \, \x_c^n (t')} & = i\kappa \delta (t-t') \n_\auxn \si^-, \\
	\avb{\x_c^n (t) \, \x_c^x (t')} & = \kappa \delta (t-t') \n_\auxn \si^+ , \\
	\avb{\x_c^n (t) \, \x_c^y (t')} & = -i\kappa \delta (t-t') \n_\auxn \si^+ .
\end{split}
\ee
Additionally, cross-correlations between neighbors develop, i.e.,
\be
\begin{split}
	\avb{\x_c^- (t) \, \x_{c,\auxn}^- (t')}& = \avb{\x_c^+ (t) \, \x_{c,\auxn}^+ (t')} = 0 , \\
	\avb{\x_c^+ (t) \, \x_{c,\auxn}^- (t')}& = 0, \\
	\avb{\x_c^- (t) \, \x_{c,\auxn}^+ (t')} & = 2\kappa \delta(t-t')    \si^+_\auxn \si^-   , \\
	\avb{\x_c^n (t) \, \x_{c,\auxn}^n (t')}& = 0, \\
	\avb{\x_c^n (t) \, \x_{c,\auxn}^+ (t')}& =  \kappa \delta(t-t') \si^+_\auxn \n ,\\
	\avb{\x_c^- (t) \, \x_{c,\auxn}^n (t')}& =  \kappa \delta(t-t') \si^- \n_\auxn ,\\
	\avb{\x_c^n (t) \, \x_{c,\auxn}^- (t')}& =  \avb{\x_c^+ (t) \, \x_{c,\auxn}^n (t')} = 0 \\
\end{split}
\ee
which, in the $(x,y)$ basis, read
\be
\begin{split}
	\avb{\x_c^x (t) \, \x_{c,\auxn}^x (t')}& = 2\kappa \delta(t-t')    \si^+_\auxn \si^-  , \\
	\avb{\x_c^y (t) \, \x_{c,\auxn}^y (t')}& = 2\kappa \delta(t-t')    \si^+_\auxn \si^-  , \\
	\avb{\x_c^x (t) \, \x_{c,\auxn}^y (t')}& = -2i\kappa \delta(t-t')    \si^+_\auxn \si^-  , \\
	\avb{\x_c^n (t) \, \x_{c,\auxn}^x (t')} & = \kappa \delta(t-t') \si^+_\auxn \n, \\
	\avb{\x_c^n (t) \, \x_{c,\auxn}^y (t')}& = -i\kappa \delta(t-t') \si^+_\auxn \n  .
\end{split}
\ee
As for decay and branching, we can introduce the vectorial notation
\be 
	\bm{\x}^{\dagger}_{c,l,m}(t)=( \x^x_{c,l}(t), \x^y_{c,l}(t) , \x^n_{c,l}(t), \x^x_{c,m}(t), \x^y_{c,m}(t) , \x^n_{c,m}(t)),
\ee
with $m$ denoting a neighbor of $l$, and write
\be
\begin{split}
	\avb{\bm{\x}_{c,l,m}(t) \bm{\x}^{\dagger}_{c,l',m'}(t')} = \kappa \delta(t - t') \delta_{l,l'} \delta_{m,m'} \times \\[2mm]
	 \times \matb[c]{c|c|c|c|c|c}   2 \n_m &   -2i \n_m (1-\n_l)  &  \n_m \si^-_l & 2\si^+_m \si^-_l & -2i \si^+_m \si_l^-  & \si_l^- \n_m   \\[0.5mm] \hline & & & & &  \\[0.1mm]
	2i \n_m (1-\n_l)  &  2\n_m  &  i \n_m \si^-_l  &  2i \si^-_l \si^+_m  &  2\si^+_m \si^-_l   &   i \si^-_l \n_m  \\[0.5mm] \hline & & & & &  \\[0.1mm]
	 \n_m \si^+_l  &  -i \n_m \si_l^+  &   \n_m  \n_l  &   \si_m^+ \n_l  & -i \si^+_m \n_l & 0 \\[0.5mm] \hline & & & & &  \\[0.1mm] 
	 2\si^+_l \si^-_m   &  -2i\si^+_l \si^-_m   &     \si_m^-  \n_l  &   2 \n_l  &   -2i \n_l (1-\n_m) &  \n_l \si^-_m \\[0.5mm] \hline & & & & &  \\[0.1mm]
	 2i\si^-_m \si^+_l &   2 \si^+_l \si^-_m   &    i \si^-_m \n_l   &  2i\n_l (1-\n_m)  &  2\n_l &  i\n_l \si^-_m  \\[0.5mm] \hline & & & & &  \\[0.1mm] 
	 \si_l^+ \n_m  &  -i\si_l^+ \n_m &   0  &    \n_l \si^+_m  &   -i\n_l \si^+_m   &  \n_l \n_m 
	    \mate,
\end{split}
\ee
where the grid is only meant as a guide to the eye, to help distinguish the various elements. Keeping only the terms which compete with decay in terms of powers of the density operator, we have
\be
\begin{split}
	\matb[c]{c|c|c|c|c|c}   0 &  0  &  0 & 2\si^+_m \si^-_l & -2i \si^+_m \si_l^-  & 0   \\[0.5mm] \hline & & & & &  \\[0.1mm]
	0  &  0  & 0  &  2i \si^-_l \si^+_m  &  2\si^+_m \si^-_l   &  0  \\[0.5mm] \hline & & & & &  \\[0.1mm]
	 0 &  0  &  0  &   0  & 0 & 0 \\[0.5mm] \hline & & & & &  \\[0.1mm] 
	 2\si^+_l \si^-_m   &  -2i\si^+_l \si^-_m   &    0  &  0  &   0 &  0  \\[0.5mm] \hline & & & & &  \\[0.1mm]
	 2i\si^-_m \si^+_l &   2 \si^+_l \si^-_m   &   0   &  0  &  0 &  0  \\[0.5mm] \hline & & & & &  \\[0.1mm] 
	 0  &  0 &   0  &   0  &   0  & 0 
	    \mate.
\end{split}
\ee

\subsubsection{Corrections to the effective action}

Here we show what corrections to the couplings in the effective action \eqref{Eq32} would ensue if we had accounted for branching and coagulation noise as well. First, we are going to disregards all terms higher than quadratic in the $\sigma^{x/y}$, $\tilde{\sigma}^{x/y}$ variables. This already makes coagulation unimportant and leaves us with just branching to analyze. Neglecting derivative terms, the covariance matrix reads
\be
	\avb{\bm{\xi}_{b,X}  \bm{\xi}_{b,Y}} = \kappa \delta (X - Y)  \matb{ccc} 2 & 0 & - \sigma^x_X \\ 0 & 2 & -\sigma^y_X \\ -\sigma^x_X & -\sigma^y_X &  2 n_X     \mate \equiv  \delta(X - Y) \bm{M}_{b,X}
\ee
and produces in the action density a correction
\be
	\Delta S_b = -\ha \bm{\tilde{\sigma}}_X^\intercal \bm{M}_{b,X} \bm{\tilde{\sigma}} = - \kappa \lqq  \lt \tilde{\sigma}^x_X \rt^2 + \lt \tilde{\sigma}^y_X \rt^2   + \tilde{n}_X^2 n_X - \tilde{\sigma^x_X} \tilde{n}_X \sigma^x_X  - \tilde{\sigma^y_X} \tilde{n}_X \sigma^y_X \rqq.
\ee
The latter two addends provide negligible fluctuations to the (gapped) quadratic parts of the $\sigma$ fields and will be disregarded. The simplified action thus reads
\begin{eqnarray}
S&=&\int_X\tilde{n}_X\left[\left(\partial_t-D\nabla^2+1-\chi\right)n_X + 2 n_XP_Xn_X-\frac{1+ 2 \chi}{2}   \tilde{n}_Xn_X  \right]\nonumber\\
&&+\int_X\tilde{\sigma}^x_X\left[ \frac{\chi+1}{2} \sigma^x_X- \frac{ 1 + 2 \chi}{2}\tilde{\sigma}^x_X\right]\nonumber\\
&&+\int_X\tilde{\sigma}^y_X\left[\frac{\chi+1}{2}
\sigma^y_X-\frac{1 + 2 \chi }{2}\tilde{\sigma}^y_X\right]\nonumber\\
&&+\int_X \sigma^y_X\left(-\frac{\omega}{\chi}\tilde{n}_XP_Xn_X
 + \frac{\omega}{\chi}\tilde{\sigma^x_X}P_X\sigma^x_X
\right)+\tilde{\sigma}^y_X\left(\frac{2\omega}{\chi}(2n_X-1)P_Xn_X-\frac{\omega}{\chi}\sigma^x_XP_X\sigma^x_X
\right).\label{eq:with_b}
\end{eqnarray}
Integration over the $\sigma^y$, $\tilde{\sigma}^y$ modes now yields, after simplifying it again along the same lines of the discussion above (i.e., disregarding all fluctuations over gapped parts, the quartic non-linearities in the $\sigma$s and derivatives),
\begin{eqnarray}
S&=&\int_X\tilde{n}_X\left[\left(\partial_t-D\nabla^2+1-\chi\right)n_X+2\left(\chi-\frac{2\omega^2}{\chi+1}\right)n_X^2 -\frac{1 + 2 \chi}{2}\tilde{n}_Xn_X\right]- \int_X\left[\frac{2\omega^2 (1 + 2\chi)}{(\chi+1)^2}\tilde{n}_X^2n_X^2-\frac{8\omega^2}{\chi+1}n^3_X\tilde{n}_X
\right]
\nonumber\\
&&+\int_X\tilde{\sigma}^x_X\left[\frac{\chi+1}{2}\sigma^x_X-\frac{1+2\chi}{2}\tilde{\sigma}^x_X\right]  - \int_X \lqq \frac{2\omega^2 (1 + 2 \chi)}{\chi + 1}\tilde{n}_Xn_X\sigma^x_X\sigma^x_X   \rqq  .
\label{eq:A4}
\end{eqnarray}
Similarly, the integration over $\sigma^x$, $\tilde{\sigma}^x$ produces
\be
	\Delta S = \ha   \int_X  \log \lqq  1 - \frac{16 \omega^2 (1 + 2 \chi)^2}{(\chi + 1)^3}  \tilde{n}_X n_X  \rqq  =  -  \frac{8 \omega^2 (1 + 2 \chi)^2}{(\chi + 1)^3}  \tilde{n}_X n_X  - \frac{64 \omega^4 (1 + 2 \chi)^4}{(\chi + 1)^6} (\tilde{n}_X n_X)^2  + \ldots.
\ee	
The final action would thus read
\begin{eqnarray}
S&=&\int_X\tilde{n}_X\left[\left(\partial_t-D\nabla^2+1-\chi-\frac{8 \omega^2 (1 + 2 \chi)^2}{(1 + \chi)^3}\right) n_X + 2\left(\chi-\frac{2\omega^2}{\chi+1}\right)n_X^2  + \frac{8\omega^2}{\chi+1}n^3_X  \right]\nonumber\\
&&-\int_X \tilde{n}_X^2 \left[ \frac{1+2\chi}{2} n_X +  \left(\frac{2\omega^2 (1 + 2\chi)}{(\chi+1)^2}+\frac{64 \omega^4 (1 + 2\chi)^4}{(\chi + 1)^6}\right) n_X^2 
\right],\label{eq:Sb}
\end{eqnarray}
from which we see that only the gap and the noise vertices get modified according to
\be
\begin{split}
	\Delta &= 1 - \chi - \frac{8 \omega^2 (1 + 2 \chi)^2}{(1 + \chi)^3},   \\
	\Xi_X &= -\frac{1 + 2\chi}{2}n_X  -  \frac{2\omega^2 ( 1+ 2 \chi)}{(\chi+1)^2}\left(1+\frac{32 ( 1 + 2 \chi)^3}{(\chi + 1)^4}\right)n_X^2.
\end{split}
\ee

\section{Discussion of the fluctuationless mean-field equations} \label{app:MF}

Rescaling time by the decay rate $t \to \gamma t$ as we have done in the main text, the mean-field equations read
\begin{subequations}
\begin{align}
	\partial_t n & =  -n +  \lqq \omega \sigma^y + \chi (1 - 2n)    \rqq n  , \label{eq:appn} \\
	\partial_t \sigma^x  & = - \lqq  \omega  \sigma^y   + \frac{\chi + 1}{2}  + \chi  n \rqq \sigma^x   , \\
	\partial_t \sigma^y &  = \omega  (\sigma^x)^2 - \lt  \frac{\chi + 1}{2} + n \chi \rt \sigma^y - 2n \omega (2n-1)  \label{eq:appy}  .
\end{align}
\end{subequations}
Due to the presence of the absorbing state, $n = 0 \,\Rightarrow\, \partial_t n = 0$. Therefore, starting from $n > 0$ (physically meaningful subspace) the dynamics cannot cross to $n<0$ (unphysical subspace) and we can thus safely restrict our considerations to the physical solutions. We first prove that no stationary solution with $\sigma^x \neq 0$ is physically acceptable. In fact, the only other way to make the middle equation vanish is to set
\be
	\omega \sigma^y = - \frac{\chi + 1}{2}  - \chi  n.	
	\label{eq:wy}
\ee
Since we must require that $n \geq 0$, we have to conclude that $\sigma^y \leq 0$. This means that, in the third equation, the first two addends are positive. Therefore, it can only vanish if the third one is negative, which implies $2n-1 > 0$, i.e., $n > 1/2$. 
However, substituting \eqref{eq:wy} into the first equation yields
\be
	n = \frac{1}{6 \chi} \lt \chi - 3  \rt < \frac{1}{6},
\ee
which is absurd. 
%
%
Hence, $\sigma^x = 0$ in the steady state. Apart from the absorbing phase $n = \sigma^y = \sigma^x = 0$, the other solutions read
\begin{subequations}
\begin{align}
	n &\equiv n(t \to \infty) = \frac{1}{4\omega^2 + 2 \chi^2} \lqq  \lt \omega^2  - \chi    \rt  \pm \sqrt{(\omega^2 - \chi)^2  + (\chi^2 + 2 \omega^2 ) (\chi^2-1)}   \rqq , \\
	\sigma^y &\equiv \sigma^y (t \to \infty) =  \frac{2 n \omega ( 1 - 2 n )}{\frac{\chi + 1}{2} + n \chi}  = \frac{1}{\omega}   \lqq  1 - \chi (1- 2 n)     \rqq  \label{eq:syss}  . 
\end{align}
\end{subequations}
Clearly, $n_+ \geq n_-$ always holds. The stability of the absorbing solution is easily checked: expanding to the leading order around it
\be
	n \to n_{\text{ss}} + \delta n = \delta n \comma  \sigma^{x/y} \to \sigma^{x/y}_{\text{ss}} + \delta \sigma^{x/y} = \delta \sigma^{x/y},
\ee
one finds $\partial_t \delta \sigma^x = -(\chi + 1)/2 \delta \sigma^x$ and
\be
	\partial_t \matb{c} \delta n \\  \delta \sigma^y   \mate = \matb{cc}  \chi - 1 & 0 \\ 2 \omega  & - \frac{\chi + 1}{2}    \mate  \matb{c} \delta n \\  \delta \sigma^y   \mate.
\ee
The eigenvalues of the stability matrix can be easily read off from the diagonal: the second one is always stable (negative), whereas the first one is stable for $\chi < 1$ and unstable for $\chi > 1$. This latter condition identifies an active phase for the system (the smallest fluctuations drives the dynamics away from the empty state).
We recall that the remaining solutions are real for $(\omega^2 - \chi)^2  + (\chi^2 + 2 \omega^2 ) (\chi^2-1) \geq 0$ (which, in particular, is guaranteed for $\chi \geq 1$). Their signs can be classified as follows:
\begin{itemize}
	\item[(I)]  $\chi > 1$ $\,\Rightarrow\,$ $n_+ \geq 0$ and $n_- \leq 0$;	
	\item[(II)]  $\chi < 1$, $\omega^2 < \chi$ $\,\Rightarrow\,$ $n_{\pm} \leq 0$; 
	\item[(III)]  $\chi < 1$, $\omega^2 > \chi$ $\,\Rightarrow\,$ $n_{\pm} \geq 0$.
\end{itemize}
The stability matrix for these solutions is
\be
	M_\pm = \matb{cc}  \omega \sigma^y_{\pm} - 1 + \chi - 4 \chi n_{\pm}   &   \omega n_{\pm}  \\   - \chi \sigma^y_{\pm} + 2 \omega - 8 \omega n_{\pm} & - \frac{\chi + 1}{2}  - \chi n_{\pm}  \mate  = \matb{cc} -2 \chi n_{\pm}   &   \omega n_{\pm}  \\   - \chi \sigma^y_{\pm} + 2 \omega - 8 \omega n_{\pm} & - \frac{\chi + 1}{2}  - \chi n_{\pm}  \mate,
\ee
where the second equality comes from applying Eq.~\eqref{eq:syss} to the first element. Again, we are only interested in the case when the solutions are real-valued and positive and therefore $M$ is a real matrix and has either real or complex-conjugate eigenvalues. Its determinant reads
\be
\begin{split}
	\det M_\pm = 2 n_{\pm}  \lqq  n_{\pm} \lt  2\chi^2 + 4 \omega^2  \rt + \chi - \omega^2      \rqq  = \pm 2 n_\pm \sqrt{(\omega^2 - \chi)^2  + (\chi^2 + 2 \omega^2 ) (\chi^2-1)} ,
\end{split}
\ee
while the trace is
\be
	\trace{M_\pm} = - \frac{\chi + 1}{2}  - 3 \chi n_{\pm}.
\ee
For $n_-$ in region (III) the determinant is negative, implying that the eigenvalues are real and one is positive, and thereby signalling an instability. The solutions $n_+$ in regions (I) and (III) feature instead a positive determinant and a negative trace, implying that both eigenvalues have negative real part, and are consequently both stable under small perturbations. All the remaining solutions are negative and can thus be discarded.
\begin{figure}[h!]
	\includegraphics[width=0.6\columnwidth]{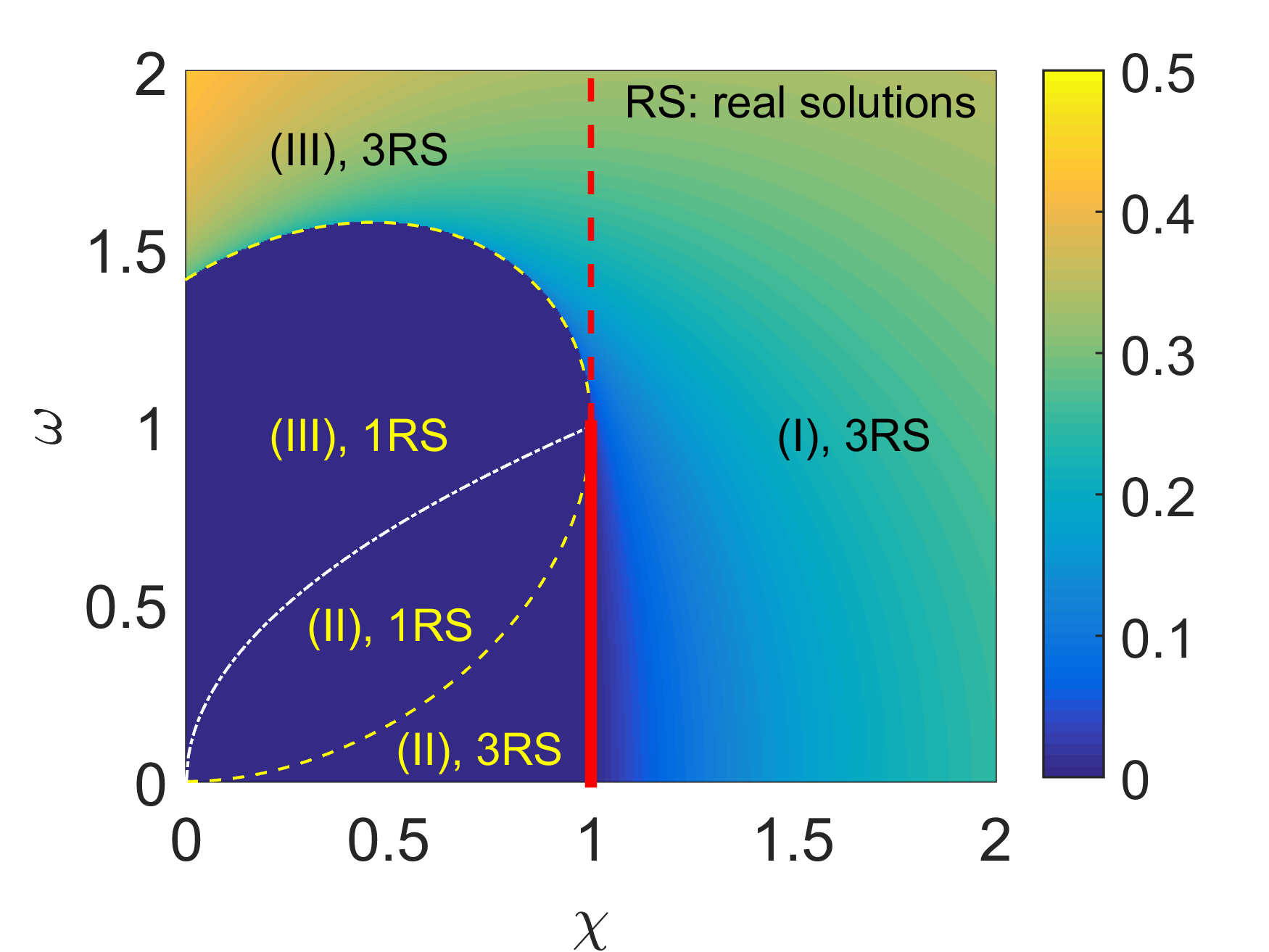}
\caption{Stationary mean-field phase diagram as extracted from Eqs.~\eqref{eq:appn}-\eqref{eq:appy}. The dark blue patch at $\chi < 1$ corresponds to the absorbing phase. The solid red line highlights where the second-order transition occurs to the active phase, whereas the dashed yellow line separates the domain of real solutions (RS), corresponding to the areas labeled with ``$3\text{RS}$'' from where the (non-absorbing) solutions are complex (one real solution, ``$1\text{RS}$'', regions). This line joins the second-order one at the bicritical point $(\omega, \chi) = (1,1)$. Its upper branch denotes the appearance of a second, physically-acceptable, attractive solution, whose density is displayed in the upper left portion of the diagram. The dashed-dotted, white line indicates $\chi = \omega^2$ and separates regions (II) and (III) defined in this Section. Finally, the vertical dashed red line separates the region of stability of the absorbing solution ($\chi < 1$) from the region in which the latter is unstable ($\chi > 1$).
}
\label{fig:app_PD}
\end{figure}
These considerations lead to the phase diagram in Fig.~\ref{fig:app_PD}, which includes the main features discussed in the main text, i.e., the presence of both a second-order and a first-order transition from the absorbing state to finite-density phases and a bicritical point where these two lines join. The regimes (I) - (III) discussed here correspond to the ones introduced in Sec.~\ref{PD}. At this level, however, we have no way to prefer the active solution over the absorbing one in regime (III), highlighting one of the advantages of employing an effective potential description.

\end{widetext}

\end{document}